\documentstyle[preprint,aps,eqsecnum]{revtex}
\begin{document}
\tightenlines \draft

\newcommand{\beq}{\begin{equation}}
\newcommand{\eeq}{\end{equation}}
\newcommand{\bea}{\begin{eqnarray}}
\newcommand{\eea}{\end{eqnarray}}
\newcommand{\cir}{{\buildrel \circ \over =}}

\title{General Covariance and the Objectivity of Space-Time Point-Events:
The Physical Role of Gravitational and Gauge Degrees of Freedom in
General Relativity}

\medskip

\author{Luca Lusanna}

\address{
Sezione INFN di Firenze\\ Polo Scientifico\\ Via Sansone 1\\ 50019
Sesto Fiorentino (FI), Italy\\ E-mail LUSANNA@FI.INFN.IT}

\author{and}

\author{Massimo Pauri}

\address{
Dipartimento di Fisica - Sezione Teorica\\ Universita' di Parma\\
Parco Area Scienze 7/A\\ 43100 Parma, Italy \\E-mail
PAURI@PR.INFN.IT}

\maketitle

\begin{abstract}

This paper deals with a number of technical achievements that are
instrumental for a dis-solution of the so-called {\it Hole
Argument} in general relativity. Such achievements include:

1) the analysis of the {\it Hole} phenomenology in strict
connection with the Hamiltonian treatment of the initial value
problem. The work is carried through in metric gravity for the
class of Christoudoulou-Klainermann space-times, in which the
temporal evolution is ruled by the {\it weak} ADM energy;

2) a re-interpretation of {\it active} diffeomorphisms as {\it
passive and metric-dependent} dynamical symmetries of Einstein's
equations, a re-interpretation which enables to disclose their (up
to now unknown) connection to gauge transformations on-shell;
understanding such connection also enlightens the real content of
the Hole Argument or, better, dis-solves it together with its
alleged "indeterminism";

3) the utilization of the Bergmann-Komar {\it intrinsic
pseudo-coordinates} \cite{1}, defined as suitable functionals of
the Weyl curvature scalars, as tools for a peculiar gauge-fixing
to the super-hamiltonian and super-momentum constraints;

4) the consequent construction of a {\it physical atlas} of
4-coordinate systems for the 4-dimensional {\it mathematical}
manifold, in terms of the highly non-local degrees of freedom of
the gravitational field (its four independent {\it Dirac
observables}). Such construction embodies the {\it physical
individuation} of the points of space-time as {\it point-events},
independently of the presence of matter, and associates a {\it
non-commutative structure} to each gauge fixing or
four-dimensional coordinate system;

5) a clarification of the multiple definition given by Peter
Bergmann \cite{2} of the concept of {\it (Bergmann) observable} in
general relativity. This clarification leads to the proposal of a
{\it main conjecture} asserting the existence of i) special
Dirac's observables which are also Bergmann's observables, ii)
gauge variables that are coordinate independent (namely they
behave like the tetradic scalar fields of the Newman-Penrose
formalism). A by-product of this achievements is the falsification
of a recently advanced argument \cite{3} asserting the absence of
(any kind of) {\it change} in the observable quantities of general
relativity.

6) a clarification of the physical role of Dirac and gauge
variables as their being related to {\it tidal-like} and {\it
inertial-like} effects, respectively. This clarification is mainly
due to the fact that, unlike the standard formulations of the
equivalence principle, the Hamiltonian formalism allows to define
notion of "force" in general relativity in a natural way;

7) a proposal showing how the physical individuation of
point-events could in principle be implemented as an experimental
setup and protocol leading to a "standard of space-time" more or
less like atomic clocks define standards of time.

We conclude that, besides being operationally essential for
building measuring apparatuses for the gravitational field, the
role of matter in the non-vacuum gravitational case is also that
of {\it participating directly} in the individuation process,
being involved in the determination of the Dirac observables. This
circumstance leads naturally to a peculiar new kind of {\it
structuralist} view of the general-relativistic concept of
space-time, a view that embodies some elements of both the
traditional {\it absolutist} and {\it relational} conceptions. In
the end, space-time point-events maintain a {\it peculiar sort of
objectivity}. Some hints following from our approach for the
quantum gravity programme are also given.

\today

\end{abstract}

\vfill\eject
\section{Introduction.}

General Relativity is commonly thought to imply that space-time
points have no {\it intrinsic physical meaning} due to the general
covariance of Einstein's equations. This feature is implicitly
described in standard modern textbooks by the statement that
solutions to the Einstein's equations related by ({\it active})
diffeomorphisms have physically identical properties. Such kind of
equivalence, which also embodies the modern understanding of
Einstein's {\it Hole Argument}, has been named as {\it Leibniz
equivalence} in the philosophical literature by Earman and Norton
\cite{4} and exploited to the effect of arguing against the {\it
substantivalist} and in defense of the {\it relational} conception
of space-time.

This paper is inspired by the belief that {\it Leibniz
equivalence} is not and cannot be the last word about the {\it
intrinsic physical properties} of space-time, well beyond the
needs of the empirical grounding of the theory. Specifically, the
content of the paper should be inscribed in the list of the
various attempts made in the literature to gain an intrinsic {\it
dynamical characterization of space-time points in terms of the
gravitational field itself}, besides and beyond the mathematical
individuation furnished to them by the coordinates. We refer in
particular to old hints offered by Synge, and to the attempts
successively sketched by Komar, Bergmann and Stachel. Actually, we
claim that we have pursued this line of thought till its natural
end.
\medskip

The Hole Argument is naturally spelled out within the
configurational Lagrangian framework of Einstein's theory. Its
very formulation, however, necessarily involve the mathematical
structure of the initial value problem which, on the other hand,
is intractable within that framework. The proper way to deal with
such problem is indeed the ADM Hamiltonian framework with its
realm of Dirac observables and gauge variables. But then the real
difficulty is just the connection between such different
frameworks, particularly from the point of view of {\it
symmetries}.

Our analysis starts off from a nearly forgotten paper by Bergmann
and Komar which enables us to enlighten this correspondence of
symmetries and, in particular, that existing between {\it active
diffeomorphisms} of the configurational approach and {\it gauge
transformations} of the Hamiltonian viewpoint. Understanding this
relation also enlightens the content of the Hole Argument
definitely or, better, dis-solves it together with its alleged
"indeterminism".
\medskip

Although we believe that the topics we discuss on the basis of the
acquired knowledge about the above correspondence of symmetries
are conceptually very significant, they are at the same time
highly technical and complicated so that it is important to keep a
firm grip on the key leading ideas, which are essentially two.
Precisely:
\medskip

\noindent 1) The Komar-Bergmann {\it intrinsic
pseudo-coordinates}, constructed in terms of the eigenvectors of
the Weyl tensor, are exploited to introduce a peculiar
gauge-fixing to the super-hamiltonian and super-momentum
constraints in the canonical reduction of general relativity. The
upshot is that, given an arbitrary coordinate system, the values
of the Dirac observables for the vacuum (i.e., the phase space
intrinsic degrees of freedom of the gravitational field), whose
dependence on space and time is indexed by the chosen coordinates,
{\it reproduces} precisely these latter as the Komar-Bergmann
intrinsic coordinates in the {\it chosen gauge}. This means that
the {\it physical individuation} of manifold's points into {\it
point-events} is realized by the {\it intrinsic components} (just
four!) of the gravitational field, in a gauge dependent fashion.
\medskip

\noindent 2) The original {\it multiple definition} offered by
Bergmann of the concept of "(Bergmann) observable"\cite{2}, a
definition that contains some ambiguities, is spelled out fully.
Such observables are required to be invariant under standard {\it
passive} diffeomorphisms and uniquely {\it predictable} from the
initial data. Once fully clarified, the concept of {\it
predictability} entails that, in order Bergmann's {\it multiple
definition} be consistent, only four of such observables can exist
for the vacuum gravitational field, and can be nothing else than
tensorial Lagrangian counterparts of the Hamiltonian Dirac
observables. We formalize this result and related consequences
into a {\it main conjecture}, which essentially amounts to
claiming the internal consistency of Bergmann's {\it multiple
definition}. Incidentally, this result helps in showing that a
recent claim about the absence of any kind of {\it change} in
general relativity is not mathematically justified.

Other achievements are consequences and refinements of these
leading ideas or reflections and hints originated by their
development.
\bigskip

Previous partial accounts of the material of this paper can be
found in Refs. \cite{5,6}.

\bigskip

The {\it Hole Argument} (see below in this Introduction for
details) has a long history which began in late 1913 and crossed
Einstein's path repeatedly. Einstein found an answer only in late
1915 in terms of the now so-called {\it point-coincidence
argument} which re-established his confidence in general
covariance, but lead him to the final conviction that space and
time must forfeit the {\it last remnant of physical objectivity}.
In Einstein's own words\cite{7}:

\begin{quotation}
{\footnotesize
\noindent
"That the requirement of general
covariance, {\it which takes away from space and time the last
remnant of physical objectivity}, is a natural one, will be seen
from the following reflexion. All our space-time verifications
invariably amount to a determination of space-time coincidences.
If, for example, events consisted merely in the motion of material
points, then ultimately nothing would be observable but the
meetings of two or more of these points. Moreover, the results of
our measurings are nothing but verifications of such meetings of
the material points of our measuring instruments with other
material points, coincidences between the hands of a clock and
points on the clock dial, and observed point-events happening at
the same place at the same time. The introduction of a system of
reference serves no other purpose than to facilitate the
description of the totality of such coincidences".}
\end{quotation}

\noindent At first sight it could seem  from these words that
Einstein simply equated general covariance with the unavoidable
{\it arbitrariness of the choice of coordinates}, a fact that, in
modern language, can be translated into {\it invariance under
passive diffeomorphisms}. The essence of the {\it
point-coincidence argument} (a terminology introduced by Stachel
in 1980), which satisfied Einstein doubts at the end of 1915,
seems to be well in tune with the Machian epistemology he shared
at the time, in particular as regards the ontological privilege of
"{\it bodies}" or "{\it fields}" versus "{\it space}". This
argument, however, offered mainly a pragmatic solution of the
issue and was based on a very idealized model of physical
measurement where all possible observations reduce to the
intersections of the world-lines of observers, measuring
instruments, and measured physical objects. Also, it does not do
full justice to the field concept which is the essence of the
theory. Furthermore, this solution left unexplored some important
aspects of the role played by the metric tensor in the Hole
Argument as well as of the related underlying full mathematical
structure of the theory. On the other hand, that Einstein was not
too much worried about such hidden properties of the metric
tensor, is also shown by the circumstance that the subsequent {\it
geometro-dynamical} re-interpretation of the metric field in more
or less {\it substantival} terms, that took place in the late
years of his life, did not lead Einstein to a re-visitation of the
Hole phenomenology.

That the Hole Argument was in fact a subtler issue that Einstein
seemingly thought in 1915 and that it consisted in much more than
mere arbitrariness in the choice of the coordinates\footnote{In
fact, however, Einstein's argument was not so naive, see below.},
has been revealed by a seminal talk given by John Stachel in 1980
\cite{8}, which gave new life to the original Hole Argument.
Stachel's rediscovery of the Hole Argument was followed by a later
important work in philosophy of science published by John Earman
and John Norton in 1987\cite{4}. This work was directed against
any possible {\it substantivalist} interpretation of the
space-time concept of general relativity and in favor of the view
that such space-time - although not {\it unreal} and not deprived
of any {\it reality} at all - has {\it no reality independent of
the bodies or fields it contains}. Earman-Norton's provocation
raised a rich philosophical debate that is still alive today (the
reader is referred to the works of John Norton for an extensive
bibliography \cite{9,10,11,12}). Getting involved in this debate
is not our aim here: after all, we believe that this debate, which
tends to reproduce, mainly {\it by analogy}, the classical
Newton-Leibniz dispute on the alternative
substantivalism/relationalism (a thorough exposition of this
unending debate can be found in Ref.\cite{13}), occasionally
oversteps the philosophical latitude allowed by the very structure
of general relativity. We shall resume these issues later on and
offer our conclusive view in the Concluding Survey.

\bigskip

This said, one could legitimately ask why are we proposing to
re-discuss the Hole Argument from a technical point of view, given
the fact that the debate alluded above is mainly of philosophical
interest and that all modern technical expositions of general
relativity forget completely the Hole Argument {\it as such}.
Thus, beside stressing the obvious circumstance that any
conceptual clarification of the foundations of general relativity
is welcome in the perspective of the quantum gravity programme, we
owe the reader from the beginning a justification for our proposal
of technical re-visitation of the issue. We shall do it, but only
after having addressed a number of preliminary topics.
\medskip

First of all, before briefly expounding the modern version of the
Hole Argument, let us recall the basic mathematical concept that
underlies it, namely the concept of {\it active diffeomorphism}
and its consequent action on the tensor fields defined on a
differentiable manifold. Our manifold will be the mathematical
manifold $M^4$, the first layer of the would-be physical
space-time of general relativity. Consider a (geometrical or {\it
active}) diffeomorphism $\phi$ which maps points of $M^4$ to
points of $M^4$: $\phi: p {\rightarrow \hspace{.2cm}}\ p' = \phi
\cdot p$, and its tangent map $\phi^{*}$ which maps tensor fields
$T {\rightarrow \hspace{.2cm}} \phi^{*} \cdot T$ in such a way
that $[T](p) {\rightarrow \hspace{.2cm}} [\phi^{*} \cdot T](p)
\equiv [T^{'}](p)$. Then $[\phi^{*} \cdot T](p) =
[T](\phi^{-1}\cdot p)$. It is seen that the transformed tensor
field $\phi^{*} \cdot T$ is a {\it new} tensor field whose
components in general will have at $p$ values that are {\it
different} from those of the components of $T$. On the other hand,
the components of $\phi^* \cdot T$ have at $p'$ - by construction
-  the same values that the components of the original tensor
field $T$ have at $p$: $T^{'}(\phi \cdot p) = T(p)$ or $T'(p) =
T(\phi^{-1}\cdot p)$. The new tensor field $\phi^* \cdot T$ is
called the {\it drag-along} of $T$.

For later use it is convenient to recall that there is another,
non-geometrical - so-called {\it dual} - way of looking at the
active diffeomorphisms, which, incidentally, is more or less the
way in which Einstein himself formulated the original Hole
Argument\footnote{ This point is very interesting from an
historical point of view \cite{10}.}. This {\it duality} is based
on the circumstance that in each region of $M^4$ covered by two or
more charts there is a one-to-one correspondence between an {\it
active} diffeomorpshism and a specific coordinate transformation
(or {\it passive} diffeomorphism). Note, incidentally, that such
{\it duality} between the two types of transformations has often
blurred the important conceptual distinction between {\it active}
and {\it passive} diffeomorphisms, which remains true and should
be kept clearly in mind. The coordinate transformation ${\cal
T}_{\phi}: x(p) {\rightarrow \hspace{.2cm}}\ x'(p) = [{\cal
T}_{\phi}x](p)$ which is {\it dual} to the active diffeomorphism
$\phi$ is defined such that $[{\cal T}_{\phi}x](\phi \cdot p) =
x(p)$. In its essence, this {\it duality} transfers the functional
dependence of the new tensor field in the new coordinate system to
the old system of coordinates. By analogy, the coordinates of the
new system $[x']$ are said to have been {\it dragged-along} with
the {\it active} diffeomorphism $\phi$. A more detailed discussion
or, better, the right mathematical way of looking {\it passively}
at the {\it active} diffeomorphisms will be expounded in Section
II, where the Bergmann-Komar \cite{14} general group $Q$ of {\it
passive dynamical symmetries} of Einstein's equations will be
introduced.
\medskip

Now, the Hole Argument, in its modern version, runs as follows.
Consider a general-relativistic space-time, as specified by the
four-dimensional mathematical manifold $M^4$ and by a metrical
tensor field $g$ which {\it represents at the same time the
chrono-geometrical and causal structure of space-time and the
potential for the gravitational field}. The metric $g$ is a
solution of the generally-covariant Einstein equations. If any
non-gravitational physical fields are present, they are
represented by tensor fields that are also dynamical fields, and
that appear as sources in the Einstein equations.

Assume now that $M^4$ contains a {\it Hole} $\mathcal{H}$: that
is, an open region where all the non-gravitational fields are
zero. On $M^4$ we can prescribe an {\it active} diffeomorphism
$\phi$ that re-maps the points inside $\mathcal{H}$, but blends
smoothly into the identity map outside $\mathcal{H}$ and on the
boundary. Now, {\it just because Einstein's equations are
generally covariant} so that they can be written down as {\it
geometrical relations}, if $g$ is one of their solutions, so is
the {\it drag-along} field $g' = \phi^* \cdot g$. By construction,
for any point $p \in \mathcal{H}$ we have (geometrically) $g'(\phi
\cdot p) = g(p)$, but of course $g'(p) \neq g(p)$ (also
geometrically). Now, what is the correct interpretation of the new
field $g'$? Clearly, the transformation entails an {\it active
redistribution of the metric over the points of the manifold}, so
the crucial question is whether, to what extent, and how the
points of the manifold are primarily {\it individuated}.
\medskip

In the mathematical literature about topological spaces, it is
always implicitly assumed that the entities of the set can be
distinguished and considered separately (provided the Hausdorff
conditions are satisfied), otherwise one could not even talk about
point mappings or homeomorphisms. It is well known, however, that
the points of a homogeneous space cannot have any intrinsic
individuality\footnote{ As Hermann Weyl \cite{15} puts it: ''There
is no distinguishing objective property by which one could tell
apart one point from all others in a homogeneous space: at this
level, fixation of a point is possible only by a {\it
demonstrative act} as indicated by terms like {\it this} and {\it
there}.''}. Quite aside from the phenomenological stance implicit
in Weyl's quoted words, there is only one way to individuate
points at the mathematical level that we are considering: namely
by {\it coordinatization}, a procedure that transfers the
individuality of $4$-tuples of real numbers to the elements of the
topological set. Precisely, we introduce by convention {\it a
standard} coordinate system for the {\it primary individuation} of
the points (like the choice of {\it standards} in metrology).
Then, we can get as many different {\it names}, for what we
consider the same primary individuation, as the coordinate charts
containing the point in the chosen atlas of the manifold. We can
say, therefore, that all the relevant transformations operated on
the manifold $M^4$ (including {\it active} diffeomorphisms which
map points to points), even if viewed in purely geometrical terms,
{\it must} be constructible in terms of coordinate
transformations. In this way we cross necessarily from the domain
of {\it geometry} to that of {\it algebra} (for a related
viewpoint based on more elaborate mathematical structures like
fibered manifolds, see Stachel \cite{16}).

Let us go back to the effect of this {\it primary} mathematical
individuation of manifold points. If we now think of the points of
$\mathcal{H}$ as also {\it physically individuated}
spatio-temporal events even before the metric is defined, then $g$
and $g'$ must be regarded as {\it physically distinct} solutions
of the Einstein equations (after all, as already noted, $g'(p)
\neq g(p)$ at the {\it same} point $p$). This, however, is a
devastating conclusion for the causality, or better, {\it
determinateness} of the theory, because it implies that, even
after we completely specify a physical solution for the
gravitational and non-gravitational fields outside the Hole - for
example, on a Cauchy surface for the initial value problem,
assuming for the sake of argument that this intuitive and
qualitative wording is mathematically correct, see Section V - we
are still {\it unable to predict uniquely the physical solution
within the Hole}. Clearly, if general relativity has to make any
sense as a physical theory, there must be a way out of this
foundational quandary, {\it independently of any philosophical
consideration}.\footnote{In this paper we prefer to avoid the term
{\it determinism}, and replace it by adopting {\it
determinateness} or {\it causality}, because we believe that the
metaphysical flavor of the former tends to overstate the issue at
hand. This is especially true if {\it determinism} is taken in
opposition to {\it indeterminism}, which is not mere absence of
{\it determinism}. The issue of determinism was however the
crucial ingredient of the quoted Earman-Norton's argument
\cite{8}, which, roughly speaking, ran as follows. Under the
substantivalist assumption, (manifold) space-time points possess
an individual reality of their own (i.e., {\it independent} of
bodies and fields of any kind), so that the rearrangement of the
metric field against their background, as envisaged in the Hole
Argument, would originate a true change in the physical state of
space-time. But now - as said before - if diffeomorphically
related metric fields represent different physical states, then
any prescription of initial data outside the Hole would fail to
determine a corresponding solution of the Einstein equations
inside the Hole, because many are equally possible. In this way,
Earman and Norton intend to confront the substantivalist with a
dire dilemma: accept {\it indeterminism}, or abandon {\it
substantivalism}. We want to add here, however, that we find it
rather arbitrary to transcribe Newtonian absolutism (or at least
part of it) into the so-called {\it manifold substantivalism}, no
less than to assert that general relativity is a {\it relational}
theory in an allegedly Leibnizian sense. As emphasized by
Rynasiewicz \cite{17}, the crucial point is that the historical
debate presupposed a clear-cut distinction between {\it matter}
and {\it space}, or between {\it content} and {\it container}; but
by now these distinctions have been blurred by the emergence of
the so-called {\it electromagnetic view of nature} in the late
nineteenth century (for a detailed model-theoretical discussion of
this point see also Friedman's book \cite{18}). Still, although
some might argue (as Earman and Norton do \cite{4}) that the
metric tensor, {\it qua} physical field, cannot be regarded as the
{\it container} of other physical fields, we argue that the metric
field and, in particular, its dynamical degrees of freedom have
{\it ontological priority over all other fields}. This preeminence
has various reasons \cite{19}, but the most important is that the
metric field tells all of the other fields how to behave causally.
We also agree with Friedman \cite{18} that, in consonance with the
general-relativistic practice of not counting the gravitational
energy induced by the metric as a component of the total energy,
{\it we should regard the manifold, endowed with its metric, as
space-time, and leave the task of representing matter to the
stress-energy tensor}. It is just because of this priority, beside
the fact that the Hole {\it is} pure gravitational field, that we
maintain, unlike other authors (see for example Ref.\cite{20}),
that the issue of the individuation of points of the manifold as
{\it physical} point-events should be discussed primarily in the
context of the vacuum gravitational field, without any recourse to
non-gravitational entities, except perhaps at the operational
level. {\it Nevertheless even matter plays peculiar role in the
process of individuation when present}. We shall come back to such
qualifications in the Concluding Survey.}

In the modern understanding, the most widely embraced escape from
the (mathematical) strictures of the Hole Argument (which is
essentially an update to current mathematical terms of the
pragmatic solution adopted by Einstein), is to {\it deny that
diffeomorphically related  mathematical solutions represent
physically distinct solutions}. With this assumption, {\it an
entire equivalence class of diffeomorphically related mathematical
solutions represents only one physical solution}. This statement,
is implicitly taken as obvious in the contemporary specialized
literature (see, e.g. Ref.\cite{21}), and, as already said, has
come to be called {\it Leibniz equivalence} in the philosophical
literature.

It is seen at this point that the conceptual content of general
covariance is far more deeper than the simple invariance under
arbitrary changes of coordinates. Stachel \cite{22,23}  has given
a very enlightening analysis of the meaning of general covariance
and of its relations with the Hole Argument, expounding the
conceptual consequences of the {\it de facto} Einstein's
acceptance of modern Leibniz equivalence through the
point-coincidence argument. Stachel stresses that asserting that
$g$ and $\phi^* \cdot g$ represent {\it one and the same
gravitational field} is to imply that {\it the mathematical
individuation of the points of the differentiable manifold by
their coordinates has no physical content until a metric tensor is
specified}. In particular, coordinates lose any {\it physical
significance whatsoever} \cite{9}. Furthermore, as Stachel
emphasizes, if $g$ and $\phi^* \cdot g$ must represent the same
gravitational field, they cannot be physically distinguishable in
any way. So when we act on $g$ with an active diffeomorphisms to
create the drag-along field $\phi^* \cdot g$, {\it no element of
physical significance} can be left behind: in particular, nothing
that could identify a point $p$ of the manifold as the {\it same}
point of space-time for both $g$ and $\phi^* \cdot g$. Instead,
when $p$ is mapped onto $p' = \phi \cdot p$, it {\it brings over
its identity}, as specified by $g'(p')= g(p)$. A further important
point made by Stachel is that simply because a theory has
generally covariant equations, it does not follow that the points
of the underlying manifold must lack any kind of physical
individuation. Indeed, what really matters is that there can be no
\emph{non-dynamical individuating field} that is specified
\emph{independently} of the dynamical fields, and in particular
independently of the metric. If this was the case, a
\emph{relative} drag-along of the metric with respect to the
(supposedly) individuating field would be physically significant
and would generate an inescapable Hole problem. Thus, the absence
of any non-dynamical individuating field, as well as of any
dynamical individuating field independent of the metric, is the
crucial feature of the purely gravitational solutions of general
relativity as well as of the very {\it concept} of {\it general
covariance}.

This conclusion led Stachel to the conviction that space-time
points {\it must} be {\it physically} individuated {\it before}
space-time itself acquires a physical bearing, and that the metric
itself plays the privileged role of {\it individuating field}: a
necessarily {\it unique role} in the case of space-time {\it
without matter}. More precisely, Stachel claimed that this
individuating role should be implemented by four invariant
functionals of the metric, already considered by Bergmann and
Komar \cite{1} (see Section IV). However, he did not follow up on
his suggestion. As a matter of fact, as we shall see, the question
is not straightforward.
\bigskip

Let us come to our program of re-visitation of the Hole Argument.
There are many reasons why one should revisit the Hole Argument
nowadays, quite apart from any {\it philosophical} interest. First
of all, this paper is inspired by the conviction that a deeper
mathematical clarification of the Hole Argument is needed anyway
because the customary statement of Leibniz equivalence is too much
synthetic and shallow. As aptly remarked by Michael Friedman
\cite{18}, if we stick to simple Leibniz equivalence, "how do we
describe this physical situation {\it intrinsically} ?". The
crucial point of the Hole issue is that the mathematical
representation of space-time provided by general relativity under
the condition of general covariance evidently contains {\it
superfluous structure} hidden behind Leibniz equivalence and that
this structure must be isolated. At the level of general
covariance, only the equivalence class is physically real so that,
on this understanding, general covariance is invariably an
unbroken symmetry and the physical world is to be described in a
diffeomorphically invariant way. Of course, the price to be paid
is that the values of all fields at manifold points as specified
by the coordinates, are not physically real. One could say
\cite{24} that only the {\it relations} among these field values
are invariant an thereby real. However, this point of view
collapses if we are able to isolate the intrinsic content of
Leibniz equivalence and to individuate physically the manifold's
point in an independent way. On the other hand, this isolation
appears to be required {\it de facto} both by any explicit
solution of Einstein's equation, which requires specification of
the arbitrariness of coordinates, and by the empirical foundation
of the theory: after all any effective kind of measurement
requires in fact a definite physical individuation of space-time
points in terms of physically meaningful coordinates. Summarizing,
it is evident that breaking general covariance is a pre-condition
for the isolation of the superfluous structure hidden within {\it
Leibniz equivalence}.

\medskip

Secondly, the program of the physical individuation of space-time
points sketched by John Stachel must be completed because, as it
will appear evident in Section IV, the mere recourse to the four
functional invariants of the metric alluded to by Stachel cannot
do, by itself, the job of physically individuating space-time
points. Above all, it is essential to realize from the beginning
that - by its very formulation - the {\it Hole Argument} is {\it
inextricably entangled} with the initial value problem although,
strangely enough, it has never been explicitly discussed in that
context in a systematic way. Possibly the reason is that most
authors have implicitly adopted the Lagrangian approach, where the
Cauchy problem is intractable because of the non-hyperbolic nature
of Einstein's equations\footnote{Actually, David Hilbert was the
first person to discuss the Cauchy problem for the Einstein
equations and to realize its connection to the Hole phenomenology
\cite{25}. He discussed the issue in the context of a
general-relativistic generalization of Mie's special relativistic
non linear electrodynamics and pointed out the necessity of fixing
a special geometrically adapted ({\it Gaussian} in his terms, or
geodesic normal as known today) coordinate system to assure the
causality of the theory. In this connection see also
Ref.\cite{26}. However, as again noted by Stachel \cite{27},
Hilbert's analysis was incomplete and neglected important related
problems.} (see Ref.\cite{28} for an updated review).

Our investigation will be based on the Hamiltonian formulation of
general relativity for it is precisely the entanglement referred
to above that provides the right point of attack of the Hole
problem. This is no surprise, after all, since the constrained
Hamiltonian approach is just the {\it only} proper way to analyze
the initial value problem of that theory and to find the {\it
deterministically predictable observables} of general relativity.
It is not by chance that the modern treatment of the initial value
problem within the Lagrangian configurational approach \cite{28}
must in fact mimic the Hamiltonian methods (see more in Section
V).

Finally, only in the Hamiltonian approach can we isolate the {\it
gauge variables}, which carry the descriptive arbitrariness of the
theory, from the {\it Dirac observables} (DO), which are gauge
invariant quantities providing a coordinatization of the reduced
phase space of general relativity, and are subjected to hyperbolic
(and therefore "{\it determinated}" or "{\it causal}" in the
customary sense) evolution equations.

\bigskip

Just in the context of the Hamiltonian formalism, we find the
tools for completing Stachel's suggestion and exploiting the old
proposal advanced by Bergmann and Komar for an intrinsic labeling
of space-time points by means of the eigenvalues of the Weyl
tensor. Precisely, Bergman and Komar, in a series of papers
\cite{29,1,30} introduced suitable invariant scalar functionals of
the metric and its first derivatives as {\it invariant
pseudo-coordinates}\footnote{ Actually, the first suggestion of
specifying space-time points {\it absolutely} in terms of
curvature invariants is due to Synge \cite{31}b}. As already
anticipated, we shall show that such proposal can be utilized in
constructing a peculiar {\it gauge-fixing to the super-hamiltonian
and super-momentum constraints} in the canonical reduction of
general relativity. This gauge-fixing makes the {\it invariant
pseudo-coordinates} into effective {\it individuating fields} by
forcing them to be {\it numerically} identical with ordinary
coordinates: in this way the individuating fields turn the {\it
mathematical} points of space-time into {\it physical
point-events}. Eventually, we discover that what really
individuates space-time points physically are the very {\it
degrees of freedom of the gravitational field}. As a
consequence,\- we advance the {\it ontological} claim that -
physically - Einstein's vacuum space-time is literally {\it
identified} with the autonomous physical degrees of freedom of the
gravitational field, while the specific functional form of the
{\it invariant pseudo-coordinates} matches these latter into the
manifold's points. The introduction of matter has the effect of
modifying the Riemann and Weyl tensors, namely the curvature of
the 4-dimensional substratum, and to allow {\it measuring} the
gravitational field in a geometric way for instance through
effects like the geodesic deviation equation. It is important to
emphasize, however, that the addition of {\it matter} does not
modify the construction leading to the individuation of
point-events, rather it makes it {\it conceptually more
appealing}.

Finally, the procedure of individuation that we have outlined
transfers, as it were, the noncommutative Poisson-Dirac structure
of the Dirac observables onto the individuated point-events. The
physical implications of this circumstance might deserve some
attention in view of the quantization of general relativity. Some
hints for the quantum gravity programme will be offered in the
final Section of the paper (Concluding Survey).

\bigskip

A Section of the paper is devoted to our second main topic: the
clarification of the concept of {\it Bergmann's observable} (BO)
\cite{2}. Bergmann's definition has various facets, namely a {\it
configurational} side having to do with invariance under {\it
passive} diffeomorphisms, an Hamiltonian side having to do with
Dirac's concept of observable, and the property of {\it
predictability} which is entangled with both sides. According to
Bergmann, (his) {\it observables} are passive diffeomorphisms
invariant quantities (PDIQ) "which can be predicted uniquely from
initial data", or "quantities that are invariant under a
coordinate transformation that leaves the initial data unchanged".
Bergmann says in addition that they are further required to be
gauge invariant, a statement that can only be interpreted as
implying that Bergmann's observables are simultaneously DO. Yet,
he offers no explicit demonstration of the compatibility of this
bundle of statements. The clarification of this entanglement leads
us to the proposal of a {\it main conjecture} asserting the i)
existence of special Dirac's observables which are also Bergmann's
observables, as well as to the ii) existence of gauge variables
that are coordinate independent (namely they behave like the
tetradic scalar fields of the Newman-Penrose formalism). A
by-product of this achievement is the falsification of a recently
advanced argument \cite{3} asserting the absence of (any kind of)
change in the observable quantities of general relativity.

Two fundamental independent but conceptually overlapping
interpretational issues of general relativity have been debated in
the recent literature, namely the question of the {\it objectivity
of temporal change}, and that of {\it change} in general. It is
well-known that many authors claim that general relativity entails
the absence of objective {\it temporal change}. Although such
claims are not {\it strictly} related to the issue of point
individuation, if sound {\it in general} they could seriously
weaken the conceptual appeal of our very program. Therefore we are
obliged to take issue against some conclusions of this sort. As to
the first point, for the sake of argument, we will restrict our
remarks to the objections raised by Belot and Earmann \cite{32}
and Earman \cite{3} (see also Refs.\cite{33,34,35} as regards the
so called {\it problem of time} or {\it frozen time}). According
to these authors, the reduced phase space of general relativity is
indeed a frozen space without evolution. We shall argue that their
claim cannot have a general ontological force, essentially because
is model dependent and it does not apply to all families of
Einstein's space-times. We show in particular that
globally-hyperbolic non-compact space-times exist - defined by
suitable boundary conditions and asymptotically flat at spatial
infinity \footnote{Precisely the Christodoulou-Klainermann
space-times \cite{36} we use in this paper.} - that provide an
explicit {\it counterexample} to the {\it frozen time argument}.
The role of the generator of real time evolution in such
space-times is played by the {\it weak} ADM energy, while the
super-hamiltonian constraint has nothing to do with temporal
change and is only the generator of gauge transformations
connecting different admissible 3+1 splittings of space-time. We
argue, therefore, that in these space-times there is neither a
frozen reduced phase space nor a possible Wheeler-De Witt
interpretation based on some local concept of time as in compact
space-times. In conclusion, we claim that our gauge-invariant
approach to general relativity is {\it perfectly adequate to
accommodate real temporal change}, so that all the consequent
developments based on it are immune to ontological criticisms like
those quoted above.

There is, however, a stronger thesis about change that has been
recently defended by John Earman in Ref.\cite{3} and must be
addressed separately. We shall call this thesis the {\it universal
no-change argument}. Indeed, Earman claims that the deep structure
of general relativity linked to Leibniz equivalence {\it and}
Bergmann's definition of observable are such that:

\begin{quotation}
\footnotesize{ \noindent "Bergmann's proposal implies that there
is no physical change, i.e., no change in the observable
quantities, at least not for those quantities that are
constructible in the most straightforward way from the materials
at hand."}
\end{quotation}

\noindent Note that Earman's argument is allegedly independent of
Hamiltonian techniques so that it too does not menace our first
topic {\it directly}. However, if technically viable, it would
imply an inner contradiction of Bergmann's multiple definition of
obervables \cite{2} and, in particular, a flaw in the relation
between Bergmann's and Dirac's notions, whose explanation is the
second of our main goals.

Earman's argument is entirely based on the configurational side of
the definition and depends crucially upon the property of {\it
predictability}. The argument exploits the implications of the
initial value problem of general relativity by referring to a
Cauchy surface $\Sigma_o$ in $M^4$ as specified by an intuitive
geometrical representation within the Lagrangian approach.
Earman's radical conclusion is that the conjunction of general
covariance and {\it predictability} in the above Cauchy sense
implies that, for any observable Bergmann's field $B(p)$, and any
{\it active} diffeomorphism $p^{'} = \phi \cdot p$ that leaves
$\Sigma_o$ and its past fixed, it follows $B(p) = B(\phi^{-1}
\cdot p) = B'(p) \equiv \phi^* B(p)$. Then, since $\Sigma_o$ is
arbitrary, the Bergmann's observable field B(p) must be {\it
constant everywhere} in $M^4$.

We shall argue that this conclusion cannot be reconciled with the
Hamiltonian side of Bergmann's definition of observable in that
concerns {\it predictability}, while the Lagrangian
configurational concept of {\it predictability} is substantially
ambiguous in the case of Einstein's equations. The flaw in the
argument should indeed be traced to the naive way in which the
Cauchy problem is usually dealt with within the configurational
Lagrangian approach.

\bigskip
\noindent

A third result -  obtained again thanks to the virtues of the
Hamiltonian approach -  is something new concerning the overall
role of gravitational and gauge degrees of freedom. Indeed, the
distinction between gauge variables and DO provided by the
Shanmugadhasan \cite{37} transformation (see Section III),
conjoined with the circumstance that the Hamiltonian point of view
brings naturally to a re-reading of geometrical features in terms
of the traditional concept of {\it force}, leads to a by-product
of our investigation that, again, would be extremely difficult to
characterize within the {\it Lagrangian viewpoint} at the level of
the Hilbert action or Einstein's equations. The additional
by-product is something that should be added to the traditional
wisdom of the equivalence principle asserting the local
impossibility of distinguishing gravitational from inertial
effects. Actually, the isolation of the gauge arbitrariness from
the true intrinsic degrees of freedom of the gravitational field
is instrumental to understand and visualize which aspects of the
local effects, showing themselves on test matter, have a {\it
genuine gravitational origin} and which aspects depend solely upon
the choice of the (local) reference frame and could therefore even
be named {\it inertial} in analogy with their non-relativistic
Newtonian counterparts. Indeed, two main differences characterize
the issue of {\it inertial effects} in general relativity with
respect to the non-relativistic situation: the existence of {\it
autonomous degrees of freedom} of the gravitational field
independently of the presence of matter sources, on the one hand,
and the {\it local nature of the general-relativistic reference
systems}, on the other. We shall show that, although the very
definition of {\it inertial forces} (and of {\it gravitational
force} in general) is rather arbitrary in general relativity, it
appears natural to characterize first of all as genuine
gravitational effects those which are directly correlated to the
DO, while the gauge variables appear to be correlated to the
general relativistic counterparts of Newtonian inertial effects.

Another aspect of the Hamiltonian connection "{\it gauge variables
- inertial effects}" is related to the 3+1 splitting of space-time
required for the canonical formalism. Each splitting is associated
with a foliation of space-time whose leaves are {\it Cauchy
simultaneity} space-like hyper-surfaces. While the field of unit
normals to these surfaces identifies a {\it surface-forming
congruence of time-like observers}, the field of the evolution
vectors identifies a {\it rotating congruence of time-like
observers}. Since a variation of the gauge variables modifies the
foliation, the identification of the two congruences of time-like
observers is connected to the fixation of the gauge, namely,
on-shell, to the choice of 4-coordinates. Then a variation of
gauge variables also modifies the inertial effects.
\medskip

It is clear by now that a complete gauge fixing within canonical
gravity has the following implications: i) the choice of a unique
3+1 splitting with its associated foliation; ii) the choice of
well-defined congruences of time-like observers; iii) the {\it
on-shell} choice of a unique 4-coordinate system. In physical
terms this set of choices amount to choosing a {\it network of
intertwined and synchronized local laboratories made up with test
matter} (obviously up to a coherent choice of chrono-geometric
standards). This interpretation shows that, unlike in ordinary
gauge theories where the gauge variables are inessential degrees
of freedom, the concept of reduced phase space is very abstract
and not directly useful in general relativity: it is nothing else
than the space of gravitational equivalence classes each of which
is described by the set of all laboratory networks living in a
gauge orbit. This makes the requirement of an {\it intrinsic}
characterization for the reduced phase space asked by Belot and
Earman \cite{32} rather meaningless.

The only weakness of the previous distinction is that the
separation of the two autonomous degrees of freedom of the
gravitational field from the gauge variables is, as yet, a
coordinate (i.e. gauge) - dependent concept. The known examples of
pairs of conjugate DO are neither coordinate-independent (they are
not PDIQ) nor tensors. Bergmann asserts that the only known method
(at the time) to build BO is based on the existence of
Bergmann-Komar invariant pseudo-coordinates. The results of this
method, however, are of difficult interpretation, so that, in
spite of the importance of this alternative non-Hamiltonian
definition of observables, no explicit determination of them has
been proposed so far. A possible starting point to attack the
problem of the connection of DO with BO seems to be a Hamiltonian
reformulation of the Newman-Penrose formalism \cite{38} (it
contains only PDIQ) employing Hamiltonian null-tetrads carried by
the time-like observers of the congruence orthogonal to the
admissible space-like hyper-surfaces. This suggests the technical
{\it conjecture} that special Darboux bases for canonical gravity
should exist in which the inertial effects (gauge variables) are
described by PDIQ while the autonomous degrees of freedom (DO) are
{\it also} BO.  Note that, since Newman-Penrose PDIQ are tetradic
quantities, the validity of the conjecture would also eliminate
the existing difference between the observables for the
gravitational field and the observables for matter, built usually
by means of the tetrads associated to some time-like observer.
Furthermore, this would also provide a starting point for defining
a metrology in general relativity in a generally covariant
way\footnote{Recall that this is the main conceptual difference
from the non-dynamical metrology of special relativity}, replacing
the empirical metrology \cite{39} used till now. It would also
enable to replace by dynamical matter the {\it test matter} of the
axiomatic approach to measurement theory (see Appendix C).
\medskip

A final step of our analysis consists in suggesting how the
physical individuation of space-time points, introduced at the
conceptual level, could in principle be implemented with {\it a
well-defined empirical procedure, an experimental set-up and
protocol for positioning and orientation}. This suggestion is
outlined in correspondence with the abstract treatment of the
empirical foundation of general relativity as exposed in the
classical paper of Ehlers, Pirani and Schild \cite{40}. The
conjunction of the Hamiltonian treatment of the initial value
problem, with the correlated physical individuation of space-time
points, and the practice of general-relativistic measurement, on
the backdrop of the axiomatic foundation closes, as it were, the
{\it coordinative circuit} of general relativity.
\bigskip

The plan of the paper is the following. In Section II the
Einstein-Hilbert Lagrangian viewpoint and the related local
symmetries are summarized. Particular emphasis is given to the
analysis of the most general group $Q$ of dynamical symmetries of
Enstein's equations (Bergmann-Komar group), and the {\it passive
view} of {\it active} diffeomorphisms is clarified. The ADM
Hamiltonian viewpoint and its related canonical local symmetries
are expounded in Section III. Building on the acquired knowledge
about the structure of $Q$, particular emphasis is given to a
discussion of the general Hamiltonian gauge group and to the
correspondence between {\it active diffeomorphisms} and {\it
on-shell gauge transformations}. Section IV is devoted to the
central issue of the individuation of the {\it mathematical
points} of $M^{4}$ as {\it physical point-events} by means of a
peculiar gauge-fixing to Bergmann-Komar {\it intrinsic
pseudo-coordinates}. A digression on the concept of BO and the
criticism of the {\it frozen time} and {\it universal no-change}
arguments are the content of Section V where our {\it main
conjecture} is advanced concerning the relations between DO and
BO. The results obtained in Sections III, IV and V about the
canonical reduction lead naturally to the physical interpretation
of the DO and the gauge variables as characterizing {\it
tidal-like} and {\it inertial-like} effects, respectively: this is
discussed in Section VI. An outline of the empirical {\it
coordinative circuit} of general relativity is sketched in Section
VII. A Concluding Survey containing some hints in view of the
quantum gravity programme and three Appendices complete the paper.
\bigskip

\vfill\eject

\section{The Einstein-Hilbert Lagrangian Viewpoint and the Related
 Local Symmetries.}

Given a pseudo-Riemannian 4-dimensional manifold $M^4$, the
Einstein-Hilbert action for pure gravity without matter \footnote{
$x^{\mu}$ is a coordinate chart in the atlas of $M^4$.}

\beq
 S_H = \int d^4x\, {\cal L}(x) = \int d^4x \, \sqrt{{}^4g}\, {}^4R,
 \label{II1}
 \eeq

\noindent defines a variational principle for the metric 2-tensor
${}^4g_{\mu\nu}(x)$ over $M^4$. The associated Euler-Lagrange
equations are Einstein's equations

\beq
 {}^4G_{\mu\nu}(x) {\buildrel {def}\over =} {}^4R_{\mu\nu}(x) - {1\over 2}\, {}^4R(x)\,
 {}^4g_{\mu\nu}(x)  = 0.
 \label{II2}
 \eeq

Although - as clarified in the Introduction - we should look at
the above equations as pure {\it mathematical} relations, we will
discuss the counting of the degrees of freedom following the
procedure applied in any classical singular field theory.

As well known, the action (\ref{II1}) is invariant under general
coordinate transformations. In other words, the {\it passive
diffeomorphisms} ${}_PDiff\, M^4$ are {\it local Noether
symmetries} (second Noether theorem) of the action. This has the
consequence that:
\bigskip

i) Einstein's equations are form invariant under coordinate
transformations (a property usually named {\it general
covariance});

ii) the Lagrangian density ${\cal L}(x)$ is singular, namely its
Hessian matrix has vanishing determinant.

\bigskip This in turn entails that:

i) four of the ten Einstein equations are {\it Lagrangian
constraints}, namely restrictions on the Cauchy data;

ii) four combinations of Einstein's equations and their gradients
vanish identically ({\it Bianchi identities}).
\bigskip

In conclusion, there are only two {\it dynamical} second-order
equations depending on the {\it accelerations} of the metric
tensor. As a consequence, the ten components ${}^4g_{\mu\nu}(x)$
of the metric tensor are functionals of two "{\it deterministic}"
dynamical degrees of freedom and eight further degrees of freedom
which are left completely undetermined by Einstein's equations
{\it even once the Lagrangian constraints are satisfied}. This
state of affairs makes the treatment of both the Cauchy problem of
the non-hyperbolic system of Einstein's equations and the
definition of observables extremely complicated within the
Lagrangian context \cite{28}. Let us stress that precisely this
arbitrariness is the source of what has been interpreted as {\it
indeterminism} of general relativity in the debate about the Hole
Argument.

Since passive diffeomorphisms play the role of Lagrangian gauge
transformations, {\it a complete Lagrangian gauge fixing amounts
to a definite choice of the coordinates  on $M^4$}, a choice
which, on the other hand, is necessary in order to explicitly
solve the Einstein partial differential equations.

On the other hand, in modern terminology, general covariance
implies that {\it a physical solution of Einstein's equations}
properly corresponds to a {\it 4-geometry}, namely the equivalence
class of all the 4-metric tensors, solutions of the equations,
written in all possible 4-coordinate systems. This equivalence
class is usually represented by the quotient ${}^4Geom = {}^4Riem
/ {}_PDiff\, M^4$, where ${}^4Riem$ denotes the space of metric
tensors solutions of Einstein's equations. Then, any two {\it
inequivalent} Einstein space-times are different 4-geometries.

\bigskip

Besides local Noether symmetries of the action, Einstein's
equations, considered as a set of partial differential equations,
have their own {\it dynamical symmetries} \cite{41} which only
partially overlap with the former ones. Let us stress that:

i) a dynamical symmetry is defined only on the space of solutions
of the equations of motion, namely it is an {\it on-shell}
concept;

ii) only a subset of such symmetries (called {\it Noether
dynamical symmetries}) can be extended {\it off-shell} in the
variational treatment of action principles. The {\it passive
diffeomorphisms} ${}_PDiff\, M^4$ are just an instantiation of
Noether dynamical symmetries of Einstein's equations.

iii) among the dynamical symmetries of Einstein's equations, there
are all the {\it active diffeomorphisms} ${}_ADiff\, M^4$, which,
as said in the Introduction, are the essential core of Einstein's
{\it Hole Argument} \footnote{Note that a subset of {\it active
diffeomorphisms} are the {\it conformal isometries}, i.e. those
conformal transformations which are also active diffeomorphisms,
namely ${}^4\tilde g = \Omega^2\, {}^4g \equiv \phi^*\, {}^4g$ for
some $\phi \in {}_ADiff\, M^4$ with $\Omega$ strictly positive.
Since the Hilbert action is not invariant under the conformal
transformations which are not {\it ordinary isometries} (i.e.
conformal isometries with $\Omega = 1$ for which ${\cal L}_X\,
{}^4g = 0$, if $X$ is the associated Killing vector field), only
these latter are Noether dynamical symmetries.}.

Yet, according to Stachel \cite{8}, it is just the dynamical
symmetry nature of {\it active diffeomorphisms} that expresses the
real {\it physically relevant} content of {\it general
covariance}. On the other hand the natural Noether symmetries of
the Hilbert action are the {\it passive diffeomorphisms}. This
dualism active-passive has been a continuous source of confusion
and ambiguity in the literature. We claim, however, that a
clarification of the issue can be drawn from a nearly forgotten
paper by Bergmann and Komar \cite{11} \footnote{ See, however,
Ref.\cite{42}.} in which it is shown that the biggest group $Q$ of
{\it passive dynamical symmetries} of Einstein's equations is not
${}_PDiff\, M^4$ [$x^{{'}\, \mu} = f^{\mu}(x^{\nu})$] but instead
a larger group of transformations of the form

\beq
 Q: x^{{'}\, \mu} = f^{\mu}(x^{\nu}, {}^4g_{\alpha\beta}(x)),
 \label{II3}
 \eeq

\noindent which, of course is thereby extended to tensors in the
standard way: ${}^4g^{'}_{\mu\nu}(x^{'}) = {{\partial
x^{\alpha}}\over {\partial f^{\mu}}}\, {{\partial x^{\beta}}\over
{\partial f^{\nu}}}\, {}^4g_{\alpha\beta}(x)$. It is clear that in
this way we allow for metric dependent coordinate systems, whose
associated 4-metrics are in general different from those
obtainable from a given 4-metric solution of Einstein's equations
by {\it passive diffeomorphisms}: actually, these transformations
map points to points, but associate with a given point $x$ an
image point $x'$ that depends also on the metric field
\footnote{Strictly speaking, Eqs.(\ref{II3}) should be defined as
transformations on the tensor bundle over $M^4$.}. It is
remarkable, however, that not only these new transformed 4-metric
tensors are still solutions of Einstein's equations, but that
indeed they belong to the {it same 4-geometry}, i.e. the same
equivalence class generated by applying all {it passive
diffeomorphisms} to the original 4-metrics: $ {}^4Geom = {}^4Riem
/ Q = {}^4Riem / {}_PDiff\, M^4$. Note, incidentally, that this
circumstance is mathematically possible only because ${}_PDiff\,
M^4$ is a {\it non-normal} sub-group of $Q$. The 4-metrics built
by using passive diffeomorphisms are, as it were, a dense sub-set
of the metrics obtainable by means of the group Q.

There is no clear statement in the literature about the dynamical
symmetry status of the group ${}_ADiff\, M^4$ of {\it active
diffeomorphisms} and their relationship with the group $Q$, a
point which is fundamental for our program. To clarify this point,
let us consider an infinitesimal transformation of the type
(\ref{II3}) connecting a 4-coordinate system $[x^{\mu}]$ to a new
one $[x^{{'}\mu}]$ by means of metric-dependent infinitesimal
descriptors:

\beq
 x^{{'}\, \mu} = x^{\mu} + \delta\, x^{\mu} = x^{\mu} + \xi^{\mu}(x, {}^4g).
 \label{II4}
 \eeq

\noindent   This will induce the usual formal variation of the
metric tensor \footnote{What is relevant here is the {\it local}
variation $\bar \delta \, {}^4g_{\mu\nu}(x) = {\cal
L}_{-\xi^{\gamma}\, \partial_{\gamma}}\, {}^4g_{\mu\nu}(x) =
{}^4g^{'}_{\mu\nu}(x) - {}^4g_{\mu\nu}(x)$ which differs from the
{\it total} variation by a {\it convective} term: $\delta\,
{}^4g_{\mu\nu}(x) = {}^4g^{'}_{\mu\nu}(x^{'}) - {}^4g_{\mu\nu}(x)
= \bar \delta\, {}^4g_{\mu\nu}(x) + \delta\, x^{\gamma}\,
\partial_{\gamma}\, {}^4g_{\mu\nu}(x)$.}

\beq
 \bar \delta\, {}^4g_{\mu\nu} = -\Big( \xi_{\mu ;\nu}(x, {}^4g) +
\xi_{\nu ;\mu}(x, {}^4g) \Big).
 \label{II5}
  \eeq

\noindent If $\bar \delta \, {}^4g_{\mu\nu}(x)$ is now identified
with the local variation of the metric tensor induced by the {\it
drag along} of the metric under an infinitesimal active
diffeomorphism ${}^4g \mapsto {}^4\tilde g$ so that

\beq
 \bar \delta\, {}^4g_{\mu\nu} = \equiv {}^4\tilde g_{\mu\nu}(x) -
{}^4g_{\mu\nu}(x) =  -\Big( \xi_{\mu ;\nu}(x, {}^4g) + \xi_{\nu
;\mu}(x, {}^4g) \Big),
 \label{II6}
 \eeq

\noindent the solution $\xi_{\mu}(x, {}^4g)$ of these Killing-type
equations identifies a corresponding {\it passive} Bergmann-Komar
dynamical symmetry belonging to $Q$. We see that, just as said in
the Introduction, the new system of coordinates $[x^{{'}\mu}]$ is
identified with the {\it drag along} of the old coordinate system.

This result should imply that {\it all the active diffeomorphisms
connected with the identity in ${}_ADiff\, M^4$ can be
reinterpreted as elements of a {\it non-normal} sub-group of
generalized passive transformations in $Q$}. Clearly this
sub-group is disjoint from the sub-group ${}_PDiff\, M^4$: this in
turn is possible because diffeomorphism groups do not possess a
canonical identity. Let us recall, however, that unfortunately
there is no viable mathematical treatment of the diffeomorphism
group in the large.

What has been defined in the Introduction as (Earman-Norton)
Leibniz equivalence of metric tensors ${}^4$ means that an {\it
Einstein (or on-shell, or dynamical) gravitational field} is an
equivalence class of solutions of Einstein's equation {\it modulo}
the dynamical symmetry transformations of ${}_ADiff\, M^4$.
Therefore, we also have

\beq
 {}^4Geom = {}^4Riem / {}_ADiff\, M^4 = {}^4Riem / Q = {}^4Riem /
 {}_PDiff\, M^4.
 \label{II7}
 \eeq

It is clear that a parametrization of the 4-geometries should be
grounded on the two independent dynamical degrees of freedom of
the gravitational field. Within the framework of the Lagrangian
dynamics, however, no algorithm is known for evaluating the
observables of the gravitational field, viz. its two independent
degrees of freedom. The only result we know of is given in
Ref.\cite{36} where, after a study of the index of Einstein's
equations, it is stated that the two degrees of freedom are
locally associated to {\it symmetric trace-free 2-tensors on
two-planes}, suggesting a connection with the Newman-Penrose
formalism \cite{38}.

On the other hand, as we shall see in the next Section, it is the
Hamiltonian framework which has the proper tools to attack these
problems. Essentially, this is due to the fact that the
Hamiltonian methods allow to work {\it off-shell}, i.e., without
immediate transition to the space of solutions of Einstein's
equations. Thus the soldering to the above results is reached {\it
only} at the end of the canonical reduction, when the {\it
on-shell} restriction is made \footnote{Note nevertheless that
even at the Lagrangian level one can define {\it off-shell (or
kinematical) gravitational fields} defined as ${}^4Riem^{'} /
{}_PDiff\, M^4$, where ${}^4Riem^{'}$ are all the possible metric
tensors on $M^4$. Of course only the subset of solutions of
Einstein equations are Einstein gravitational fields.}.

\vfill\eject

\section{The ADM Hamiltonian Viewpoint and The Related Canonical Local
Symmetries.}

This Section provides the technical analysis of the Cauchy problem
and the counting of degrees of freedom within the framework of the
ADM canonical formulation of metric gravity \cite{43}. Recall that
this formulation holds for globally hyperbolic pseudo-Riemannian
4-manifolds $M^4$ which are asynptotically flat at spatial
infinity \footnote{As shown in Ref.\cite{44} and said in the
Introduction, in this case the so-called {\it problem of time} can
be treated in such a way that in presence of matter and in the
special-relativistic limit of vanishing Newton constant, one
recovers the parametrized Minkowski theories equipped with a {\it
global time}: in such theories, however, space-time points are
individuated as point-events in a kinematical and absolute way. Of
course, this result is precluded if space-time is {\it spatially
compact without boundary (or closed)}. Let us remark that
parametrized Minkowski theories give the reformulation of the
dynamics of isolated systems in special relativity on arbitrary
space-like hyper-surfaces, leaves of the foliation associated with
an arbitrary 3+1 splitting and defining a surface-forming
congruence of accelerated time-like observers. In these theories
the embeddings $z^{\mu}(\tau ,\sigma )$ of the space-like
hyper-surfaces are new configuration variables at the Lagrangian
level. However they are {\it gauge variables} because the
Lagrangian is invariant under separate $\tau$- and $\vec
\sigma$-reparametrizations (which are diffeomorphisms). This form
of {\it special relativistic general covariance} implies the
existence of four first class constraints analogous to the
super-hamitonian and super-momentum constraints of ADM canonical
gravity, playing the role of assuring the independence of the
description from the choice of the 3+1 splitting (there is no Hole
phenomenology)}. Unlike the Lagrangian formulation, the
Hamiltonian formalism requires a 3+1 splitting of $M^4$ and a
global time function $\tau$. This entails in turn a foliation of
$M^4$ by space-like hyper-surfaces $\Sigma_{\tau}$ ({\it
simultaneity Cauchy surfaces}), to be coordinatized by {\it
adapted} 3-coordinates $\vec \sigma$ \footnote{An improper vector
notation is used throughout for the sake of simplicity.}. A
canonical formulation with well-defined Poisson brackets requires
in addition the specification of suitable {\it boundary
conditions} at spatial infinity, viz. a definite choice of the
functional space for the fields\footnote{See Ref.\cite{44} for a
detailed discussion of this point. It is pointed out there that in
order to have well defined asymptotic {\it weak and strong ADM
Poincare' charges} (generators of the asymptotic Poincare' group)
all fields must have a suitable direction-independent limit at
spatial infinity. In presence of matter, switching off the Newton
constant reduces these charges to the conserved generators of the
Poincare' group for the isolated system with the same matter. Of
course, in closed space-times, the ADM Poincare' charges do not
exist and the special relativistic limit is lost. }. Finally, one
should not forget the fact that the problem of the boundary
conditions constitutes an intriguing issue within the Lagrangian
approach.

The reader is referred to Appendix A for the relevant notations
and technical developments of the Hamiltonian description of
metric gravity, which requires the use of Dirac-Bergmann
\cite{45,46,47,48,49} theory of constraints (see Refs.\cite{50,51}
for updated reviews).

We start off with replacement of the ten components
${}^4g_{\mu\nu}$ of the 4-metric tensor by the configuration
variables of ADM canonical gravity: the {\it lapse} $N(\tau , \vec
\sigma )$ and {\it shift} $N_r(\tau ,\vec \sigma )$ functions and
the six components of the {\it 3-metric tensor} on
$\Sigma_{\tau}$, ${}^3g_{rs}(\tau , \vec \sigma )$. Einstein's
equations are then recovered as the Euler-Lagrange equations of
the ADM action

\bea
  S_{ADM}&=&\int d\tau \, L_{ADM}(\tau )= \int d\tau d^3\sigma
{\cal L}_{ADM}(\tau ,\vec \sigma )=\nonumber \\
 &=&- \epsilon k\int_{\triangle \tau}d\tau  \, \int d^3\sigma \, \lbrace
\sqrt{\gamma} N\, [{}^3R+{}^3K_{rs}\, {}^3K^{rs}-({}^3K)^2]\rbrace
(\tau ,\vec \sigma ),
 \label{III1}
 \eea

\noindent which differs from Einstein-Hilbert action (\ref{II1})
by a suitable surface term. Here ${}^3K_{rs}$ is the extrinsic
curvature of $\Sigma_{\tau}$, ${}^3K$ its trace, and ${}^3R$ the
3-curvature scalar.

Besides the ten configuration variables listed above, the ADM
functional phase space $\Gamma_{20}$ is {\it coordinatized} by ten
canonical momenta ${\tilde \pi}^N(\tau ,\vec \sigma )$, ${\tilde
\pi}^r_{\vec N}(\tau ,\vec \sigma )$, ${}^3{\tilde \Pi}^{rs}(\tau
, \vec \sigma )$ \footnote{As shown in Ref.\cite{44}, a consistent
treatment of the boundary conditions at spatial infinity requires
the explicit separation of the {\it asymptotic} part of the lapse
and shift functions from their {\it bulk} part: $N(\tau ,\vec
\sigma ) = N_{(as)}(\tau ,\vec \sigma ) + n(\tau , \vec \sigma )$,
$N_r(\tau ,\vec \sigma ) = N_{(as)r}(\tau ,\vec \sigma ) +
n_r(\tau , \vec \sigma )$, with $n$ and $n_r$ tending to zero at
spatial infinity in a direction-independent way. On the contrary,
$N_{(as)}(\tau ,\vec \sigma ) = - \lambda_{\tau}(\tau ) - {1\over
2}\, \lambda_{\tau u}(\tau )\, \sigma^u$ and $N_{(as)r}(\tau ,\vec
\sigma ) = - \lambda_{r}(\tau ) - {1\over 2}\, \lambda_{r u}(\tau
)\, \sigma^u$. The {\it Christodoulou-Klainermann space-times}
\cite{36}, with their {\it rest-frame} condition of zero ADM
3-momentum and absence of super-translations, are singled out by
these considerations. The allowed foliations of these space-times
tend asymptotically to Minkowski hyper-planes in a
direction-independent way and are asymptotically orthogonal to the
ADM four-momentum. They have $N_{(as)}(\tau ,\vec \sigma ) =
\epsilon$, $N_{(as) r}(\tau ,\vec \sigma ) = 0$. Therefore, in
these space-times there are {\it asymptotic inertial time-like
observers} (the {\it fixed stars} or the {\it CMB rest frame}) and
the global mathematical time labeling the Cauchy surfaces can be
identified with their rest time. For the sake of simplicity we
shall ignore these aspects of the theory, with the caveat that the
canonical pairs $N$, ${\tilde \pi}^N$, $N_r$, ${\tilde
\pi}^r_{\vec N}$ should be always replaced by the pairs $n$,
${\tilde \pi}^n$, $n_r$, ${\tilde \pi}^r_{\vec n}$. }. Such
canonical variables, however, are not independent since they are
restricted to the {\it constraint sub-manifold} $\Gamma_{12}$ by
the eight {\it first class} constraints

\bea
 {\tilde \pi}^N(\tau ,\vec \sigma ) &\approx& 0 ,\nonumber \\
 {\tilde \pi}^r_{\vec N}(\tau ,\vec \sigma ) &\approx& 0,\nonumber
 \\
 &&{}\nonumber \\
 {\tilde {\cal H}}(\tau ,\vec \sigma )&=&\epsilon [k\sqrt{\gamma}\,
{}^3R-{1\over {2k \sqrt{\gamma}}} {}^3G_{rsuv}\, {}^3{\tilde
\Pi}^{rs}\, {}^3{\tilde \Pi}^{uv}] (\tau ,\vec \sigma ) \approx
0,\nonumber \\
 {}^3{\tilde {\cal H}}^r(\tau ,\vec \sigma )&=&-2\, {}^3{\tilde
\Pi}^{rs}{}_{| s} (\tau ,\vec \sigma )=-2[\partial_s\, {}^3{\tilde
\Pi}^{rs}+{}^3\Gamma^r_{su} {}^3{\tilde \Pi}^{su}](\tau ,\vec
\sigma ) \approx 0.
 \label{III2}
  \eea

While the first four are {\it primary} constraints, the remaining
four are the super-hamiltonian and super-momentum  {\it secondary}
constraints arising from the requirement that the primary
constraints be constant in $\tau$. More precisely, this
requirement guarantees that, once we have chosen the initial data
inside the constraint sub-manifold $\Gamma_{12}(\tau_o)$
corresponding to a given initial Cauchy surface $\Sigma_{\tau_o}$,
the time evolution does not take them out of the constraint
sub-manifolds $\Gamma_{12}(\tau)$, for $\tau > \tau_o$.

The evolution in $\tau$ is ruled by the Hamilton-Dirac Hamiltonian

\beq
 H_{(D)ADM}=\int d^3\sigma \, \Big[ N\, {\tilde {\cal H}}+N_r\,
{}^3{\tilde {\cal H}}^r + \lambda_N\, {\tilde \pi}^N + \lambda
^{\vec N}_r\, {\tilde \pi}^r_{\vec N}\Big](\tau ,\vec \sigma )
\approx 0,
 \label{III3}
 \eeq

\noindent where $\lambda_N(\tau , \vec \sigma )$ and
$\lambda^r_{\vec N}(\tau ,\vec \sigma )$ are {\it arbitrary Dirac
multipliers} in front of the primary constraints\footnote{These
are four {\it velocity functions} (gradients of the metric tensor)
which are not determined by Einstein's equations. As shown in
Ref.\cite{44}, the correct treatment of the boundary conditions
leads to rewrite Eq.(\ref{III3}) in terms of $n$ and $n_r$ (see
footnotes 17 and 18).}. The resulting hyperbolic system of
Hamilton-Dirac equations has the same solutions of the
non-hyperbolic system of (Lagrangian) Einstein's equations with
the same boundary conditions. Let us stress that the Hamiltonian
hyperbolicity is explicitly paid by the arbitrariness of the Dirac
multipliers. Of course this is just the Hamiltonian counterpart of
the "{\it indeterminateness}" or the so-called "{\it
indeterminism}" surfacing in the Hole Argument.

At this point a number of important questions must be clarified.
When used as generators of canonical transformations, the eight
first class constraints will map points of the constraint surface
to points on the same surface. We shall say that they generate the
infinitesimal transformations of the {\it off-shell Hamiltonian
gauge group} ${\cal G}_8$ \footnote{Note that the off-shell
Hamiltonian gauge transformations are {\it local Noether
transformations} (second Noether theorem) under which the ADM
Lagrangian (\ref{III1}) is {\it quasi-invariant}.}. The action of
${\cal G}_8$ gives rise to a {\it Hamiltonian gauge orbit} through
each point of the constraint sub-manifold $\Gamma_{12}$. Every
such orbit is parametrized by eight phase space functions - namely
the independent {\it off-shell Hamiltonian gauge variables} -
conjugated to the first class constraints. We are left thereby
with a pair of conjugate canonical variables, the {\it off-shell
DO}, which are the only {\it Hamiltonian gauge-invariant and
"determinatedly" ruled} quantities. The same counting of degrees
of freedom of the Lagrangian approach is thus obtained. Finally,
let us stress here, in view of the later discussion, that both the
off-shell Christoffel symbols and the off-shell Riemann tensor can
be read as functions of both the off-shell DO and the Hamiltonian
gauge variables. Likewise, the on-shell Christoffel symbols and
the on-shell Riemann tensor will depend on both the on-shell DO
and the on-shell Hamiltonian gauge variables.

\bigskip

The eight infinitesimal off-shell Hamiltonian gauge
transformations have the following interpretation\cite{44}:

i) those generated by the four primary constraints modify the
lapse and shift functions: these in turn determine how densely the
space-like hyper-surfaces $\Sigma_{\tau}$ are distributed in
space-time and also the conventions to be made on each
$\Sigma_{\tau}$ about simultaneity (the choice of clocks
synchronization) and gravito-magnetism;

ii) those generated by the three super-momentum constraints induce
a transition on $\Sigma_{\tau}$ from a given 3-coordinate system
to another one;

iii) that generated by the super-hamiltonian constraint induces a
transition from a given 3+1 splitting of $M^4$ to another one, by
operating normal deformations \cite{52} of the space-like
hyper-surfaces\footnote{Note that in {\it compact} space-times the
super-hamiltonian constraint is usually interpreted as generator
of the evolution in some {\it internal time}, either like York's
internal {\it extrinsic} time or like Misner's internal {\it
intrinsic} time. Here instead the super-hamiltonian constraint is
the generator of those Hamiltonian gauge transformations which
imply that the description is independent of the choice of the
allowed 3+1 splitting of space-time: {\it this is the correct
answer to the criticisms raised against the phase space approach
on the basis of its lack of manifest covariance}. }.

Making the quotient of the constraint hyper-surface with respect
to the off-shell Hamiltonian gauge transformations by defining
$\Gamma_4 = \Gamma_{12} / {\cal G}_8$, we obtain the so-called
{\it reduced off-shell conformal super-space}. Each of its points,
i.e. a {\it Hamiltonian off-shell (or kinematical) gravitational
field}, is an off-shell equivalence class, called an {\it
off-shell conformal 3-geometry}, for the space-like hyper-surfaces
$\Sigma_{\tau}$: note that, since it contains all the off-shell
4-geometries connected by Hamiltonian gauge transformations, {\it
it is not a 4-geometry}.

\bigskip

An important digression is in order here. The space of parameters
of the off-shell gauge group ${\cal G}_8$ contains eight arbitrary
functions. Four of them are the Dirac multipliers $\lambda_N(\tau
,\vec \sigma )$, $\lambda_r^{\vec N}(\tau ,\vec \sigma )$ of
Eqs.(\ref{III3}), while the other four are functions $ \alpha
(\tau ,\vec \sigma )$, $\alpha_r(\tau ,\vec \sigma )$ which
generalize the lapse and shift functions in front of the secondary
constraints in Eqs.(\ref{III3}) \footnote{In Ref. \cite{14} they
are called {\it descriptors} and written in the form $\alpha = N\,
\xi$, $\alpha^r = {}^3g^{rs}\, \alpha_s = \xi^r \pm N^r\, \xi$. }.
These arbitrary functions correspond to the eight local Noether
symmetries under which the ADM action is quasi-invariant.

On the other hand, from the analysis of the dynamical symmetries
of the Hamilton equations (equivalent to Einstein's equations), it
turns out (see Refs.\cite{53,54}) that {\it on-shell} only a
sub-group ${\cal G}_{4\, dyn}$ of ${\cal G}_8$ survives, depending
on  four arbitrary functions only. But in the present context, a
crucial result for our subsequent discussion is that a further
subset, that we will denote by ${\cal G}_{4\, P} \subset {\cal
G}_{4\, dyn}$, can be identified within the sub-group ${\cal
G}_{4\, dyn}$: precisely the subset corresponding to the phase
space counterparts of those passive diffeomorphisms which are {\it
projectable} to phase space. On the other hand, as already said,
Einstein's equations have $Q$ as the largest group of dynamical
symmetries and, even if irrelevant to the local Noether symmetries
of the ADM action, this larger group is of fundamental importance
for our considerations. In order to take it into account in the
present context, the parameter space of ${\cal G}_8$ must be
enlarged to arbitrary functions depending also on the 3-metric,
$\lambda_N(\tau ,\vec \sigma ) \mapsto \lambda_N(\tau ,\vec \sigma
, {}^3g_{rs}(\tau ,\vec \sigma ))$, ... , $\alpha_r(\tau ,\vec
\sigma ) \mapsto \alpha_r(\tau ,\vec \sigma , {}^3g_{rs}(\tau
,\vec \sigma ))$. Then, the restriction of this enlarged gauge
group to the {\it dynamical symmetries} of the Hamilton equations
defines an extended group ${\tilde {\cal G}}_{4\, dyn}$ which,
under inverse Legendre transformation, defines a new {\it
non-normal} sub-group $Q_{can}$ of the group $Q$ (see
Ref.\cite{14}). But now, the remarkable and fundamental point is
that $Q_{can}$ {\it contains both active and passive
diffeomorphisms}. In particular:
\bigskip

i) the intersection $Q_{can} \cap {}_PDiff\, M^4$ identifies the
space-time passive diffeomorphisms which, respecting the 3+1
splitting of space-time, are {\it projectable} to ${\cal G}_{4\,
P}$ in phase space;

ii) the remaining elements of $Q_{can}$ are the {\it projectable}
subset of active diffeomorphisms in their passive view.\bigskip

\noindent This entails that, as said in Ref.\cite{14},
Eq.(\ref{II4}) may be completed with ${}^4Geom = {}^4Riem /
Q_{can}$.\bigskip

In conclusion, the real gauge group acting on the space of the
solutions of the Hamilton-Dirac equations (\ref{a16}) is the {\it
on-shell extended Hamiltonian gauge group} ${\tilde {\cal G}}_{4\,
dyn}$ and the on-shell equivalence classes obtained by making the
quotient with respect to it eventually coincide with the on-shell
4-geometries of the Lagrangian theory. Therefore, the {\it
Hamiltonian Einstein (or on-shell, or dynamical) gravitational
fields} coincide with the Lagrangian Einstein (or on-shell, or
dynamical) gravitational fields.\bigskip

This is the way in which {\it passive} space-time diffeomorphisms,
under which the Hilbert action is invariant, are reconciled {\it
on-shell} with the allowed Hamiltonian gauge transformations
adapted to the 3+1 splittings of the ADM formalism. Furthermore,
our analysis of the Hamiltonian gauge transformations and their
Legendre counterparts gives an extra {\it bonus}: namely that the
on-shell phase space extended gauge transformations include also
symmetries that are images of {\it active} space-time
diffeomorphsms. The basic relevance of this result for a deep
understanding of the Hole Argument will appear fully in Section
IV.

\bigskip

Having clarified these important issues, let us come back to the
canonical reduction. The off-shell freedom corresponding to the
eight independent types of Hamiltonian gauge transformations is
reduced on-shell to four types like in the case of ${}_PDiff\,
M^4$: precisely the transformations in [$Q_{can} \cap {}_PDiff\,
M^4$] . At the off-shell level, this property is manifest by the
circumstance that the original Dirac Hamiltonian contains {\it
only} 4 arbitrary Dirac multipliers and that the {\it correct
gauge-fixing procedure} \cite{55,44} starts by giving {\it only}
the four gauge fixing constraints for the secondary constraints.
The requirement of time constancy then generates the four gauge
fixing constraints to the primary constraints, while time
constancy of such secondary gauge fixings leads to the
determination of the four Dirac multipliers\footnote{This agrees
with the results of Ref.\cite{56} according to which the {\it
projectable} space-time diffeomorphisms depend only on four
arbitrary functions and their time derivatives.}. Since the
original constraints plus the above eight gauge fixing constraints
form a second class set, it is possible to introduce the
associated {\it Dirac brackets} and conclude the canonical
reduction by realizing an off-shell reduced phase space
$\Gamma_4$. Of course, once we reach a {\it completely fixed
Hamiltonian gauge} (a copy of $\Gamma_4$), general covariance is
completely broken. Finally, note that a completely fixed
Hamiltonian gauge {\it on-shell} is equivalent to a {\it definite
choice of the space-time 4-coordinates} on $M^4$ within the
Lagrangian viewpoint.

\bigskip

In order to visualize the meaning of the various types of degrees
of freedom \footnote{This visualization remains only implicit in
the conformal Lichnerowicz-York approach \cite{57,58,59,60}.} we
need a determination of a {\it Shanmugadhasan canonical basis}
\cite{37} of metric gravity \cite{44} having the following
structure ($\bar a =1,2$ are non-tensorial indices of the DO
\footnote{Let us stress that the DO are in general neither tensors
nor invariants under space-time diffeomorphisms. Therefore their
(unknown) functional dependence on the original variables changes
(off-shell) with the gauge and, therefore, (on-shell) with the
4-coordinate system.} $r_{\bar a}$, $\pi_{\bar a}$) with

 \bea
\begin{minipage}[t]{3cm}
\begin{tabular}{|l|l|l|} \hline
$N$ & $N_r$ & ${}^3g_{rs}$ \\ \hline ${\tilde \pi}^N \approx 0$ &
${\tilde \pi}_{\vec N}^r \approx 0$ & ${}^3{\tilde \Pi}^{rs}$ \\
\hline
\end{tabular}
\end{minipage} &&\hspace{2cm} {\longrightarrow \hspace{.2cm}} \
\begin{minipage}[t]{4 cm}
\begin{tabular}{|ll|l|l|l|} \hline
$N$ & $N_r$ & $\xi^{r}$ & $\phi$ & $r_{\bar a}$\\ \hline
 ${\tilde \pi}^N \approx 0$ & ${\tilde \pi}_{\vec N}^r \approx 0$
& ${\tilde \pi}^{{\vec {\cal H}}}_r \approx 0$ &
 $\pi_{\phi}$ & $\pi_{\bar a}$ \\ \hline
\end{tabular}
\end{minipage} \nonumber \\
 &&{}\nonumber \\
&& {\longrightarrow \hspace{.2cm}} \
\begin{minipage}[t]{4 cm}
\begin{tabular}{|ll|l|l|l|} \hline
$N$ & $N_r$ & $\xi^{r}$ & $Q_{\cal H} \approx 0$ & $r^{'}_{\bar
a}$\\ \hline
 ${\tilde \pi}^N \approx 0$ & ${\tilde \pi}^r_{\vec N} \approx 0$
& ${\tilde \pi}^{{\vec {\cal H}}}_r \approx 0$ &
 $\Pi_{\cal H}$ & $\pi^{'}_{\bar a}$ \\ \hline
\end{tabular}
\end{minipage}.
 \label{III4}
 \eea

\noindent It is seen that we need a sequence of two canonical
transformations.\bigskip

a) The first one replaces seven first-class constraints with as
many Abelian momenta ($\xi^r$ are the gauge parameters of the
passive 3-diffeomorphisms generated by the super-momentum
constraints) and introduces the conformal factor $\phi$ of the
3-metric as the configuration variable to be determined by the
super-hamiltonian constraint  \footnote{Recall that the {\it
strong} ADM energy is the flux through the surface at spatial
infinity of a function of the 3-metric only, and it is weakly
equal to the {\it weak} ADM energy (volume form) which contains
all the dependence on the ADM momenta. This implies \cite{44} that
the super-hamiltonian constraint must be interpreted as the
equation ({\it Lichnerowicz equation}) that uniquely determines
the {\it conformal factor} $\phi = ( det\, {}^3g )^{1/12}$ of the
3-metric as a functional of the other variables. This means that
the associated gauge variable is the {\it canonical momentum
$\pi_{\phi}$ conjugate to the conformal factor}: this latter
carries information about the extrinsic curvature of
$\Sigma_{\tau}$. It is just this variable, and {\it not} York's
time, which parametrizes the {\it normal} deformation of the
embeddable space-like hyper-surfaces $\Sigma_{\tau}$. As said in
footnote 15, the theory is independent of the choice of the 3+1
splitting like in parametrized Minkowski theories. As a matter of
fact, a gauge fixing for the super-hamiltonian constraint, i.e. a
choice of a particular 3+1 splitting, is done by fixing the
momentum $\pi_{\phi}$ conjugate to the conformal factor. This
shows the dominant role of the conformal 3-geometries in the
determination of the physical degrees of freedom, just as in the
Lichnerowicz-York conformal approach.}. Note that the final gauge
variable, namely the momentum $\pi_{\phi}$ conjugate to the
conformal factor, is the only gauge variable of momentum type: it
plays the role of a {\it time} variable, so that the Lorentz
signature of space-time is made manifest by the Shanmugadhasan
transformation in the set of gauge variables $(\pi_{\phi};
\xi^r)$; this makes the difference with respect to the proposals
of Refs.\cite{33,61}. More precisely, the first canonical
transformation should be called a {\it quasi-Shanmugadhasan }
transformation, because nobody has succeeded so far in
Abelianizing the super-hamiltonian constraint. Note furthermore
that this transformation is a {\it point} canonical
transformation.

b) The second canonical transformation would be instead a {\it
complete Shanmugadhasan} transformation, where $Q_{{\cal H}}(\tau
,\vec \sigma ) \approx 0$ would denote the Abelianization of the
super-hamiltonian constraint \footnote{If $\tilde \phi [r_{\bar
a}, \pi_{\bar a}, \xi^r, \pi_{\phi}]$ is the solution of the
Lichnerowicz equation, then $Q_{{\cal H}}=\phi - \tilde \phi
\approx 0$. Other forms of this canonical transformation should
correspond to the extension of the York map \cite{62} to
asymptotically flat space-times: in this case the momentum
conjugate to the conformal factor would be just York time and one
could add the maximal slicing condition as a gauge fixing. Again,
however, nobody has been able so far to build a York map
explicitly.}. The variables $N$, $N_r$, $\xi^r$, $\Pi_{\cal H}$
are the final {\it Abelianized Hamiltonian gauge variables} and
$r^{'}_{\bar a}$, $\pi^{'}_{\bar a}$ the final DO.
\bigskip

In absence of explicit solutions of the Lichnerowicz equation, the
best we can do is to construct the {\it quasi-Shanmugadhasan}
transformation. On the other hand, such transformation has the
remarkable property that, in the {\it special gauge}
$\pi_{\phi}(\tau ,\vec \sigma ) \approx 0$, the variables $r_{\bar
a}$, $\pi_{\bar a}$ form a canonical basis of off-shell DO for the
gravitational field {\it even if} the solution of the Lichnerowicz
equation is not known.

\bigskip

The four gauge fixings to the secondary constraints, when written
in the quasi-Shanmugadhasan canonical basis, have the following
meaning:\hfill\break

 i) the three gauge fixings for the parameters $\xi^r$ of the
spatial passive diffeomorphisms generated by the super-momentum
constraints correspond to the choice of a system of 3-coordinates
on $\Sigma_{\tau}$\footnote{Since the diffeomorphism group has no
canonical identity, this gauge fixing has to be done in the
following way. We choose a 3-coordinate system by choosing a
parametrization of the six components ${}^3g_{rs}(\tau ,\vec
\sigma )$ of the 3-metric in terms of {\it only three} independent
functions. This amounts to fix the three functional degrees of
freedom associated with the diffeomorphism parameters $\xi^r(\tau
,\vec \sigma )$. For instance, a 3-orthogonal coordinate system is
identified by ${}^3g_{rs}(\tau ,\vec \sigma ) = 0$ for $r \not= s$
and ${}^3g_{rr} = \phi^2\,  exp(\sum_{\bar a = 1}^2 \gamma_{r\bar
a} r_{\bar a})$. Then, we impose the gauge fixing constraints
$\xi^r(\tau ,\vec \sigma ) - \sigma^r \approx 0$ as a way of
identifying this system of 3-coordinates with a conventional
origin of the diffeomorphism group manifold. }. The time constancy
of these gauge fixings generates the gauge fixings for the shift
functions $N_r$ while the time constancy of the latter leads to
the fixation of the Dirac multipliers $\lambda^{\vec N}_r$;
\hfill\break

ii) the gauge fixing to the super-hamiltonian constraint
determines $\pi_{\phi}$: it is a fixation of the form of
$\Sigma_{\tau}$ and amounts to the choice of one particular 3+1
splitting of $M^4$. Since the time constancy of the gauge fixing
on $\pi_{\phi}$ determines the gauge fixing for the lapse function
$N$ (and then of the Dirac multiplier $\lambda_N$), it follows a
connection with the choice of the standard of local proper time.
\bigskip

All this entails that, after such a fixation of the gauge $G$, the
functional form of the DO in terms of the original variables
becomes gauge-dependent. At this point it is convenient to denote
them as $r^G_{\bar a}$, $\pi^G_{\bar a}$.
\bigskip

In conclusion, a representative of a {\it Hamiltonian kinematical
or off-shell gravitational field}, in a given gauge equivalence
class, is parametrized by $r_{\bar a}$, $\pi_{\bar a}$ and is an
element of a {\it conformal gauge orbit} (it contains all the
3-metrics in a conformal 3-geometry) spanned by the gauge
variables $\xi^r$, $\pi_{\phi}$, $N$, $N_r$. Therefore, according
to the gauge interpretation based on constraint theory, a {\it
Hamiltonian kinematical or off-shell gravitational field} is an
equivalence class of 4-metrics modulo the Hamiltonian group of
gauge transformations, which contains a well defined conformal
3-geometry. Clearly, this is a consequence of the different
invariance properties of the ADM and Hilbert actions, even if they
generate the same equations of motion.

Recall that the evolution is parametrized  by the mathematical
parameter $\tau$ of the adapted coordinate system $(\tau ,\vec
\sigma )$ on $M^4$, which labels the surfaces $\Sigma_{\tau}$. As
shown in Ref.\cite{44}, the Hamiltonian ruling the evolution is
the {\it weak ADM energy} \cite{63} (the volume form), which, in
every completely fixed gauge, is a functional of only the DO of
that gauge. As shown by DeWitt \cite{64}, this is a consequence of
the fact that {\it in non-compact space-times the weakly vanishing
ADM Dirac Hamiltonian (\ref{III3}) has to be modified with a
suitable surface term in order to have functional derivatives,
Poisson brackets and Hamilton equations mathematically well
defined mathematically}.

Finally, let us stress the important fact that the Shanmugadhasan
canonical transformation is a {\it highly non-local}
transformation. Since it is not known how to build a global atlas
of coordinate charts for the group manifold of diffeomorphism
groups, it is not known either how to express the $\xi^r$'s,
$\pi_{\phi}$ and the DO in terms of the original ADM canonical
variables.\footnote{This should be compared to the Yang-Mills
theory in case of a trivial principal bundle, where the
corresponding variables are defined by a path integral over the
original canonical variables \cite{65,50,51}.}.
\bigskip

\vfill\eject

\section{Physical Individuation of Space-Time Points by means of
Gauge Fixings to Bergmann-Komar Intrinsic Coordinates.}

It's time to pursue our main task, namely the clarification of the
physical individuality of space-time point-events in general
relativity.

Let us begin by recalling again that the ADM formulation assumes
the existence of a mathematical 4-manifold, the space-time $M^4$,
admitting 3+1 splittings with space-like leaves
$\Sigma_{\tau}\approx R^3$. All fields (also matter fields when
present) depend on $\Sigma_{\tau}$-adapted coordinates $(\tau
,\vec \sigma )$ for $M^4$. As emphasized in the Introduction,
unlike the case of special relativity, the {\it mathematical
points} of $M^4$ have no intrinsic physical meaning, a
circumstance that originates the {\it Hole} phenomenology. This
has the effect that, as stressed in particular by Stachel
\cite{23}, the space-time manifold points must be {\it physically}
individuated {\it before} space-time itself acquires a physical
meaning. In the case of general relativity there is {\it no
non-dynamical individuating field} like the distribution of rods
and clocks in Minkowsky space-time, that can be specified
independently of the dynamical fields, in particular independently
of the metric. Thus, Stachel claimed that, for the pure
gravitational solutions, {\it the metric itself plays the role of
individuating field} and that this role should be implemented by
four invariant functionals of the metric; however, he didn't
pursue this program further. We must insist again that a crucial
component of the individuation issue is the inextricable
entanglement of the Hole Argument with the initial value problem.
Now, however, we have at our disposal the right framework for
dealing with the initial value problem, so our main task should be
to put all things together and develop Stachel's suggestion.
Finally, another fundamental tool at our disposal is the
clarification we gained in Section II concerning the relation
between active diffeomorphisms in their passive view and the
dynamical gauge symmetries of Einstein's equations in the
Hamiltonian approach.

We are then ready to move forward by conjoining Stachel's
suggestion with the proposal advanced by Bergmann and Komar
\cite{1} that, in the absence of matter fields, the values of four
invariant scalar fields built from the contractions of the Weyl
tensor (actually its eigenvalues) can be used to build  {\it
intrinsic pseudo-coordinates}\footnote{As shown in Ref.\cite{66}
in general space-times with matter there are 14 algebraically
independent curvature scalars for $M^4$.}.

The four invariant scalar eigenvalues $\Lambda^{(k)}_W(\tau ,\vec
\sigma )$, $k=1,..,4$, of the Weyl tensor, written in Petrov
compressed notations, are

\bea
 &&\Lambda^{(1)}_W = Tr\, ({}^4C\, {}^4g\, {}^4C\,
 {}^4g),\nonumber \\
 &&\Lambda^{(2)}_W = Tr\, ({}^4C\, {}^4g\, {}^4C\, {}^4\epsilon ),
 \nonumber \\
 &&\Lambda^{(3)}_W = Tr\, ({}^4C\, {}^4g\, {}^4C\, {}^4g\, {}^4C\,
 {}^4g),\nonumber \\
 &&\Lambda^{(4)}_W = Tr\, ({}^4C\, {}^4g\, {}^4C\, {}^4g\, {}^4C\,
 {}^4\epsilon ),
 \label{IV1}
 \eea

\noindent where ${}^4C$ is the Weyl tensor, $^4g$ the metric, and
${}^4\epsilon$ the Levi-Civita totally antisymmetric tensor.

Bergmann and Komar then propose that we build a set of (off-shell)
{\it intrinsic coordinates} for the point-events of space-time as
four suitable functions of the $\Lambda^{(k)}_W$'s,

\beq
 {\bar \sigma}^{\bar A}(\sigma ) = F^{\bar A}[\Lambda^{(k)}_W
 [{}^4g(\sigma), \partial {}^4g(\sigma)]],
\,(\bar A = 1,2,...,4).
 \label{IV2}
  \eeq

\noindent Indeed, under the hypothesis of no space-time
symmetries,\footnote{Our attempt to use intrinsic coordinates to
provide a physical individuation of point-events would {\it prima
facie} fail in the presence of symmetries (with or without
matter), when the $F^{\bar A}[\Lambda^{(k)}_W [{}^4g(\sigma ),
\partial {}^4g(\sigma )]]$ become degenerate. This objection was
originally raised by Norton \cite{12} as a critique to
manifold-plus-further-structure (MPFS) substantivalism (according
to which the points of the manifold, conjoined with additional
local structure such as the metric field, can be considered
physically real; see for instance \cite{67}). Several responses
are possible. First, although most of  the {\it known} exact
solutions of the Einstein equations admit one or more symmetries,
these mathematical models are very idealized and simplified; in a
realistic situation (for instance, even with two masses)
space-time is filled with the excitations of the gravitational
degrees of freedom, and admits no symmetries at all. A case study
is furnished by the non-symmetric and non-singular space-times of
Christodoulou-Klainermann \cite{36}. Second, the parameters of the
symmetry transformations can be used as supplementary
individuating fields, since, as noticed by Komar \cite{30} and
Stachel \cite{23} they also depend upon metric field, through its
isometries. To this move it has been objected \cite{24} that these
parameters are purely mathematical artifacts, but a simple
rejoinder is that the symmetric models too are mathematical
artifacts. Third, and most important, in our analysis of the
physical individuation of points we are arguing a question of
principle, and therefore we must consider {\it generic} solutions
of the Einstein equations rather than the null-measure set of
solutions with symmetries.} we would be tempted (like Stachel) to
use the $F^{\bar A}[\Lambda^{(k)}_W]$ as individuating fields to
{\it label the points of space-time}, at least
locally.\footnote{Problems might arise if we try to extend the
labels to the entire space-time: for instance, the coordinates
might turn out to be multi-valued.}

Of course, since they are invariant functionals, the $F^{\bar
A}[\Lambda^{(k)}_W]$'s are quantities invariant under passive
diffeomorphisms (PDIQ), therefore, as such, they do not define a
coordinate chart for the atlas of the mathematical Riemannian
4-manifold $M^4$ in the usual sense (hence the name of {\it
pseudo-coordinates} and the superior bar we used in $F^{\bar A}$).
Moreover, the tetradic 4-metric which can be built by means of the
intrinsic pseudo-coordinates (see the next Section) is a formal
object invariant under passive diffeomorphisms that does not
satisfy Einstein's equations (but possibly much more complex
derived equations). Therefore, the action of active
diffeomorphisms on the tetradic metric is not {\it directly}
connected to the Hole argument. All this leads to the conclusion
that the proposal advanced by Bergmann \cite{2} ("we might then
identify a {\it world point} (location-plus-instant-in-time) by
the values assumed by (the four intrinsic pseudo-coordinates)") to
the effect of individuating point-events in terms of {\it
intrinsic pseudo-coordinates} is not - as it stands - {\it
physically viable} in a tractable way. This is not the final
verdict, however, and we must find a dynamical bridge between the
intrinsic pseudo-coordinates and the ordinary 4-coordinate systems
which provide the primary identification of the points of the
mathematical manifold.
\medskip

Our procedure starts when we recall that, within the Hamiltonian
approach, Bergmann and Komar \cite{1} proved the fundamental
result that the Weyl eigenvalues $\Lambda^{(A)}_W$, once
re-expressed as functionals of the Dirac (i.e. ADM) canonical
variables, {\it do not depend on the lapse and shift functions}
but only on the 3-metric and its conjugate canonical momentum,
$\Lambda^{(k)}_W[{}^4g(\tau ,\vec \sigma),
\partial {}^4g(\tau ,\vec \sigma)] = {\tilde \Lambda}^{(k)}_W[{}^3g(\tau
,\vec \sigma), {}^3\Pi(\tau ,\vec \sigma)]$. This result is
crucial since it entails that just the {\it intrinsic
pseudo-coordinates} ${\bar \sigma}^{\bar A}$ can be exploited as
natural and peculiar {\it coordinate gauge conditions} in the
canonical reduction procedure.

Taking into account the results of Section III, we know that, in a
completely fixed (either {\it off}- or {\it on-shell}) gauge, both
the four intrinsic {\it pseudo-coordinates} and the ten {\it
tetradic} components of the metric field (see Eq.(\ref{V2}) of the
next Section)  become gauge dependent functions of the four DO of
that gauge. For the Weyl scalars in particular we can write:

\beq
 \Lambda^{(k)}_W(\tau ,\vec \sigma ){|}_G =  {\tilde \Lambda}^{(k)}_W[{}^3g(\tau
,\vec \sigma), {}^3\Pi(\tau ,\vec \sigma)]{|}_G =
\Lambda_G^{(k)}[r^G_{\bar a}(\tau ,\vec \sigma ), \pi^G_{\bar
a}(\tau ,\vec \sigma )].
 \label{IV3}
  \eeq

\noindent where ${|}_G$ denotes the specific gauge (see footnote
25). Conversely, by the inverse function theorem, in each gauge,
the DO of that gauge can be expressed as functions of the 4
eigenvalues restricted to that gauge: $\Lambda^{(k)}_W (\tau ,\vec
\sigma ){|}_G $.

Our program is implemented in the following way: after having
selected a {\it completely arbitrary mathematical} coordinate
system $\sigma^A \equiv [\tau,\sigma^a]$ adapted to the
$\Sigma_\tau$ surfaces, we choose {\it as physical individuating
fields} four suitable functions $F^{\bar A}[\Lambda^{(k)}_W(\tau
,\vec \sigma )]$, and express them as functionals $\tilde F^{\bar
A}$ of the ADM variables

\beq
 F^{\bar A}[\Lambda^{(k)}_W(\tau ,\vec \sigma )] = F^{\bar
A}[{\tilde \Lambda}^{(k)}_W[{}^3g(\tau ,\vec \sigma), {}^3\Pi(\tau
,\vec \sigma)]] = \tilde F^{\bar A}[{}^3g(\tau ,\vec \sigma),
{}^3\Pi(\tau ,\vec \sigma)].
 \label{IV4}
  \eeq

\noindent The space-time points, {\it mathematically individuated}
by the quadruples of real numbers $\sigma^A$, become now {\it
physically individuated point-events} through the imposition of
the following gauge fixings to the four secondary constraints

\beq
 {\bar \chi}^A(\tau ,\vec \sigma )\- {\buildrel {def} \over =}\-  \sigma^A -
 {\bar \sigma}^{\bar A}(\tau ,\vec \sigma ) = \sigma^A -
 F^{\bar A}\Big[{\tilde \Lambda}^{(k)}_W[{}^3g(\tau ,\vec \sigma), {}^3\Pi(\tau ,\vec
\sigma)]\Big] \approx 0.
 \label{IV5}
 \eeq

\noindent Then, following the standard procedure we end with a
completely fixed Hamiltonian gauge, say $G$. This will be a
correct gauge fixing provided the functions $F^{\bar A}[{
\Lambda}^{(k)}_W(\tau ,\vec \sigma )]$ are chosen so that the
${\bar \chi}^A(\tau ,\vec \sigma )$'s satisfy the {\it orbit
conditions}

\beq
 det\, | \{ {\bar \chi}^A(\tau ,\vec \sigma ), {\tilde {\cal
 H}}^B(\tau ,{\vec \sigma}^{'}) \} | \not= 0,
 \label{IV6}
 \eeq

\noindent where ${\tilde {\cal H}}^B(\tau ,\vec \sigma ) = \Big(
{\tilde {\cal H}}(\tau ,\vec \sigma ); {}^3{\tilde {\cal
H}}^r(\tau ,\vec \sigma ) \Big) \approx 0$ are the
super-hamiltonian and super-momentum constraints of
Eqs.(\ref{III2}). These conditions enforce the Lorentz signature
on Eq.(\ref{IV5}), namely the requirement that $F^{\bar \tau}$ be
a {\it time} variable, and imply that {\it the $F^{\bar A}$'s are
not DO}.

The above gauge fixings allow in turn the determination of the
four Hamiltonian gauge variables $\xi^r(\tau ,\vec \sigma )$,
$\pi_{\phi}(\tau ,\vec \sigma )$ of Eqs.(\ref{III4}). Then, their
time constancy induces the further gauge fixings ${\bar
\psi}^A(\tau ,\vec \sigma ) \approx 0$ for the determination of
the remaining gauge variables, i.e., the lapse and shift functions
in terms of the DO in that gauge as

\bea
 {\dot {\bar \chi}}^A(\tau ,\vec \sigma ) &=& {{\partial {\bar
 \chi}^A(\tau ,\vec \sigma )} \over {\partial \tau}} + \{
 {\bar \sigma}^{\bar A}(\tau ,\vec \sigma ), {\bar H}_D \} = \delta^{A\tau}
 +\nonumber \\
 &+& \int d^3\sigma_1\, \Big[ N(\tau ,{\vec \sigma}_1)\, \{
  \sigma^{\bar A}(\tau ,\vec \sigma ), {\cal H}(\tau ,{\vec
  \sigma}_1) \} + N_r(\tau ,{\vec \sigma}_1)\, \{  \sigma^{\bar A}(\tau ,\vec \sigma ),
  {\cal H}^r(\tau ,{\vec \sigma}_1) \}\Big] =\nonumber \\
  &=& {\bar \psi}^A(\tau ,\vec \sigma ) \approx 0.
  \label{IV7}
  \eea

\noindent Finally, ${\dot {\bar \psi}}^A(\tau ,\vec \sigma )
\approx 0$ determines the Dirac multipliers $\lambda^A(\tau ,\vec
\sigma )$.\bigskip

In conclusion, the gauge fixings (\ref{IV5}) ({\it which break
general covariance}) constitute the crucial bridge that transforms
the {\it intrinsic pseudo-coordinates} into {\it true physical
individuating coordinates}.

As a matter of fact, after going to Dirac brackets, we enforce the
point-events individuation in the form of the {\it identity}

\beq
 \sigma^A \equiv {\bar \sigma}^{\bar A} = {\tilde F}^{\bar A}_{G}[
r^G_{\bar a}(\tau ,\vec \sigma ), \pi^G_{\bar a}(\tau , \vec
\sigma)] = F^{\bar A}[\Lambda^{(k)}_W(\tau ,\vec \sigma )]{|}_G.
 \label{IV8}
  \eeq

\medskip

In this {\it physical 4-coordinate grid}, the 4-metric, as well as
other  fundamental physical entities, like e.g. the space-time
interval $ds^{2}$ with its associated causal structure, and the
lapse and shift functions, depend entirely on the DO in that
gauge. The same is true, in particular, for the solutions of the
eikonal equation \cite{36} ${}^4g^{AB}(\sigma^D)\, {{\partial
U(\sigma^D)}\over {\partial \sigma^A}}\,  {{\partial
U(\sigma^D)}\over {\partial \sigma^B}} = 0$, which define
generalized wave fronts and, therefore, through the envelope of
the null surfaces $U(\sigma^D) = const.$ at a point,  the light
cone at that point.
\bigskip

Let us stress that, according to the previous Sections, only on
the solutions of Einstein's equations the completely fixed gauge
$G$ is equivalent to the fixation of a definite 4-coordinate
system $\sigma^A_G$. Our gauge fixing (\ref{IV5}) ensures that
{\it on-shell} we get $\sigma^A = \sigma^A_G$. In this way we get
a physical 4-coordinate grid on the mathematical 4-manifold $M^4$
dynamically determined by tensors over $M^4$ with a rule which is
invariant under ${}_PDiff\, M^4$ but such that the functional form
of the map {\it $\sigma^A \mapsto \, physical\,\,\,
4-coordinates$} depends on the complete chosen gauge $G$: we see
that what is usually called the {\it local point of view}
\cite{68} (see later on) is justified a posteriori in every
completely fixed gauge.

\bigskip

Summarizing, the effect of the whole procedure is that {\it the
values of the {\it DO}, whose dependence on space (and on {\it
parameter} time) is indexed by the chosen coordinates $(\tau ,\vec
\sigma )$, reproduces precisely such $(\tau ,\vec \sigma )$ as the
Bergmann--Komar {\it intrinsic coordinates} in the chosen gauge
$G$}. In this way {\it mathematical points have become physical
individuated point-events} by means of the highly non-local
structure of the DO. If we read the identity (\ref{IV8}) as
$\sigma^A \equiv f^{\bar A}_G(r^G_{\bar a}, \pi^G_{\bar a})$, we
see that each coordinate system $\sigma^A$ is determined on-shell
by the values of the 4 canonical degrees of freedom of the
gravitational field in that gauge. This is tantamount to claiming
that {\it the physical role and content of the gravitational field
in absence of matter is just the very identification of the points
of Einstein space-times into physical point-events by means of its
four independent phase space degrees of freedom}. The existence of
physical point-events in general relativity appears here as a
synonym of the existence of the DO, i.e. of the true physical
degrees of freedom of the gravitational field.

As said in the Introduction, the addition of matter does not
change this ontological conclusion, because we can continue to use
the gauge fixing (\ref{IV5}). However, matter changes the Weyl
tensor through Einstein's equations and contributes to the
separation of gauge variables from DO in the quasi-Shanmugadhasan
canonical transformation through the presence of its own DO. In
this case we have Dirac observables both for the gravitational
field and for the matter fields, but {\it the former are modified
in their functional form} with respect to the vacuum case by the
presence of matter. Since the gravitational Dirac observables will
still provide the individuating fields for point-events according
to our procedure, {\it matter will come to influence the very
physical individuation of points}.

\hfill\break

We have seen that, once the orbit conditions are satisfied, the
Bergmann-Komar {\it intrinsic pseudo-coordinates} $F^{\bar
A}[{\tilde \Lambda}^{(k)}_W[{}^3g(\tau ,\vec \sigma), {}^3\Pi(\tau
,\vec \sigma)]{|}_G$ become just the {\it individuating fields}
Stachel was looking for. Indeed, by construction, the intrinsic
pseudo-coordinates are both PDIQ and also {\it numerically
invariant} under the drag along induced by active diffeomorphisms
(in the notations of the Introduction we have $[\phi^*F^{\bar
A}](p) \equiv [F^{\bar A\, '}](p) = [F^{\bar A}](\phi^{-1} \cdot
p)\,\,$), a fact that is also essential for maintaining a
connection to the {\it Hole Argument}.

A better understanding of the real connection between Stachel
point of view and ours can be achieved by exploiting
Bergmann-Komar's group of passive transformations $Q$ discussed in
Section II. We can argue in the following way. Given a
4-coordinate system $\sigma^A$, the passive view of each active
diffeomorphism $\phi$ defines a new 4-coordinate system
$\sigma^A_{\phi}$ (drag-along coordinates produced by a
generalized Bergmann-Komar transformation (\ref{II3})). This means
that there will be two functions $F^{\bar A}$ and $F^{\bar
A}_{\phi}$ realizing these two coordinates systems through the
gauge fixings

\bea
 &&\sigma^A - F^{\bar A}[\Lambda^{(k)}_W(\sigma)] \approx
 0,\nonumber \\
 &&\sigma_{\phi}^A - F_{\phi}^{\bar A}[\Lambda^{(k)}_W(\sigma_{\phi})] \approx
 0,
 \label{IV9}
 \eea

It is explicitly seen in this way that the functional freedom in
the choice of the four functions $F^{\bar A}$ allows to cover all
those coordinates charts $\sigma^A$ in the atlas of the
mathematical space-time $M^4$ which are adapted to any allowed 3+1
splitting. By using gauge fixing constraints more general than
those in Eq.(\ref{IV5}) (like the standard gauge fixings used in
ADM metric gravity) we can reach all the 4-coordinates systems of
$M^4$. Here, however we wanted to restrict to the class of gauge
fixings (\ref{IV5}) for the sake of clarifying the
interpretational issues. \hfill\break

Let us conclude by noting that the gauge fixings (\ref{IV5}),
(\ref{IV7}) induce a {\it coordinate-dependent non-commutative
Poisson bracket structure} upon the {\it physical point-events} of
space-time by means of the associated Dirac brackets implying
Eqs.(\ref{IV8}). More exactly, on-shell, each coordinate system
gets a well defined non-commutative structure determined by the
associated functions ${\tilde F}^{\bar A}_G(r^G_{\bar a},
\pi^G_{\bar a})$, for which we have $\{ {\tilde F}^{\bar
A}_G(r^G_{\bar a}(\tau , \vec \sigma ), \pi^G_{\bar a}(\tau ,\vec
\sigma )), {\tilde F}^{\bar B}_G(r^G_{\bar a}(\tau ,{\vec
\sigma}_1), \pi^G_{\bar a}(\tau ,{\vec \sigma}_1)) \}^* \not= 0$.
The meaning of this structure in view of quantization is worth
investigating (see the Concluding Survey).\hfill\break

\vfill\break

\section{A Digression on Bergmann Observables: the Main Conjecture and the Issue of
the Objectivity of Change.}

This Section is devoted to some crucial aspects of the definition
of {\it observable} in general relativity. While, for instance in
astrophysics, {\it matter observables} are usually defined as {\it
tetradic quantities} evaluated with respect to the tetrads of a
time-like observer so that they are obviously invariant under
${}_PDiff\, M^4$ (PDIQ), the definition of the notion of
observable for the gravitational field without matter faces a
dilemma. Two fundamental definitions of {\it observable} have been
proposed in the literature.
\bigskip

1) The {\it off-shell and on-shell Hamiltonian non-local Dirac
observables} (DO) \footnote{ For other approaches to the
observables of general relativity see Refs.\cite{69}: the {\it
perennials} introduced in this Reference are essentially our DO.
See Ref.\cite{70} for the difficulties in observing {\it
perennials} experimentally at the classical and quantum levels as
well as for their quantization. See Ref.\cite{71} about the
non-existence of observables built as spatial integrals of local
functions of Cauchy data and their first derivatives, in the case
of vacuum gravitational field in a closed universe. Also,
Rovelli's evolving constants of motion and partial observables
\cite{72} are related with DO; however, the holonomy loops used in
loop quantum gravity \cite{73} are PDIQ but not DO. On the other
hand, even recently Ashtekar \cite{74} noted that "The issue of
diffeomorphism invariant observables and practical methods of
computing their properties" is one among the relevant challenges.}
which, by construction, satisfy hyperbolic Hamilton equations of
motion and are, therefore, deterministically {\it predictable}. In
general, as already said, they are neither tensorial quantities
nor invariant under ${}_PDiff\, M^4$ (PDIQ).
\bigskip

2) The {\it configurational Bergmann observables} ({\it BO})
\cite{2}: they are quantities defined on $M^4$ {\it which not only
are independent of the choice of the coordinates}, i.e. they are
either scalars or invariants\footnote{In Ref.\cite{29} Bergmann
defines: i) a {\it scalar} as a local field variable which retains
its numerical value at the same world point under coordinate
transformations (passive diffeomorphisms), $\varphi^{'}(x^{'}) =
\varphi (x)$; ii) an {\it invariant} $I$ as a functional of the
given fields which has been constructed so that if we substitute
the coordinate transforms of the field variables into the argument
of $I$ instead of the originally given field variables, then the
numerical value of $I$ remains unchanged. } under ${}_PDiff\, M^4$
(PDIQ), but are also "{\it uniquely predictable from the initial
data}". An equivalent, but according to Bergmann more useful,
definition of a (PIDQ) BO, is "{\it a quantity that is invariant
under a coordinate transformation that leaves the initial data
unchanged}".

\medskip

Let us note, first of all, that PDIQ's are {\it not} in general
DO, because they may also depend on the eight gauge variables $N$,
$N_r$, $\xi^r$, $\pi_{\phi}$. Thus most, if not all, of the
curvature scalars are gauge dependent quantities at least at the
kinematical off-shell level. For example, each 3-metric in the
conformal gauge orbit has a different 3-Riemann tensor and
different 3-curvature scalars. Since 4-tensors and 4-curvature
scalars depend: i) on the lapse and shift functions (and their
gradients); ii) on $\pi_{\phi}$, both implicitly and explicitly
through the solution of the Lichnerowicz equation (and this
affects the 3-curvature scalars), most of these objects are in
general gauge dependent variables from the Hamiltonian point of
view, at least at the off-shell kinematical level. The simplest
relevant off-shell scalars with respect to ${}_PDiff\, M^4$, which
exhibit such gauge dependence, are the bilinears
${}^4R_{\mu\nu\rho\sigma}\, {}^4R^{\mu\nu\rho\sigma}$, ${}^4R
_{\mu\nu\rho\sigma}\, \epsilon^{\mu\nu\alpha\beta}\,
{}^4R_{\alpha\beta}{} ^{\rho\sigma}$  and the {\it four
eigenvalues of the Weyl tensor} exploited in Section IV. What said
here {\it does hold, in particular, for the line element $ds^{2}$
and, therefore, for the causal structure}.

On the other hand, BO are those special PDIQ which are
simultaneously {\it predictable}. Yet, the crucial question is now
"{\it what does it precisely mean to be predictable within the
configurational Lagrangian framework} ?". Bergmann, gave in fact a
third definition of BO or, better, a {\it third part} of the
original definition, as "a dynamical variable that (from the
Hamiltonian point of view) has vanishing Poisson brackets with all
the constraints", i.e., essentially, is also a DO. This means that
Bergmann thought, though only implicitly and without proof, that
{\it predictability} implies that a BO must {\it also} be {\it
projectable to phase space to a special subset of DO that are also
PDIQ}.

The unresolved multiplicity of Bergmann's definitions leads to an
entangled net of problems. First of all, as shown at length in
Ref.\cite{28}, in order to tackle the Cauchy problem at the
Lagrangian configuration level \footnote{ In the theory of systems
of partial differential equations this is done in a {\it passive}
way in a given coordinate system and then extended to all
coordinate systems} one has firstly to disentangle the Lagrangian
constraints from Einstein's equations, then to take into account
the Bianchi identities, and finally to write down a system of
hyperbolic equations. As a matter of fact one has to mimic the
Hamiltonian approach, but with the additional burden of lacking an
algorithm for selecting those predictable configurational field
variables whose Hamiltonian counterparts are just the DO. The only
thing one might do is to adopt an inverse Legendre transformation,
to be performed after a Shanmugadhasan canonical transformation
characterizing a possible complete set of DO. Yet, this
corresponds just to the inverse of Bergmann's statement that the
BO are projectable to special (PDIQ) DO. In conclusion {\it
Lagrangian configurational predictability must be equivalent to
the statement of off-shell gauge invariance}. The moral is that
the complexity of the issue should warn against any unconstrained
utilization of geometric intuitions in dealing with the initial
value problem of general relativity.

This Hamiltonian predictability of BO entails in turn that only
{\it four functionally independent BO can exist for the vacuum
gravitational field}, since the latter has only two pairs of
conjugate independent degrees of freedom. Let us see now why
Bergmann's multiple definition of BO raises additional subtle
problems.

Bergmann himself proposed a constructive procedure for the BO.
This is essentially based on his re-interpretation of Einstein's
{\it coincidence argument} in terms of the individuation of
space-time points as point-events by using {\it intrinsic
pseudo-coordinates}. In his - already quoted - words\cite{2}: "we
might then identify a {\it world point}
(location-plus-instant-in-time) by the values assumed by (the four
intrinsic pseudo-coordinates) and ask for the value, there and
then, of a fifth field". As an instantiation of this procedure,
Bergmann refers to Komar's \cite{30} pseudo-tensorial
transformation of the 4-metric tensor to the {\it intrinsic
pseudo-coordinate} system [$\sigma^A = \sigma^A ({\bar
\sigma}^{\bar A})$ is the inversion of Eqs.(\ref{IV2})]

\beq
 {}^4{\bar g}_{\bar A\bar B}({\bar \sigma}^{\bar C} ) = {{\partial \sigma^C}\over
{\partial {\bar \sigma}^{\bar A} }}\, {{\partial \sigma^D}\over
{\partial {\bar \sigma}^{\bar B}}}\, {}^4g_{CD}(\sigma ).
 \label{V1}
 \eeq

\noindent The ${}^4{\bar g}_{\bar A\bar B}$ represent {\it ten
invariant scalar} \,(PDIQ or {\it tetradic}) {\it components} of
the metric; of course, they are not all independent since must
satisfy eight functional restrictions following from Einstein's
equations.

Now, Bergmann {\it claims} that the ten components $ {}^4{\bar
g}_{\bar A\bar B}({\bar \sigma}^{\bar C} )$ are a complete, but
non-minimal, set of BO. {\it This claim, however, cannot be true}.
As already pointed out, since BO are predictable they must in fact
be equivalent to (PDIQ) DO so that, for the vacuum gravitational
field, exactly four functions at most, out of the ten components
${}^4{\bar g}_{\bar A\bar B}({\bar \sigma}^{\bar C} )$, can be
simultaneously BO and DO, while the remaining components must be
{\it non-predictable} PDIQ, counterparts of ordinary Hamiltonian
gauge variables.

On the other hand, as shown in Section IV, the four independent
degrees of freedom of the pure vacuum gravitational field, even
for Bergmann, {\it have allegedly already been exploited for the
individuation of point-events}. Besides, as Bergmann explicitly
asserts in his purely {\it passive} interpretation, the PDIQ $
{}^4{\bar g}_{\bar A\bar B}({\bar \sigma}^{\bar C} )$ identify
on-shell a 4-geometry, i.e. an equivalence class in ${}^4Geom =
{}^4Riem / {}_PDiff\, M^4$. Furthermore, as shown in Section II,
Eq.(\ref{II7}), the identification of the same 4-geometry starting
from active diffeomorphisms can be done by using their passive
re-formulation (the group $Q$). Finally, let us remark that
Bergmann's intention to exploit first the intrinsic
pseudo-coordinates and then "ask for the value, there and then, of
a fifth field" makes sense only if such "{\it fifth field}" is a
{\it matter field}. Asking the question for purely gravitational
quantities like ${}^4{\bar g}_{\bar A\bar B}({\bar \sigma}^{\bar
C} )$ would be at least tautological since, as we have seen, only
four of them can be independent and have already been exploited.
If the individuation procedure is intended to be effective, it
would make little sense to assert that point-events have such and
such values in terms of point-events.

\medskip

But now, Bergmann's incorrect claim is relevant also to another
interesting quandary. Indeed, Bergmann's {\it main}
configurational notion of {\it observable} and its implications
are accepted as they stand in the already quoted paper of John
Earmann, Ref.\cite{3}. In particular Earman notes that the
intrinsic coordinates [${\bar \sigma}^{\bar C}$] can be used to
support Bergmann's observables and says "one can speak of the
event of the metric - components - $ {}^4{\bar g}_{\bar A\bar
B}({\bar \sigma}^{\bar C} )$ - having - such - and - such - values
- in - the - coordinate - system - $\{ {\bar \sigma}^{\bar C}  \}$
- at - the - location - where - the - ${\bar \sigma}^{\bar C}$ -
take - on - values - such - and - so" and (aptly) calls such an
item a {\it Komar event}, adding moreover that "the fact that a
given Komar event occurs (or fails to occur) is an observable
matter in Bergmann's sense, albeit in an abstract sense because
how the occurrence of a Komar event is to be observed/measured is
an unresolved issue" \footnote{ For our point of view about this
issue, see Section VII}. Earman's principal aim, however, is to
exploit Bergmann's definition of BO to show that "it implies that
there is no physical change, i.e., no change in the observable
quantities, at least not for those quantities that are
constructible in the most straightforward way from the materials
at hand". Although we are not committed here to object to what
Earman calls "modern Mc-Taggart argument" about change, we are
obliged to take issue against Earman's radical {\it universal
no-change argument} because, if sound, it would contradict the
substance of Bergmann's definition of predictability and would
jeopardize the relation between BO and DO which is fundamental to
our program.

In order to scrutinize this point, let us resume, for the sake of
clarity, the essential basic ingredients of the present
discussion.

{\it One}: the equations of motion derived from Einstein-Hilbert
Action and those derived from ADM Action have {\it exactly the
same physical content}: the ADM Lagrangian leads, through the
Legendre transformation, to equations equivalent to Einstein's
equations. {\it Two}: Hamiltonian predictability must, therefore,
be equivalent to Lagrangian predictability: specification of the
latter, however, is awkward. {\it Three}: the only functionally
independent Hamiltonian predictable quantities for the vacuum
gravitational field, are 4 DO. {\it Four}: by inverse Legendre
transformation, every DO has {\it a Lagrangian predictable
counterpart}. Then, a priori, one among the following three
possibilities might be true: i) all the existing BO must also be
DO; this means however that only four functionally independent BO
can exist; ii) some of the existing BO are also DO while other are
not; iii) no one of the existing BO is also a DO. Possibilities
ii) and iii) entail that Bergmann's multiple definition (that
including the third part) of BO is inconsistent, so that no BO
satisfying such multiple definition would exist. Yet, the third
part of Bergmann definition is essential for the overall meaning
of it since no Lagrangian definition of predictability {\it
independent of its Hamiltonian counterpart} can exist because of
{\it Two}. Thus cases ii) and iii) imply inconsistency of the very
concept of Bergmann's observabiity. Of course, it could be that
even i) is false since, after all, Bergmann did not prove the
self-consistency of his multiple definition: but this would mean
that no Lagrangian predictable quantity can exist which is
simultaneously a PDIQ. Here, we are {\it assuming} that Bergmann's
multiple definition is consistent and that i) is true. We will
formalize this assumption into a definite {\it constructive
conjecture} later on in this Section.

Let us take up again the discussion about the reality of change.
As already noted, the discussion in terms of BO in the language of
{\it Komar events} (or {\it coincidences}) must be restricted to
the properties of matter fields because, consistently with the
multiple Bergmann's definition, only four of the BO can be purely
gravitational in nature. And, if these latter have already been
exploited for the individuation procedure, it would again make
little sense to ask whether point-events do or do not change.
Therefore, let us consider Earman's argument by examining his
interpretation of predictability and the consequent implications
for a BO, say $B(p)$, $p \in M^4$, which, besides depending on the
4-metric and its derivatives up to some finite order, also depends
on matter variables, and is of course a PDIQ. In order to simplify
the argument, Earman concentrates on the special case of the
vacuum solutions to the Einstein's field equations, asserting
however that the argument easily generalizes to non-vacuum
solutions. Since we have already excluded the case of vacuum
solutions, let us take for granted that this generalization is
sound. Earman argues essentially in the following way: 1) There
are existence and uniqueness proofs for the initial value problem
of Einstein's equations, which show that for appropriate initial
data associated to a three manifold $\Sigma_o \subset M^4$, there
is a unique {\it up to diffeomorphism} (obviously to be intended
{\it active}) \footnote{ Note that Earman deliberately deviates
here from the purely {\it passive viewpoint} of Bergmann (and of
the standard Cauchy problem for partial differential equations) by
resorting to {\it active} diffeomorphisms in place of the {\it
coordinate transformations that leave the initial data unchanged}
or, possibly, in place of their extension in terms of the passive
re-interpretation of active diffeomorphisms ($Q$ group).} maximal
development for which $\Sigma_o$ is a Cauchy surface; 2) By
definition, a BO is a PDIQ whose value $B(p)$ at some point $p$ in
the future of $\Sigma_o$ is predictable from initial data on
$\Sigma_o$. If $\phi : M^4 \mapsto M^4$ is an {\it active}
diffeomorphism that leaves $\Sigma_o$ and its past fixed, the
point $p$ will be sent to the point $p^{'} = \phi \cdot p$. Then,
the general covariance of Einstein's equations, {\it conjoined
with predictability}, is interpreted to imply $B(p) = B(\phi \cdot
p)$. This result, together with the definition $B'(p) =
B(\phi^{-1}\cdot p)$ of the drag along of $B$ under the active
diffeomorphism $\phi$, entails $\phi^*B = B$ for a BO. In
conclusion, since $\Sigma_o$ is arbitrary, a matter BO should be
{\it constant everywhere} in $M^4$.

\bigskip

It is clear that this conclusion cannot hold true for any matter
dependent BO that is projectable to the DO of the gravitational
field {\it cum} matter, if only for the fact that such BO are in
fact ruled by the weak {\it ADM energy} which generates real
temporal change (see Section III and Eq.(\ref{V3}) below). The
crucial point in Earman's argument is the assertion that
predictability implies $B(p) = B(\phi \cdot p)$. But this does
{\it not} correspond to the property of {\it off-shell gauge
invariance} spelled above as the main qualification of predictable
quantities, except of course for the trivial case of quantities
everywhere constant. As clarified in Section II, the relations
between active diffeomophisms and gauge transformation (which are
necessarily involved by the DO) is not straightforward. Precisely,
because of the properties of the group $Q$ of Bergmann and Komar,
we have to distinguish between the {\it active} diffeomorphisms in
Q that {\it do belong} to $Q_{can}$ and those that {\it do not
belong} to $Q_{can}$ \footnote{Recall that:
\medskip
i) the intersection $Q_{can} \cap {}_PDiff\, M^4$ identifies the
space-time {\it passive} diffeomorphisms which, respecting the 3+1
splitting of space-time, are {\it projectable} to ${\cal G}_{4\,
P}$ in phase space;

ii) the remaining elements of $Q_{can}$ are the {\it projectable}
subset of {\it active diffeomorphisms} in their passive view.).

iii) the elements of Q which do not belong to $Q_{can}$ are not
projectable to phase space at all.}.

It is clear that, lacking projectability to phase space, these
latter do not correspond to gauge transformations and have,
therefore, nothing to do with Lagrangian {\it predictability}.
Thus, for most {\it active} diffeomorphisms\footnote{ In
particular, the special $\phi$'s considered by Earman}, the
conclusion $B(p) = B(\phi \cdot p)$ cannot hold true. This
erroneous conclusion seems to be just an instantiation of how
misleading may be any loose geometrical and non-algorithmic
interpretation of $\Sigma_o$ as a Cauchy surface within the
Lagrangian configuration approach to the initial value problem of
general relativity.

\bigskip

Having settled this important point, let us come back to tetradic
fields. Besides the tetradic components (\ref{V1}) of the
4-metric, we have to take into account the extrinsic curvature
tensor ${}^3K^{AB}(\sigma ) = {{\partial \sigma^A}\over {\partial
x^{\mu}}}\, {{\partial \sigma^B}\over {\partial x^{\nu}}}\,
{}^3K^{\mu\nu}(x)$. In the coordinates $\sigma^A$ adapted to
$\Sigma_{\tau}$, it has the components ${}^3K^{\tau\tau}(\sigma )
= {}^3K^{\tau r}(\sigma ) = 0$ and ${}^3K^{rs}(\sigma )$ [see
Eq.(\ref{a6})] and we can rewrite it as

\beq
 {}^3{\bar K}^{\bar A\bar B}({\bar \sigma}^{\bar C}) = {{\partial
 {\bar \sigma}^{\bar A}}\over {\partial \sigma^A}}\, {{\partial
 {\bar \sigma}^{\bar B}}\over {\partial \sigma^B}}\, {}^3K^{AB}(\sigma )
 =  {{\partial
 {\bar \sigma}^{\bar A}}\over {\partial \sigma^r}}\, {{\partial
 {\bar \sigma}^{\bar B}}\over {\partial \sigma^s}}\, {}^3K^{rs}(\sigma ).
 \label{V2}
 \eeq

\noindent In this way we get 10 additional scalar (tetradic)
quantities (only two of which are independent) replacing
${}^3K^{rs}(\sigma )$ and, therefore, the ADM momenta ${}^3{\tilde
\Pi}^{rs}(\sigma )$ of Eqs.(\ref{a10}).

In each intrinsic coordinate system ${\bar \sigma}^{\bar A} =
F^{\bar A}[\Lambda^{(k)}_W(\sigma )]$, we get consequently the 20
scalar (tetradic) components ${}^3{\bar g}_{\bar A\bar B}({\bar
\sigma}^{\bar C})$ and ${}^3{\bar K}^{\bar A\bar B}({\bar
\sigma}^{\bar C})$ of Eqs.(\ref{V1}), (\ref{V2}). However, only 16
out of them are functionally independent because of the 4 scalar
intrinsic constraint ${\bar {\cal H}}^{\bar A}({\bar \sigma}^{\bar
C}) = {{\partial {\bar \sigma}^{\bar A}}\over {\partial
\sigma^A}}\, {\cal H}^A(\sigma ) \approx 0$ that replace the
super-hamiltonian and super-momentum constraints ${\cal
H}^a(\sigma ) = \Big( {\cal H}(\sigma ); {\cal H}^r(\sigma ) \Big)
\approx 0$.

\bigskip

The various aspects of the discussion given above strongly suggest
that, in order to give consistency to Bergmann's unresolved
multiple definition of BO and, in particular, to his (strictly
speaking unproven) claim \cite{2} of {\it existence} of DO that
are simultaneously (PDIQ) BO, the following {\it conjecture}
should be true:

\bigskip
\noindent {\bf Main Conjecture}: "The Darboux basis whose 16 ADM
variables consist of the 8 Hamiltonian gauge variables $N$, $N_r$,
$\xi^r$, $\pi_{\phi}$, the 3 Abelianized constraints ${\tilde
\pi}_r^{{\vec {\cal H}}} \approx 0$, the conformal factor $\phi$
(to be determined by the super-hamiltonian constraint) and the
(non-tensorial) DO $r_{\bar a}$, $\pi_{\bar a}$, appearing in the
quasi-Shanmugadhasan canonical basis (\ref{III4}) {\it can be
replaced} by a Darboux basis whose 16 variables are all PDIQ (or
tetradic variables), such that four of them are simultaneous DO
and BO, eight vanish because of the first class constraints, and
the other 8 are coordinate-independent gauge variables.\footnote{
Note that Bergmann's constructive method based on tetradic
4-metric is not by itself conclusive in this respect !}"
\bigskip

If this conjecture is sound, it would be possible to construct an
{\it intrinsic} Darboux basis  of the Shanmugadhasan type
(Eq.(\ref{III4})). Then a $F$-dependent transformation performed
off-shell before adding the gauge fixings $\sigma^A - {\bar
\sigma}^{\bar A}(\sigma ) \approx 0$, should exist bringing from
the non-tensorial Darboux basis (\ref{III4}) to this new {\it
intrinsic} basis. Since the final result would be a representation
of the gauge variables as coordinate-independent (PIDQ) gauge
variables and of the DO as {\it Dirac-and-Bergmann observables},
the only $F$-dependence should consist in the possibility of
mixing the PDIQ gauge variables among themselves and in making
canonical transformations in the subspace of the Dirac-Bergmann
observables.

More precisely, we would have a family of quasi-Shanmugadhasan
canonical bases in which all the variables are PDIQ and include 8
PDIQ first class constraints that play the role of momenta. It
would be interesting to check the form of the eighth constraint
replacing the standard super-hamiltonian constraint. By
re-expressing the 4 Weyl eigenvalues in terms of anyone of these
PDIQ canonical bases, we could still define a Hamiltonian gauge,
namely an on-shell 4-coordinate system and then derive the
associated individuation of point-events by means of gauge-fixings
of the type (\ref{IV5}). {\it Note that this would break general
covariance even if the canonical basis is PDIQ} ! The only
difference with respect to the standard bases would be that,
instead of being non-tensorial quantities, both $r^G_a$, $\pi^G_a$
and ${\tilde F}^{\bar A}_G$ in Eq.(\ref{IV8}) would be PDIQ.

\medskip

As anticipated in the Introduction, further strong support to the
conjecture comes from Newman-Penrose formalism \cite{38} where the
basic tetradic fields are the 20 Weyl and Ricci scalars which are
PDIQ by construction . While the vanishing of the Ricci scalars is
equivalent to Einstein's equations (and therefore to a scalar form
of the super-hamiltonian and super-momentum constraints), the 10
Weyl scalars plus 10 scalars describing the ADM momenta
(restricted by the four primary constraints) should lead to the
construction of a Darboux basis spanned only by PDIQ restricted by
eight PDIQ first class constraints. Again, a quasi-Shanmugadhasan
transformation should produce the Darboux basis of the conjecture.
The problem of the phase space re-formulation of Newman-Penrose
formalism is now under investigation \cite{75}.
\bigskip

A final important logical component of the issue of the
objectivity of change is the particular question of {\it temporal}
change. This aspect of the issue is not usually tackled as a
sub-case of Earman's {\it no-universal-change} argument discussed
above in terms of BO, so it should be answered separately. We
shall restrict our remarks to the objections raised by Belot and
Earmann \cite{32} and Earman \cite{3} (see also
Refs.\cite{33,34,35} for the so called {\it problem of time} in
general). According to these authors, the reduced phase space of
general relativity is a frozen space without evolution (see in
particular the penetrating and thorough discussion about time and
change given in Ref.\cite{3}). Belot and Earman draw ontological
conclusions about the absence of real (temporal) change in general
relativity from the circumstance that, in spatially compact models
of general relativity, the Hamiltonian temporal evolution boils
down to a mere gauge transformation and is, therefore, physically
meaningless. We want to stress, however, that this result {\it
does not apply to all families of Einstein space-times}. In
particular, there exist space-times like the
Christodoulou-Klainermann space-times \cite{36} we are using in
this paper that constitute a {\it counterexample to the frozen
time argument}. They are defined by suitable boundary conditions,
are globally hyperbolic, non-compact, and asymptotically flat at
spatial infinity. The existence of such meaningful counterexamples
entails, of course, that we are not allowed to draw negative {\it
ontological} conclusions in general about the issue of temporal
change in general relativity.

\medskip

More precisely, it is possible to show \cite{44} (see also Section
III) that:

\bigskip

1) The imposition of suitable boundary conditions on the fields
and the gauge transformations of canonical ADM metric gravity
eliminates the super-translations and reduces the asymptotic
symmetries at spatial infinity to the asymptotic ADM Poincar\'e
group. The asymptotic implementation of Poincar\'e group makes
possible the general-relativistic definition of angular momentum
and the matching of general relativity with particle physics.

\bigskip

2) The boundary conditions of point 1) require that the leaves of
the foliations associated with the admissible 3+1 splittings of
space-time must tend to Minkowski space-like hyper-planes
asymptotically orthogonal to the ADM 4-momentum in a
direction-independent way. This property is concretely enforced by
using a technique introduced by Dirac \cite{45} for thye selection
of space-times admitting asymptotically flat 4-coordinates at
spatial infinity. \footnote{ Dirac's method brings to an
enlargement of ADM canonical metric gravity with non-vanishing ADM
Poincar\'e charges. Such space-times admit preferred asymptotic
inertial observers, interpretable as fixed stars (the standard for
measuring rotations). Such non-Machian properties allow to merge
the standard model of elementary particles in general relativity
with all the (gravitational and non-gravitational) fields
belonging to the same function space (suitable weighted Sobolev
spaces). Besides the existence of a realization of the Poincar\'e
group, only one additional property is required: namely that the
space-like hyper-surfaces admit an involution \cite{76} allowing
the definition of a generalized Fourier transform with its
associated concepts of positive and negative energy. This
disproves the claimed impossibility of defining particles in
curved space-times \cite{77}.}

\bigskip

3) The super-hamiltonian constraint is the generator of the gauge
transformations connecting different admissible 3+1 splittings of
space-time and {\it has nothing to do with the temporal
evolution}.

\bigskip

4) As shown by DeWitt \cite{64}, and already stressed by us, the
weakly vanishing ADM Dirac Hamiltonian has to be modified with a
suitable surface term in order that functional derivatives,
Poisson brackets and Hamilton equations be mathematically well-
defined in such non-compact space-times. This fact, in conjunction
with the points 1), 2), 3) above, entails that {\it there is an
effective evolution in the mathematical time} which parametrizes
the leaves of the foliation associated with any 3+1 splitting.
Such evolution is ruled by the weak {\it ADM energy} \cite{44,63},
i.e. by a non-vanishing Hamiltonian which exists also in the
reduced phase space. This is the {\it rest-frame instant form of
metric gravity}. As seen in Section III, each gauge fixing creates
a realization of $\Gamma_4$ and the {\it weak {\it ADM} energy} is
a functional of only the DO of that gauge. Then, the DO themselves
(as any other function of them) satisfy the Hamilton equations

 \beq
 \dot r^G_{\bar a} = \{ r^G_{\bar a}, E_{\mathrm{ADM}}\}^*,
\quad \dot \pi^G_{\bar a} = \{\pi^G_{\bar a},
E_{\mathrm{ADM}}\}^*,
 \label{V3}
  \eeq

\noindent where $E_{\mathrm{ADM}}$ is intended as the restriction
of the weak {\it ADM} Energy to $\Gamma_4$ and where the
$\{\cdot,\cdot\}^*$ are Dirac Brackets.

\bigskip

5) When matter is present in this family of space-times, switching
off Newton's constant  ($G \mapsto 0$) yields the description of
matter in Minkowski space-time foliated with the space-like
hyper-planes orthogonal to the total matter 4-momentum (Wigner
hyper-planes intrinsically defined by matter isolated system). In
this way one gets the rest-frame instant form of dynamics
reachable from parametrized Minkowski theories. Incidentally, this
is the first example of consistent deparametrization of general
relativity in which the ADM Poincar\'e group tends to the
Poincar\'e group of the isolated matter system.

\bigskip

We can conclude that in these space-times there is neither a
frozen reduced phase space nor a Wheeler-DeWitt interpretation
based on some local concept of time like in compact space-times.
Therefore, {\it our gauge-invariant approach to general relativity
is perfectly adequate to accommodate objective temporal change}.

\bigskip

A further objection raised by Belot and Earmann \cite{32} is that
the {\it non-locality} of the gauge-invariant DO for the
gravitational field is "philosophically unappealing". It may be,
but, first of all, we can easily retort that this property is a
unavoidable consequence\footnote{ Within the Hamiltonian approach;
see for instance Ref.\cite{65} for the simpler case of Yang-mills
equations} of establishing a well-posed Cauchy problem for any
system of non-hyperbolic under-determined partial differential
equations, like Einstein's equations.\footnote{ This feature has a
Machian flavor, although in a non-Machian context: with or without
matter, the whole 3-space is involved in the definition of the
observables. Furthermore, these space-times allow the separation
\cite{44} of the 4-center of mass of the universe ({\it decoupled
point particle clock}) reminding the Machian statement that only
relative motions are dynamically relevant. See Ref.\cite{78} for a
thorough discussion of Mach's influence on Einstein and for
comments on the ontology of space-time (following from the hole
argument) consistent with our interpretation.} Second, just to the
contrary, we believe that the specific property of DO for a pure
gravitational field of being {\it non local in terms of manifold's
mathematical points} is indeed philosophically appealing and
displays the real physical content of Leibniz equivalence (see
more in the Concluding Survey).

\vfill\eject

\section{On the Physical Interpretation of Dirac Observables and
Gauge Variables: Tidal-like and Inertial-like Effects.}

Having settled the problem of the physical individuation of
space-time points and that of the observables, let us discuss with
a greater detail the physical meaning of the Hamiltonian gauge
variables and DO.

As shown in Section III, the 20 off-shell canonical variables of
the ADM Hamiltonian description are naturally subdivided into {\it
two sets} by the quasi-Shanmugadhasan transformation:
\bigskip

i) The first set contains {\it seven off-shell Abelian Hamiltonian
gauge variables} whose conjugate momenta are seven Abelianized
{\it first class constraints}. The eighth canonical pair comprises
the variable in which the super-hamiltonian constraint has to be
solved (the conformal factor of the 3-metric, $\phi =
({}^3g)^{1/12}$) and its conjugate momentum as the eighth gauge
variable. Precisely, the gauge variables are:

\bea
 &&\pi_{\phi},\quad momentum\, conjugate\, to\, the\, conformal\, factor\,
 (primary\, gauge\, variable),\nonumber \\
 &&{}\nonumber \\
 &&N = \sqrt{ {{{}^4g}\over {{}^3g}} } = \sqrt{{}^4g_{\tau\tau} -
 {}^3g^{rs}\, {}^4g_{\tau r}\, {}^4g_{\tau s}},\quad lapse\, function\,
 (its\, secondary\, gauge\, variable),\nonumber \\
 &&{}\nonumber \\
 &&\xi^r,\quad (non-local)\, parameters\, of\, Diff_P\, \Sigma_{\tau}\,
 (primary\, gauge\, variables),\nonumber \\
 &&{}\nonumber \\
 &&N_r = - {}^4g_{\tau r},\quad shift\, functions\, (their\, secondary\,
 gauge\, variables).
 \label{VI1}
 \eea

\noindent Note that a "primary" gauge variable has its
arbitrariness described by a Dirac multiplier, while a "secondary"
gauge variable inherits the arbitrariness of the Dirac multipliers
through the Hamilton equations.

\bigskip

ii) The second set contains the {\it off-shell gauge invariant
(non-local and in general non tensorial) DO}: $r_{\bar a}(\tau
,\vec \sigma )$, $\pi_{\bar a}(\tau ,\vec \sigma )$, $\bar a =
1,2$. They satisfy hyperbolic Hamilton equations and are not BO in
general.\vskip 0.5cm

Let us stress again that the above subdivision of canonical
variables in two sets is a peculiar outcome of the
quasi-Shanmugadhasan canonical transformation which has no simple
counterpart within the {\it Lagrangian viewpoint} at the level of
the Hilbert action and/or of Einstein's equations: at this level
the only clear statement is whether or not the curvature vanishes.
As anticipated in the Introduction this subdivision amounts to an
extra piece of (non-local) information which should be added to
the traditional wisdom of the {\it equivalence principle}
asserting the local impossibility of distinguishing gravitational
from inertial effects. Indeed, we shall presently see that it
allows to distinguish and visualize which aspects of the local
physical effects on test matter contain a {\it genuine
gravitational} component and which aspects depend solely upon the
choice of the (local) reference frame and/or coordinate system:
these latter could then be called {\it inertial}, in analogy with
the non-relativistic Newtonian situation. Recall that when a
complete choice of gauge is made, the gauge variables become fixed
uniquely by the gauge-fixing procedure to functions of DO in that
gauge.

One should be careful in discussing this subject because the very
definition of {\it inertial force} (and {\it gravitational} as
well) seems rather unnatural in general relativity. We can take
advantage, however, from the circumstance that Hamiltonian point
of view leads naturally to a re-reading of geometrical features in
terms of the traditional concept of {\it force}.

First of all, recall that we are still considering here the case
of pure gravitational field without matter. We know from Section
III that on-shell, in any chosen 4-coordinate system, the
mathematical representation of any physical effects is given in
terms of functionals of the non-local DO in the completely fixed
Hamiltonian gauge that corresponds to the chosen 4-coordinates. It
is then natural first of all to characterize as genuine
gravitational effects those which are directly correlated to the
{\it DO}, and crucial to stress that such purely gravitational
effects {\it are absent in Newtonian gravity}, where there are no
autonomous gravitational fields, i.e., fields not generated by
matter sources. It seems therefore plausible to trace {\it
inertial} (much better than {\it fictitious}, in the relativistic
case) effects to a pure dependence on the {\it Hamiltonian gauge
variables}\footnote{By introducing dynamical matter the
Hamiltonian procedure leads to distinguish among {\it
action-at-a-distance, gravitational, and inertial} effects, with
consequent relevant implications upon concepts like gravitational
passive and active masses and, more generally, upon the problem of
the origin of inertia. See Ref.\cite{79} for other attempts of
separating inertial from tidal effects in the equations of motion
in configuration space for test particles, in a framework in which
asymptotic inertial observers are refuted. In this reference one
finds also the following version (named Mach 11) of the Mach
principle: "The so-called {\it inertial effects}, occurring in a
non-inertial frame, are gravitational effects caused by the
distribution and motion of the distant matter in the universe,
relative to the frame". Thus {\it inertial} means here non-tidal +
true gravitational fields generated by cosmic matter. In the above
reference it is also suggested that superfluid Helium II may be an
alternative to fixed stars as a standard of non rotation. Of
course all these interpretations are questionable. On the other
hand, the Hamiltonian framework offers the tools for making such a
distinction while distant matter effects are hidden in the
non-locality of DO and gauge variables. Since in a fixed gauge the
gauge variables are functions of the Dirac obervables in that
gauge, tidal effects are clearly mixed with inertial ones. For a
recent critical discussion about the origin of inertia and its
connection with inertial effects in accelerated and rotating
frames see Ref.\cite{80}.}. Recall also that, at the
non-relativistic level, Newtonian gravity is fully described by
{\it action-at-a-distance forces} and, in absence of matter, there
are no {\it tidal} forces among test particles. Tidal-like forces
are {\it entirely} determined by the variation of the
action-at-a-distance force created by the Newton potential of a
massive body on the test particles. In vacuum general relativity
instead the {\it geodesic deviation equation} shows that tidal
forces, locally described by the Riemann tensor, act on test
particles even in absence of any kind of matter.

\bigskip

We can also say, therefore, that the role of the Hamiltonian gauge
variables, whose form change from one gauge fixing to another, is
that of describing the {\it form} in which physical gravitational
effects determined by the DO show themselves. Such {\it
appearances} undergo {\it inertial} changes upon going (on-shell)
from one coordinate system (or gauge fixing) to another and above
all from one reference frame to another (see later on).
Furthermore, while {\it purely inertial} effects could be manifest
even in case the DO are null, genuine gravitational effects are
always necessarily dressed by inertial-like appearances. Thus, the
situation is only vaguely analogous to the phenomenology of
non-relativistic inertial forces. These latter describe {\it
purely apparent} (or really {\it fictitious}) mechanical effects
which show up in accelerated Galilean reference frames
\footnote{With arbitrary {\it global} translational and rotational
3-accelerations.} and can be eliminated by going to (global)
inertial reference frames \footnote{See Ref.\cite{81} for the
determination of {\it quasi-inertial reference frames} in
astronomy as those frames in which rotational and linear
acceleration effects lie under the sensibility threshold of the
measuring instruments.}. Besides the existence of autonomous
gravitational degrees of freedom, it is therefore clear that the
further deep difference concerning inertial-like forces in the
general-relativistic case with respect to Newtonian gravity rests
upon the purely {\it local nature} of general relativistic
inertial frames of reference and/or coordinate systems.

\bigskip

For the sake of clarity, consider the non-relativistic Galilean
framework in greater detail. If a {\it global non-inertial
reference frame} has translational acceleration $\vec w(t)$ and
angular velocity $\vec \omega (t)$ with respect to a given
inertial frame, a particle with free motion ($\vec a = {\ddot
{\vec x}} = 0$) in the inertial frame has the following
acceleration as seen from the non-inertial frame

\beq
 {\vec a}_{NI} = - \vec w(t) + \vec x \times {\dot {\vec
\omega}}(t) + 2 {\dot {\vec x}} \times \vec \omega (t) + \vec
\omega (t) \times [\vec x \times \vec \omega (t)].
 \label{VI2}
 \eeq

\noindent After multiplication of this equation by the particle
mass, the second term on the right hand side is the {\it Jacobi
force}, the third is the {\it Coriolis force} and the fourth the
{\it centrifugal force}.

We have given in Ref.\cite{82} a description of non-relativistic
gravity which is generally covariant under {\it arbitrary} passive
Galilean coordinate transformations [$t^{'} = T(t)$, ${\vec x}^{'}
= \vec f (t, \vec x)$]. The analogue of Eq.(\ref{VI2}) in this
case contains more general {\it apparent} forces, which reduce to
those appearing in Eq.(\ref{VI2}) in particular rigid coordinate
systems. The discussion given in Ref.\cite{82} is a good
introduction to the relativistic case, just because in general
relativity there are no {\it global inertial reference frames}.
\bigskip

Two different approaches have been considered in the literature in
the general relativistic case concerning the choice of reference
frames, namely using either

\bigskip

i) a single accelerated time-like observer with an arbitrary
associated tetrad,

\vskip 0.5cm or \vskip 0.5cm

ii) a congruence of accelerated time-like observers\footnote{While
in Newtonian physics an {\it absolute reference frame} is an
imagined extension of a rigid body and a clock (with any
coordinate systems attached), in general relativity \cite{83} we
must replace the rigid body either by a cloud of test particles in
free fall ({\it geodesic congruence}) or by a test fluid ({\it
non-geodesic congruence} for non-vanishing pressure). Therefore a
{\it reference frame} is schematized as a future-pointing
time-like congruence with all the possible associated 4-coordinate
systems. This is called a {\it platform} in Ref.\cite{84}, where
there is a classification of the possible types of platforms and
the definition of the position vector of a neighboring observer in
the local rest frame of a given observer of the platform. Then,
the Fermi-Walker covariant derivative (applied to a vector in the
rest frame it produces a new vector still in the rest frame
\cite{85}) is used to define the 3-velocity (and then the
3-acceleration) of a neighboring observer in the rest frame of the
given observer, as the natural generalization of the Newtonian
relative 3-velocity (and 3-acceleration). See Ref.\cite{86} for a
definition, based on these techniques, of the 3-acceleration of a
test particle in the local rest frame of an observer crossing the
particle geodesics, with the further introduction of the Lie and
co-rotating Fermi-Walker derivatives.} with a conventionally
chosen associated field of tetrads\footnote{The time-like tetrad
field is the 4-velocity field of the congruence. The conventional
choice of the spatial triad is equivalent to a choice of a
specific system of gyroscopes (see footnote 78 in Appendix B for
the definition of a Fermi-Walker transported triad). See the local
interpretation \cite{86} of {\it inertial} forces as effects
depending on the {\it choice of a congruence of time-like
observers with their associated tetrad fields} as a reference
standard for their description. Note that, in gravitational fields
without matter, gravito-magnetic effects as described by
${}^4g_{\tau r}$ are purely inertial effects in our sense, since
are determined by the {\it shift} gauge variables. While in metric
gravity the tetrad fields are used only to rebuild the 4-metric,
the complete theory taking into account all the properties of the
tetrad fields is {\it tetrad gravity} \cite{87}.}.

\bigskip

Usually, in both approaches the observers are {\it test
observers}, which describe phenomena from their kinematical point
of view without generating any dynamical effect on the system.
\bigskip

i) Consider first the case of a single {\it test observer} with
his tetrad (see Ref.\cite{88,89}).

After the choice of the associated local Minkowskian system of
(Riemann-Gaussian) 4-coordinates, the line element
becomes\footnote{If the test observer is in free fall ({\it
geodesic observer}) we have $\vec a = 0$. If the triad of the test
observer is Fermi-Walker transported (standard of {\it
non-rotation} of the gyroscope) we have $\vec \omega = 0$. } $ds^2
= - \delta_{ij} dx^i dx^j + 2 \epsilon_{ijk} x^j {{\omega^k}\over
c} dx^o dx^i + [1 + {{2 \vec a \cdot \vec x}\over {c^2}}
(dx^o)^2]$. The {\it test observer} describes a nearby time-like
geodesics $y^{\mu}(\lambda )$ ($\lambda$ is the affine parameter
or proper time) followed by a test particle in {\it free fall} in
a given gravitational field by means of the following spatial
equation: ${{d^2 \vec y}\over {(dy^o)^2}} = - \vec a - 2 \vec
\omega \times {{d \vec y}\over {dy^o}} + {2\over {c^2}} \Big( \vec
a \cdot {{d\vec y}\over {dy^o}}\Big)\, {{d\vec y}\over {dy^o}}$.
Thus, the relative acceleration of the particle with respect to
the observer with this special system of coordinates\footnote{It
replaces the global non-inertial non-relativistic reference frame.
With other coordinate systems, other terms would of course
appear.} is composed by the observer 3-acceleration plus a
relativistic correction and by a Coriolis acceleration\footnote{
This is caused by the rotation of the spatial triad carried by the
observer relative to a Fermi-Walker transported triad. The
vanishing of the Coriolis term justifies the statement that for an
observer which is not in free fall ($\vec a \not= 0$) a local
coordinate system produced by Fermi-Walker transport of the
spatial triad of vectors is the best possible realization of a
non-rotating system.}.  Note that, from the Hamiltonian point of
view, the constants $\vec a$ and $\vec \omega$ are constant
functionals of the DO of the gravitational field in this
particular gauge.

\medskip

As said above, different Hamiltonian gauge fixings on-shell,
corresponding to on-shell variations of the Hamiltonian gauge
variables, give rise to different appearances of the physical
effects as gauge-dependent functionals of the DO in that gauge of
the type ${\cal F}_{G}(r_{\bar a}, \pi_{\bar a})$ (like $\vec a$
and $\vec \omega$ in the previous example). \vskip 0.5truecm

In absence of matter we can consider the {\it zero curvature
limit}, which is obtained by putting the DO to zero. In this way
we get Minkowski space-time (a solution of Einstein's equations)
equipped with those kinds of coordinates systems which are
compatible with Einstein's theory \footnote{As shown in
Ref.\cite{44} this implies the vanishing of the Cotton-York
3-conformal tensor, namely the condition that the allowed 3+1
splittings of Minkowski space-time compatible with Einstein's
equations have the leaves {\it 3-conformally flat} in absence of
matter. This solution of Einstein's equations, has been named {\it
void space-time} in Ref.\cite{87}: Minkowski space-time in
Cartesian 4-coordinates is just a gauge representative of it. Note
that, even if Einstein always rejected this concept, a void
space-time corresponds to the description of a special class of
4-coordinate systems for Minkowski space-time without matter.}. In
particular, the quantities ${\cal F}_G = lim_{r_{\bar a},
\pi_{\bar a} \rightarrow 0}\, {\cal F}_G(r_{\bar a}, \pi_{\bar
a})$ describe inertial effects in those 4-coordinate systems for
Minkowski space-time which have a counterpart in Einstein general
relativity. Note finally that special relativity, considered as an
autonomous theory, admits much more general inertial effects
associated with the 3+1 splittings of Minkowski space-time whose
leaves are not 3-conformally flat.
\medskip

In presence of matter Newtonian gravity is recovered with a double
limit:\medskip

a) the limit in which DO are restricted to the solutions of the
Hamilton equations (\ref{V3}), $r_{\bar a} \rightarrow f_{\bar
a}(matter)$, $\pi_{\bar a} \rightarrow g_{\bar a}(matter)$;

b) the  $c \rightarrow \infty$ limit, in which curvature effects,
described by matter after the limit a), disappear. \vskip
0.5truecm

This implies that the functionals ${\cal F}_G(r_{\bar a},
\pi_{\bar a})$ must be restricted to the limit ${\cal F}_{Newton}
= lim_{c \rightarrow \infty}\,\, lim_{r_{\bar a} \rightarrow
f_{\bar a}, \pi_{\bar a} \rightarrow g_{\bar a}}\,\,\Big( {\cal
F}_{G\,o} + {1\over c} {\cal F}_{G\, 1} + ...\Big) = {\cal F}_{G\,
o} {|}_{r_{\bar a} = f_{\bar a}, \pi_{\bar a} =  g_{\bar a}}$.
Then ${\cal F}_{Newton}$, which may be coordinate dependent,
becomes the {\it Newtonian inertial force} in the corresponding
general Galilean coordinate system.
\bigskip

ii) Consider then the more general case of a {\it congruence of
accelerated time-like observers}. In this way it is possible to
get a much more accurate and elaborate description of the relative
3-acceleration, as seen in his own local rest frame by each
observer of the congruence which intersects the geodesic of a test
particle in free fall (see Ref.\cite{86}). The identification of
various types of 3-forces depends upon:\medskip

a) the gravitational field (the form of the geodesics obviously
depends on the metric tensor; usually the effects of the
gravitational field are classified as gravito-electric and
gravito-magnetic, even if this is strictly valid only in harmonic
coordinates)
\medskip
b) the properties (acceleration, vorticity, expansion, shear) of
the congruence of observers,
\medskip
c) the choice of the time-parameter used to describe the particle
3-trajectory in the local observer rest frame.
\bigskip

\noindent There are, therefore, many possibilities for defining
the relative 3-acceleration (see Ref.\cite{86}) and its separation
in various types of inertial-like accelerations (See Appendix B
for a more complete discussion of the properties of the
congruences of time-like observers).
\bigskip

Summarizing, once a local reference frame has been chosen, in
every 4-coordinate system we can consider: \medskip

a) the {\it genuine tidal gravitational effects} which show up in
the geodesic deviation equation: they are well defined
gauge-dependent functionals of the DO associated to that gauge: DO
could then be called {\it non-local tidal-like degrees of
freedom};

b) the fact that {\it geodesic curves} will have different
geometrical descriptions corresponding to different gauges,
although they will be again functionally dependent only on the DO
in the relevant gauge;

c) the issue of the description of the {\it relative
3-acceleration of a free particle in free fall}, as given in the
local rest frame of a generic observer of the congruence, which
will contain various terms. Such terms are identifiable with the
general relativistic extension of the various non-relativistic
kinds of inertial accelerations and all will again depend on the
DO in the chosen gauge, both directly and through the Hamiltonian
gauge variables of that gauge.
\bigskip

Three general remarks: \medskip

First of all, the picture we have presented is not altered by the
presence of matter. The only new phenomenon besides the above
purely gravitational, {\it inertial and tidal} effects, is that
from the solution of the super-hamiltonian and super-momentum
constraints emerge {\it action-at-a-distance, Newtonian-like and
gravito-magnetic}, effects among matter elements, as already noted
in footnote 44.\medskip

Secondly, we would like to recall that Bergmann \cite{2} made the
following critique of general covariance: it would be desirable to
restrict the group of coordinate transformations (spacetime
diffeomorphisms) in such a way that it could contain an invariant
sub-group describing the coordinate transformations that change
the frame of reference of an outside observer; the remaining
coordinate transformations would be like the gauge transformations
of electromagnetism. Yet, this is just what we have done with the
redefinition of the lapse and shift functions after separating out
their asymptotic part (see footnote 18). In this way {\it
preferred} inertial asymptotic coordinate systems are selected
which can be identified as fixed stars.\medskip

Thirdly, it should be stressed that the {\it reference standards}
of time and length correspond to units of {\it coordinate time and
length} and not to proper times and proper lengths \cite{39}: this
is not in contradiction with general covariance, because an
extended {\it laboratory}, in which one defines the reference
standards, corresponds to a particular {\it completely fixed
on-shell Hamiltonian gauge} plus a local congruence of time-like
observers. For instance, in astronomy and in the theory of
satellites, the unit of time is replaced by a unit of coordinate
length ({\it ephemerides time}). This leads to the necessity of
taking into account the theory of measurement in general
relativity. This will be done in the next Section. \medskip

Fourthly, let us remark that the definitions given in Section V
for the {\it notion of observable} in general relativity are in
correspondence with the following two different points of view,
existing in the physical literature, that are clearly spelled out
in Ref.\cite{90} and related references, namely:\hfill\break

i) The {\it non-local point of view} of Dirac \cite{45}, according
to which causality implies that only gauge-invariant quantities,
i.e., DO, can be measured. As we have shown, this point of view is
consistent with general covariance. For instance, ${}^4R(\tau
,\vec \sigma )$ is a scalar under diffeomorphisms, and therefore a
BO,  but it is not a DO - at least the kinematical level - and
therefore, according to Dirac, not an observable quantity. Even if
${}^4R(\tau ,\vec \sigma )\, {\buildrel \circ \over =}\, 0$ in
absence of matter, the other curvature scalars do not vanish in
force of Einstein's equations and, lacking known solutions without
Killing vectors, it is not clear  their connection with the DO.
The 4-metric tensor ${}^4g_{\mu\nu}$ itself as well as the line
element $ds^2$ are not DO so a completely fixed gauge is needed to
get a definite functional form for them in terms of the DO in that
gauge. This means that all standard procedures for defining
measures of length and time \cite{20,85,60} and the very
definition of angle and distance properties of the material bodies
forming the reference system, are gauge dependent. Then they are
determined only after a complete gauge fixing and after the
restriction to the solutions of Einstein's equations has been
made\footnote{Note that in textbooks these procedures are always
defined without any reference to Einstein's equations.}. Likewise,
it is only after the gauge fixing that the procedure for measuring
the Riemann tensor with $n \geq 5$ test particles described in
Ref.\cite{60} (see also Ref.\cite{31}) becomes completely
meaningful, just as it happens for the electro-magnetic vector
potential in the radiation gauge.

Finally note that, after the introduction of matter, even the
measuring apparatuses should be described by the gauge invariant
matter DO associated with the given gauge. \hfill\break

ii) The {\it local point of view}, according to which the
space-time manifold $M^4$ is a kind of postulated (often without
any explicit statement) background manifold of physically
determinated {\it events}, like it happens in special relativity
with its absolute chrono-geometric structure. Space-time points
are assumed physically distinguishable, because any measurement is
performed in the frame of a given reference system interpreted as
a physical laboratory. In this view the gauge freedom of generally
covariant theories is reduced to mere passive coordinate
transformations. See for instance Ref. \cite{91} for a refusal of
DO in general relativity based on the local point of view. However
this point of view ignores completely the Hole Argument and must
renounce to a deterministic evolution, so that it is ruled out
definitely by our results.
\bigskip

In Ref.\cite{90} the non-local point of view is accepted and there
is a proposal for using some special kind of matter to define a
{\it material reference system} (not to be confused with a
coordinate system) to localize points in $M^4$, so to recover the
local point of view in some approximate way \footnote{The main
approximations are: 1) to neglect, in Einstein equations, the
energy-momentum tensor of the matter forming the material
reference system (it's similar to what happens for test
particles); 2) to neglect, in the system of dynamical equations,
the entire set of equations determining the motion of the matter
of the reference system (this introduces some {\it indeterminism}
in the evolution of the entire system).} because in the analysis
of classical experiments both approaches lead to the same
conclusions. This approach relies therefore upon {\it matter} to
solve the problem of the individuation of space-time points as
point-events, at the expense of loosing determinism. The emphasis
on the fundamental role of matter for the individuation issue
points is present also in Refs.\cite{92,93,33,35}, where {\it
material clocks} and {\it reference fluids} are exploited as test
matter.

As we have shown, the problem of the individuation can be solved
{\it before and without} the introduction of matter. Matter only
contributes to a deformed individuation and, obviously, is
fundamental in trying to establish a general-relativistic theory
of measurement.

\vfill\eject

\section{Implementing the Physical Individuation of Point-Events with
Well-Defined Empirical Procedures: a Realization of the Axiomatic
Structure of Ehlers, Pirani and Schild.}

The problem of the individuation of space-time points as
point-events cannot be methodologically separated from the problem
of defining a theory of measurement consistent with general
covariance. This means that we should not employ the absolute
chrono-geometric structures of special relativity, like it happens
in all the formulations on a given background (gravitational waves
as a spin two field over Minkowski space-time, string theory,...).
Moreover matter (either test or dynamical) is now an essential
ingredient for defining the experimental setup.

At present we do not have such a theory, but only preliminary
attempts and an empirical metrology \cite{39}, in which the
standard unit of time is a {\it coordinate time} and not a proper
time. As already said, a laboratory network with its standards
corresponds to a description givn in a completely fixed
Hamiltonian gauge viz., being on-shell, in a uniquely determined
4-coordinate system. We shall take into account the following
pieces of knowledge.

A) Ehlers, Pirani and Schild \cite{40} developed an axiomatic
framework for the foundations of general relativity and
measurements (reviewed in Appendix C for completeness). These
authors exploit the notions of {\it test objects} as {\it
idealizations} to the effect of approximating the {\it conformal,
projective, affine} and {\it metric} structures of Lorentzian
manifolds; such structures are then used to define {ideal geodesic
clocks} \cite{88}. The axiomatic structure refers to basic objects
such as {\it test light rays} and {\it freely falling test
particles}. The first ones are used in principle to reveal the
{\it conformal} structure of space-time, the second ones the {\it
projective} structure. Under an axiom of compatibility which is
well corroborated by experiment (see Ref.\cite{94}), it can be
shown that these two independent classes of observations determine
completely the structure of space-time. Let us remark that one
should extend this axiomatic theory to tetrad gravity (space-times
with frames) in order to include objects like {\it test
gyroscopes} needed to detect gravito-magnetic
effects.\footnote{Stachel \cite{16}, stresses the dynamical (not
axiomatic) aspect of the general relativistic space-times
structures associated to the behavior of ideal measuring rods
({\it geometry}) and clocks ({\it chronometry}) and free test
particles ({\it inertial structures})}.

B) De Witt \cite{92} introduced a procedure for measuring the
gravitational field based on a reference fluid (a stiff elastic
medium) equipped with material clocks. This phenomenological test-
fluid is then exploited to bring in Bergmann-Komar invariant
pseudo-coordinates $\zeta^{\bar A}$, $\bar A =1,..,4$, as a method
for coordinatizing the space-time where to do measurements and
also for grounding space-time geometry operationally, at least in
the weak field regime. De Witt essentially proposes to simulate a
mesh of local clocks and rods. Even if De Witt considers the
measurement of a weak quantum gravitational field smeared over
such a region, his procedure could even be adopted classically. In
this perspective, our approach furnishes the ingredients of the
Hamiltonian description of the gravitational field, which were
lacking at the time De Witt developed his preferred covariant
approach.

C) Antennas and interferometers are the tools used to detect
gravitational waves on the earth. The mechanical prototype of
these measurements are test springs with end masses feeling the
gravitational field as the tidal effect described by the geodesic
deviation equation \cite{88,95}. Usually, however, one works on
the Minkowski background in the limit of weak field and
non-relativistic velocities. See Ref.\cite{96} for the extension
of this method to a regime of weak field but with relativistic
velocities in the framework of a background-independent
Hamiltonian linearization of tetrad gravity.

\bigskip

Lacking solutions to Einstein's equations with matter
corresponding to simple systems to be used as idealizations for a
measuring apparatuses described by matter DO (hopefully also BO),
a generally covariant theory of measurement as yet does not exist.
We hope, however, that some of the clarifications achieved in this
paper of the existing ambiguities about observables will help in
developing such a theory.\bigskip

In the meanwhile we want to sketch in this last Section a scheme
for implementing - at least in principle - the physical
individuation of points as an experimental setup and protocol for
positioning and orientation. Our construction should be viewed in
parallel to the axiomatic treatment of Ehlers, Pirani and Schild.
We could reproduce the logical scheme of this axiomatic approach
in the following way.

a) A {\it radar-gauge} system of coordinates can be defined in a
finite four-dimensional volume by means of a network of artificial
satellites similar to the Global Position System \cite{97}. Let us
consider a family of spacecrafts, whose navigation is controlled
from the Earth by means of the standard GPS. Note that the GPS
receivers are able to determine their actual position and velocity
because the GPS system is based on the advanced knowledge of the
gravitational field of the Earth and of the satellites'
trajectories, which in turn allows the {\it coordinate}
synchronization of the satellite clocks . During the navigation
the spacecrafts are test objects. Once the spacecrafts have
arrived in regions with non weak fields, like near the Sun or
Jupiter, they become the (non test) elements of an experimental
setup and protocol for the determination of a local 4-coordinate
system and of the associated 4-metric.

Each satellite, endowed with an atomic clock and a system of
gyroscopes, may be thought as a time-like observer (the satellite
world-line) with a tetrad (the time-like vector is the satellite
4-velocity and the spatial triad is built with gyroscopes) and one
of them is chosen as the origin of the radar-4-coordinates we want
to define. This means that the natural framework should be tetrad
gravity instead of metric gravity.

Since the geometry of space-time and the motion of the satellites
are not known in advance in our case, we must think of the
receivers as obtaining four, so to speak, {\it conventional}
coordinates by operating a full-ranging protocol involving
bi-directional communication to four {\it super-GPS} that
broadcast the time of their standard unsynchronized clocks (see
the discussion given in Ref.\cite{5} and Refs.\cite{98} for other
proposals in the same perspective). This first step parallels the
axiomatic construction of the {\it conformal structure} of
space-time.
\medskip

b) Then, choose Einstein's simultaneity convention to synchronize
the atomic clocks by means of radar signals \cite{99}, with
respect to the satellite chosen as origin: this allows
establishing a {\it radar-gauge system of 4-coordinates} lacking
any direct metrical content

\beq
 \sigma^A_{(R)} = ( \tau_{(R)}; \sigma^r_{(R)}),
 \label{VII1}
 \eeq

\noindent in a finite region ($\tau_{(R)} = const$ defines the
radar simultaneity surfaces). Then the navigation system provides
determination of the 4-velocities (time-like tetrads) of the
satellites and the ${}^4g_{(R)\tau\tau}$ component of the 4-metric
in these coordinates. \medskip

Note that by replacing {\it test radar signals} (conformal
structure) with {\it test particles} (projective structure) in the
measurements, we would define a different 4-coordinate system.
\medskip

In the framework of metric gravity suitable spacecrafts make
repeated measurements of the motion of {\it four} test particles.
In this way we test also the {\it projective structure} in a
region of space-time with a vacuum gravitational field). By the
motion of gyroscopes we measure the {\it shift} components
${}^4g_{(R)\tau r}$ of the 4-metric) and end up (in principle)
with the determination of all the {\it components of the
four-metric} with respect to the {\it radar-gauge} coordinate
system:

\beq
 {}^4g_{(R)AB}(\tau_{(R)}, \sigma^r_{(R)}).
  \label{VII2}
 \eeq

The tetrad gravity alternative, employing test gyroscopes and
light signals (i.e. only the {\it conformal structure}), is the
following. By means of exchanges (two-ways signals) of {\it
polarized} light it should be possible to determine how the
spatial triads of the satellites are rotated with respect to the
triad of the satellite chosen as origin (see also Ref.\cite{100}).
Once we have the tetrads ${}^4E^A_{(r)(\alpha )}(\tau_{(R)}, {\vec
\sigma}_{(R)})$ in radar coordinates, we can build from them the
inverse 4-metric ${}^4g_{(R)}^{AB}(\tau_{(R)}, {\vec
\sigma}_{(R)}) = {}^4E^A_{(r)(\alpha )}(\tau_{(R)}, {\vec
\sigma}_{(R)})\, {}^4\eta^{(\alpha )(\beta )}\, {}^4E^B_{(r)(\beta
)}(\tau_{(R)}, {\vec \sigma}_{(R)})$ in radar coordinates.
\bigskip

c) By measuring the spatial and temporal variation of
${}^4{g}_{(R)AB}$, the components of the Weyl tensor and the Weyl
eigenvalues can in principle be determined.
\bigskip

d) Points a), b) and c) furnishes {\it operationally} a slicing of
space-time into surfaces $\tau_{(R)} = const$, a system of
coordinates $\sigma^r_{(R)}$ on the surfaces, as well as a
determination of the components of the metric ${}^4{g}_{(R)AB}$.
The components of the Weyl tensor (= Riemann in void) and the
local value of the Weyl eigenvalues, with respect to the
radar-gauge coordinates $(\tau_{(R)}, \sigma^r_{(R)})$ are also
thereby determined. It is then a matter of computation: \bigskip

i) to check whether Einstein's equation in radar-gauge coordinates
are satisfied;\bigskip

ii) to get a numerical determination of the intrinsic coordinate
functions ${\bar \sigma}^{\bar A}_R$ defining the radar gauge by
the gauge fixings $\sigma^A_R - {\bar \sigma}_R^{\bar A}(\sigma_R)
\approx 0$. Since we know the eigenvalues of the Weyl tensor in
the radar gauge, it is possible to solve in principle for the
functions $F^{\bar A}$ that reproduce the {\it radar-gauge}
coordinates as {\it radar-gauge intrinsic} coordinates

\beq \sigma^A_{(R)} = F^{\bar A}[{\tilde
\Lambda}^{(k)}_W[{}^3g(\tau ,\vec \sigma), {}^3\Pi(\tau ,\vec
\sigma)]],
 \label{VII3}
  \eeq

\bigskip

\noindent consistently with the {\it gauge-fixing} that enforces
just this particular system of coordinates:

\beq
 {\bar \chi}^A(\tau ,\vec \sigma ) {\buildrel {def} \over =}  \sigma^A -
 {\bar \sigma}^{\bar A}(\tau ,\vec \sigma ) = \sigma^A -
 F^{\bar A}\Big[{\tilde \Lambda}^{(k)}_W[{}^3g(\tau ,\vec \sigma), {}^3\Pi(\tau ,\vec
\sigma)]\Big] \approx 0.
 \label{VII4}
 \eeq

\noindent Finally, the {\it intrinsic coordinates} are
reconstructed as functions of the {\it DO} of the radar gauge, at
each point-event of space-time, as the identity

\beq \sigma^A \equiv {\bar \sigma}^{\bar A} = {\tilde F}^{\bar
A}_{G}[ r^{(R)}_{\bar a}(\tau ,\vec \sigma ), \pi^{(R)}_{\bar
a}(\tau , \vec \sigma)],
 \label{VII5}
 \eeq

\bigskip

This procedure of principle would close the {\it coordinate
circuit} of general relativity, linking individuation to
operational procedures \cite{5}.

\eject

\section{Concluding Survey}

\noindent The main results of our investigation are:
\medskip
\hfill\break \noindent 1) A definite procedure for the {\it
physical individuation} of the mathematical points of the would-be
space-time manifold $M^4$ into physical {\it point-events},
through a gauge-fixing identifying the mathematical 4-coordinates
with the {\it intrinsic pseudo-coordinates} of Komar and Bergmann
(defined as suitable functionals of the Weyl scalars). This has
led to the conclusion that each of the point-events of space-time
is endowed with its own physical individuation as the value, as it
were, at that point, of the four canonical coordinates or Dirac
observables (just four!), which describe the dynamical degrees of
freedom of the gravitational field. Since such degrees of freedom
are non-local functionals of the metric and curvature, they are
unresolveably entangled with the whole texture of the metric
manifold in a way that is strongly both gauge-dependent and highly
non-local with respect to the background mathematical
coordinatization. We can also say, on the other hand, that {\it
any coordinatization} of the manifold can be seen as embodying the
physical individuation of points, because it can be implemented as
the Komar-Bergmann {\it intrinsic pseudo-coordinates}, after a
suitable choice of the functions of the Weyl scalars and of the
gauge-fixing. We claim that our results bring the
Synge-Bergmann-Komar-Stachel program of the physical individuation
of space-time points to its natural end.

\medskip
\hfill\break \noindent 2) A clarification of Bergmann's multiple
ambiguous definition of {\it observable} in general relativity.
This has led to formulate our {\it main conjecture} concerning the
unification of Bergmann's and Dirac's concepts of observable, as
well as to restate the issue of change and, in particular and
independently, of temporal change, within the Hamiltonian approach
to Einstein equations. When concretely carried out, this program
would provide even {\it explicitly} evidence for the {\it
invariant objectivity} of point-events. Furthermore, the existence
of simultaneous {\it Bergmann-Dirac observables} and {\it PDIQ
gauge variables} would lead to a description of tidal-like and
inertial-like effects in a coordinate independent way, while the
{\it Dirac-Bergmann} observables only would remain as the only
quantities subjected to a causal evolution. If the conjecture
about the existence of simultaneous DO-BO observables is sound, it
would open the possibility of a new type of coordinate-independent
canonical quantization of the gravitational field. Only the DO
should be quantized in this approach, while the gauge variables,
i.e., the {\it appearances} of inertial effects, should be treated
as c-number fields (a prototype of this quantization procedure is
under investigation \cite{101} in the case of special relativistic
and non-relativistic quantum mechanics in non-inertial frames).
This would permit to preserve causality (the space-like character
of the simultaneity Cauchy 3-surfaces), the property of having
only the 3-metric quantized (with implications similar to loop
quantum gravity for the quantization of spatial quantities), and
to avoid any talk of {\it quantization of the 4-geometry} (see
more below), a talk we believe to be deeply misleading (in this
connection see Ref. \cite{102})

\bigskip

The key technical tool which made such results possible has been
the re-visitation of a nearly forgotten paper by Bergmann and
Komar concerning the most general {\it Coordinate Group Symmetries
of General Relativity} \cite{14}. We have made explicit that this
paper establishes the general framework for characterizing the
correspondence between the {\it active diffeomorphisms} operating
in the configurational manifold $M^4$, on the one hand, and the
{\it on-shell gauge transformations} of the ADM canonical approach
to general relativity, on the other. Understanding such
correspondence is fundamental for fully disclosing the connection
of the Hole phenomenology, at the Lagrangian level, with the
correct formulation of the initial value problem of the theory and
its gauge invariance, at the Hamiltonian level. The upshot is the
discovery that, as concerns both the Hole Argument and the issue
of {\it predictability}, not all active diffeomorphisms of $M^4$
can play an effective role, because not all of them satisfy the
bounds imposed by a correct mathematical setting of the initial
value problem at the Lagrangian level.
\bigskip
\hfill\break \noindent 3) An illustration of the physical role
played by the Dirac observables and the gauge variables to the
effect of characterizing intrinsic {\it tidal effects} and {\it
inertial effects}, respectively.
\bigskip
\hfill\break \noindent 4) An outline of the implementation (in
principle) of the physical individuation of point-events as an
experimental setup and protocol for positioning and orientation,
which closes, as it were, the empirical {\it coordinative circuit}
of general relativity.

\bigskip

We want to conclude our discussion with some general conceptual
remarks about the foundation of general relativity and some
venturesome suggestions concerning quantum gravity. First of all,
our program is substantially grounded upon the {\it gauge nature}
of general relativity. Such property of the theory, however, is
far from being a simple matter and we believe that it is not
usually spelled out in a sufficiently explicit and clear fashion.
The crucial point is that general relativity {\it is not} a
standard gauge theory like, e.g., electromagnetism or Yang-Mills
theories in some relevant respects. Let us recall the very general
definition of gauge theory given by Henneaux and Teitelboim
\cite{48}

\begin{quotation}
{\footnotesize \noindent These are theories in which the physical
system being dealt with is described by more variables than there
are physically independent degrees of freedom. The physically
meaningful degrees of freedom then re-emerge as being those
invariant under a transformation connecting the variables (gauge
trasformation). Thus, one introduces extra variables to make the
description more transparent, and brings in at the same time a
gauge symmetry to extract the physically relevant content. }
\end{quotation}

The relevant fact is that, while from the point of view of the
constrained Hamiltonian formalism general relativity is a gauge
theory like any other, it is radically different from the physical
point of view. For, in addition to creating the distinction
between what is observable\footnote{In the Dirac or Bergmann
sense.} and what is not, the gauge freedom of general relativity
is unavoidably entangled with the definition--constitution of the
very {\it stage}, space--time, where the {\it play} of physics is
enacted. In other words, the gauge mechanism has the double role
of making the dynamics unique (as in all gauge theories), and of
fixing the spatio-temporal reference background. It is only after
a complete gauge-fixing (i.e. after the individuation of a {\it
well defined} physical laboratory network) is specified, and after
going on shell, that even the mathematical manifold $M^4$ gets a
{\it physical individuation} and becomes the spatio-temporal
carrier of well defined physical generalized {\it tidal-like} (DO)
and {\it inertial-like} (gauge variables) effects.

In gauge theories such as electromagnetism (or even Yang-Mills),
we can rely from the beginning on empirically validated,
gauge-invariant dynamical equations for the {\it local} fields.
This is not the case for general relativity: in order to get
dynamical equations for the basic field in a {\it local} form, we
must pay the price of Einstein's general covariance which, by
ruling out any background structure at the outset, weakens the
objectivity that the spatiotemporal description could have had
{\it a priori}.

With reference to the definition of Henneaux and Teitelboim, we
could say, therefore, that the introduction of extra variables
does make the mathematical description of general relativity more
transparent, but it also makes its physical interpretation more
obscure and intriguing, at least at first sight.

The isolation of the superfluous structure hidden behind Leibniz
equivalence, which surface in the physical individuation of
point-events, renders even more glaring the ontological diversity
and prominence of the gravitational field with respect to all
other fields, as well as the difficulty of reconciling the deep
nature of the gravitational field with the standard wisdom of
theories based on background space-time. Any procedure of
linearizing these latter unavoidably leads to looking at gravity
as to a spin-2 theory in which the graviton stands on the same
ontological level of other quanta: in the standard approach,
photons, gluons and gravitons all live on the stage on equal
footing. From the point of view we gained in this paper, however,
{\it non-linear gravitons} do in fact constitute the stage for the
causal play of photons, gluons as well as of other matter actors
like electrons and quarks. More precisely, if our main conjecture
is sound, the {\it non-linear graviton} would be represented by a
pair of {\it scalar fields}. Finally, concerning the "background"
philosophy, it's worth recalling that Steven Weinstein \cite{102}
has aptly remarked that what ultimately makes unattractive viewing
the gravitational field as a simple distribution of properties
(the field {\it strengths}) in {\it flat space-time}, on the same
footing of all other fields, is the fact that, because of the
universal nature of gravitation, the distinctive properties of
this background space-time would be {\it completely unobservable}.
\medskip

It should be clear by now that the Hole Argument has little to do
with an alleged \emph{indeterminism} of general relativity as a
dynamical theory. For, in our analysis of the initial-value
problem within the Hamiltonian framework, we have shown that {\it
on shell} a complete gauge-fixing (which could in theory concern
the whole space-time) is equivalent to the choice of an atlas of
coordinate charts on the space-time manifold, and in particular
{\it within the Hole}. At the same time, we have shown that a
peculiar subset of the active diffeomorphisms of the manifold can
be interpreted as passive Hamiltonian gauge transformations.
Because the gauge must be fixed {\it before} the initial-value
problem can be solved to obtain a solution (outside and inside the
Hole), it makes little sense to apply active diffeomorphisms to an
already generated solution to obtain an allegedly ``different''
space-time. Conversely, it is possible to generate these
``different'' solutions by appropriate choices of the initial
gauge fixing. Of course, the price to be payed for the physical
individuation of the {\it stage} is the breaking of general
covariance.

We can, therefore, say that general covariance represents a
horizon of {\it a priori} possibilities for the physical
constitution of the space-time, possibilities that must be
actualized within any given solution of the dynamical equations.
Of course, what we call here {\it physical constitution} embodies
at the same time the {\it chrono-geometrical}, the {\it
gravitational}, and the {\it causal} properties of the space-time
stage.

We believe in conclusion that these results cast some light over
the \emph{intrinsic structure} of the general relativistic
space-time that had disappeared within Leibniz equivalence and
that was the object of Michael Friedman's non-trivial
question\footnote{ It is rather curious to recall here the
following passage of Leibniz: "Space being uniform, there can be
neither any external nor internal reason, by which to distinguish
its parts, and to make any choice between them. For, any external
reason to discern between them, can only be grounded upon some
internal one. Otherwise we should discern what is indiscernible,
or choose without discerning" \cite{103}. Clearly, if the parts of
space were real, the Principle of Sufficient Reason would be
violated. Therefore, for Leibniz, space is not real. The upshot,
however, is that space (space-time) in general relativity is not
{\it uniform} and this is just the reason why - in our sense - it
is {\it real}. Thus, Leibniz equivalence called upon for general
relativity happens to hide the very nature of space-time, instead
of disclosing it.}.

\bigskip

In 1972, Bergmann and Komar wrote \cite{14}:
\begin{quotation}
{\footnotesize \noindent [...] in general relativity the identity
of a world point is not preserved under the theory's widest
invariance group. This assertion forms the basis for the
conjecture that some physical theory of the future may teach us
how to dispense with world points as the ultimate constituents of
space-time altogether.}
\end{quotation}
Indeed, would it be possible to build a fundamental theory that is
grounded in the reduced phase space parametrized by the Dirac
observables? This would be an abstract and highly nonlocal theory
of gravitation that would admit an infinity of gauge-related,
spatio-temporally local realizations. From the mathematical point
of view, this theory would be just an especially perspicuous
instantiation of the relation between canonical structure and
locality that pervades contemporary theoretical physics nearly
everywhere.

On the other hand, beyond the mathematical transparency and the
latitude of choices guaranteed by general covariance, we need to
remember that {\it local} spatio-temporal realizations of the
abstract theory would still be needed for implementation of
measurements in practice; conversely, any real-world experimental
setting entails the choice of a definite {\it local realization},
with a corresponding gauge fixing that breaks general covariance.

\bigskip

Can this basic freedom in the choice of the {\it local
realizations} be equated with a ``taking away from space and time
the last remnant of physical objectivity,'' as Einstein suggested?
We believe that if we strip the physical situation from Einstein's
``spatial obsession'' about {\it realism as locality (and
separability)}, a significant kind of spatio-temporal objectivity
survives. It is true that the {\it functional dependence} of the
Dirac observables upon the spatio-temporal coordinates depends on
the particular choice of the latter (or equivalently, of the
gauge); yet, there is no a-priori {\it physical} individuation of
the points independently of the metric field, so we cannot say
that the physical-individuation procedures corresponding to
different gauges individuate physical point-events that are {\it
really} different. Given the conventional nature of the primary
standard mathematical individuation of manifold points through
$n$-tuples of real numbers, we could say instead that the {\it
real point-events} are constituted by the non-local values of
gravitational degrees of freedom, while the underlying point
structure of the mathematical manifold may be changed at will.

\bigskip

Taking into account our results on the whole, we want to spend a
few additional words about the consequences that the acquired
knowledge entail for the concept of space-time of general
relativity, as seen in the wider context of the traditional debate
on the absolutist/relationist dichotomy.

First of all, let us recall that, in remarkable diversity with
respect to the traditional historical presentation of Newton's
absolutism due to the influence of Leibniz, Newton himself had in
fact a deeper understanding about the reality of space and time
with respect to what has been traditionally ascribed to his
absolutism. In a well-known passage of {\it De Gravitatione}
\cite{104}, Newton expounds what could be defined an original {\it
structuralist view} of space-time (see also Ref.\cite{105}. He
writes:
\begin{quotation}
{\footnotesize \noindent Perhaps now it is maybe expected that I
should define extension as substance or accident or else nothing
at all. But by no means, for it has its own manner of existence
which fits neither substance nor accidents [\ldots] Moreover the
immobility of space will be best exemplified by duration. For just
as the parts of duration derive their individuality from their
order, so that (for example) if yesterday could change places with
today and become the latter of the two, it would lose its
individuality and would no longer be yesterday, but today; so the
parts of space derive their character from their positions, so
that if any two could change their positions, they would change
their character at the same time and each would be converted
numerically into the other \emph{qua} individuals.  The parts of
duration and space are only understood to be the same as they
really are because of their mutual order and positions
(\emph{propter solum ordinem et positiones inter se}); nor do they
have any other {\it principle of individuation} besides this order
and position which consequently cannot be altered.}
\end{quotation}

We have just disclosed the fact that the points of
general-relativistic space-times, quite unlike the points of the
homogeneous Newtonian space, are endowed with a remarkably rich
{\it non-point-like} texture furnished by the metric field.
Therefore, the general-relativistic metric field itself or,
better, its independent degrees of freedom, have the capacity of
characterizing the "mutual order and positions" of points {\it
dynamically}, and in fact much more than this.

Two important remarks are in order. First: such dynamical degrees
of freedom are non-local functionals of the 3-metric and curvature
\footnote{Admittedly, at least at the classical level, we don't
know of any detailed analysis of the relationship between the
notion of {\it non-local observable} (the predictable degrees of
freedom of a gauge system), on one hand, and the notion of a
quantity which has to be operationally {\it measurable} by means
of {\it local apparatuses}, on the other. Note that this is true
even for the simple case of the electro-magnetic field where the
Dirac observables are defined by the {\it transverse} vector
potential and the {\it transverse} electric field. Knowledge of
such fields at a definite mathematical time involves data on the
whole Cauchy surface at that time. Even more complex is the
situation in the case of Yang-Mills theories \cite{65}} so that
they are unresolveably entangled with the whole texture of the
metric manifold in a way that is strongly both gauge-dependent and
highly non-local. Still, once they are calculated, they appear as
{\it local fields} in terms of the background {\it mathematical}
coordinatization, a fact that makes the identity Eq.(\ref{IV8})
possible. Contrary to the opinion of Belot and Earman \cite{32},
from our point of view the peculiar non-locality of the Dirac
observables is therefore philosophically appealing and constitutes
an asset rather than a liability of our approach; and shows, in a
sense, a Machian flavor within a non-Machian environment.

Second, consider the ADM approach to Einstein's equations for the
gravitational field {\it cum} matter. In this case we have Dirac
observables both for the gravitational field and for the matter
fields, but {\it the former are modified in their functional form}
with respect to the vacuum case by the {\it presence of matter}.
Since the gravitational Dirac observables will still provide the
individuating fields for point-events according to the conceptual
procedure presented in this paper, {\it matter will come to
influence the very physical individuation of points}.

In conclusion, we agree with Earman and Norton that the Hole
phenomenology constitutes a decisive argument against strict
manifold substantivalism. However, the isolation of the intrinsic
structure hidden within Leibniz equivalence does not support the
standard relationist view either. With reference to the third
criterion ($R_{3}$) stated by Earman for relationism (see
Ref.\cite{13}, p.14): "No irreducible, monadic, spatiotemporal
properties, like 'is located at space-time point p' appears in a
correct analysis of the spatiotemporal idiom", we observe that if
by 'space-time point' we mean our physically individuated
point-events instead of the naked manifold's point, then - because
of the autonomous existence of the intrinsic degrees of freedom of
the gravitational field (an essential ingredient of general
relativity) - the quoted spatiotemporal property should be
admitted in our spatiotemporal idiom.

In conclusion, what emerges from our analysis is rather a kind of
{\it new structuralist} conception of space-time. Such new
structuralism is not only richer than that of Newton, as it could
be expected because of the dynamical structure of Einstein
space-time, but richer in an even deeper sense. For this {\it new
structuralist} conception turns out to include elements common to
the tradition of both {\it absolutism} (space has an autonomous
existence independently of other bodies or matter fields) {\it and
relationism} (the physical meaning of space depends upon the
relations between bodies or, in modern language, the specific
reality of space depends (also) upon the (matter) fields it
contains).
\bigskip

Let us close this survey with some hints that our results tend to
suggest for the quantum gravity programme. As well-known this
programme is documented nowadays by two inequivalent quantization
methods: i) the perturbative background-dependent {\it string}
formulation, on a Fock space containing elementary particles; ii)
the non-perturbative background-independent {\it loop} quantum
gravity approach, based on the non-Fock {\it polimer} Hilbert
space. In this connection, see Ref.\cite{105} for an attempt to
define a {\it coarse-grained structure} as a bridge between
standard {\it coherent states} in Fock space and some {\it shadow
states} of the discrete quantum geometry associated to a {\it
polimer} Hilbert space. As well-known, this approach still fails
to accomodate elementary particles.
\medskip

Now, the individuation procedure we have proposed transfers, as it
were, the non-commutative Poisson-Dirac structure of the Dirac
observables onto the individuated point-events even if, of course,
the coordinates on the l.h.s. of the identity Eq.(\ref{IV8}) are
c-numbers quantities. Of course, no direct physical meaning can be
attributed to this circumstance at the classical level. One could
guess, however, that such feature might deserve some attention in
view of quantization, for instance by maintaining that the
identity (Eq.(\ref{IV8})) could still play some role at the
quantum level. We will assume here that the {\it main conjecture}
is verified so that all the quantities we consider are manifesttly
covariant.

Let us first lay down some qualitative premises concerning the
status of Minkowski space-time in relativistic quantum field
theory (call it {\it micro space-time}, see Ref.\cite{106}). Such
status is quite peculiar. From the chrono-geometric point of view,
the micro space-time is a universal, classical, non-dynamical
space-time, just Minkowski's space-time of the special theory of
relativity, utilized without any scale limitation from below.
However, it is introduced into the theory through the
group-theoretical requirement of relativistic invariance of the
{\it statistical} results of measurements with respect to the
choice of {\it macroscopic reference frames}. The micro space-time
is therefore {\it anchored} to the macroscopic medium-seized
objects that asymptotically define the experimental conditions in
the laboratory. It is, in fact, in this asymptotic sense that a
physical meaning is attributed to the classical spatiotemporal
{\it coordinates} upon which the quantum fields' operators depend
as {\it parameters}. Thus, the spatiotemporal properties of the
{\it micro Minkowski manifold}, including its basic causal
structure, are, as it were, projected on it {\it from
outside}\footnote{ In view of these circumstances, it could appear
{\it prima facie} miraculous that the quantum-relativistic
micro-causality conditions and the standard gauge theories whose
symmetries depend on the points of the {\it micro space-time},
have had such an extraordinary empirical success. But this should
appear less miraculous if one notes that the most consistent
quantum field theories seem to be the so-called
asymptotically-free non-Abelian gauge theories. If we interpret
{\it physical observability} in this context as possibility of
probing micro-structure by means of interactions, we should
conclude that such gauge theories are, in fact, consistent to the
extent that they are insensitive to short-scale space-time
behavior, a circumstance that does not hold in the case of
gravitational interaction. In this sense, each point of the {\it
micro-space-time} could be coherently understood, so to speak, as
a representative {\it compendium} of something spatiotemporal but
not continuously extended in the traditional sense of the manifold
$M^4$. Also, one should not forget that the Minkowski structure of
the {\it micro-space-time} has been probed down to the scale of
$10^{-18}$ m., yet only from the point of view of {\it scattering}
experiments, involving a limited number of {\it real} particles.}.

In classical field theories space-time points play the role of
individuals and we have seen how the latter can be individuated
dynamically. No such possibility, however, is consistently left
open in a non-metaphoric way in relativistic quantum field theory.
From this point of view, Minkowski's {\it micro space-time} is in
a worse position than general relativistic space-time: it lacks
the existence of Riemannian {\it intrinsic pseudo-coordinates}, as
well as all of the non-dynamical (better, operational and
pragmatical) additional elements that are being used for the
individuation of its points, like rigid rods and clocks in rigid
and un-accelerated motion, or various combinations of {\it
genidentical} world-lines of free test particles, light rays,
clocks, and other devices. Summarizing, the role of {\it
Minkowski's micro space-time} seems to be essentially that of an
instrumental {\it external translator} of the symbolic structure
of quantum theory into the {\it causal} language of the
macroscopic irreversible traces that constitute the experimental
findings within {\it macro space-time}. The conceptual status of
this {\it external translator} fits then very well with that of
epistemic precondition for the formulation of relativistic quantum
field theory in the sense of Bohr, independently of one's attitude
towards the interpretation of quantum theory of measurement. Thus,
barring macroscopic Schr\"odinger Cat states of the would-be
quantum space-time, any conceivable formulation of a quantum
theory of gravity would have to respect, at the {\it operational}
level, the {\it epistemic priority} of a classical spatiotemporal
continuum. Talking about the quantum structure of space-time needs
overcoming a serious conceptual difficulty concerning the
localization of the gravitational field: indeed, what does it even
mean to talk about the {\it values} of the gravitational field
{\it at a point}, to the effect of points individuation, if the
field itself is subject to quantum fluctuations ? One needs in
principle some sort of reference structure in order to give
physical operational meaning to the spatiotemporal language, one
way or the other. This instrumental background, mathematically
represented by a manifold structure, should play, more or less,
the role of the Wittgenstein's staircase. It is likely, therefore,
that in order to attribute some meaning to the individuality of
points that lend themselves to the basic structure of standard
quantum theory, one should split, as it were, the individuation of
point-events from the true quantum properties, i.e., from the
fluctuations of the gravitational field and the micro-causal
structure. Now, it seems that our canonical analysis of the
individuation issue, tends to prefigure a new approach to
quantization having in view a Fock space formulation. Accordingly,
unlike loop quantum gravity, this approach could even lead to a
background-independent incorporation of the standard model of
elementary particles (provided the Cauchy surfaces admit Fourier
transforms). Two options present themselves for a quantization
program respecting relativistic causality \footnote{ Recall that a
3+1 splitting of the mathematical space-time, including the
notions of space-like, light-like, and time-like directions, is
presupposed from the beginning.}:
\medskip

\noindent 1) The procedure for the individuation outlined in
Section IV suggests to quantize the DO=BO of each Hamiltonian
gauge, as well as all the matter DO, and to use the weak ADM
energy of that gauge as Hamiltonian for the functional
Schr\"odinger equation (of course there might be ordering
problems). This quantization would yield as many Hilbert spaces as
4-coordinate systems, which would likely be grouped in unitary
equivalence classes (we leave aside asking what could be the
meaning of inequivalent classes, if any). In each Hilbert space
the DO=BO quantum operators would be distribution-valued quantum
fields on a {\it mathematical micro space-time} parametrized by
the 4-coordinates $\tau$, $\vec \sigma$ associated to the chosen
gauge. Strictly speaking, due to the non-commutativity of the
operators ${\hat {\tilde F}}{}^{\bar A}$ associated to the
classical gauge-fixing (\ref{IV5}) $\sigma^A - F^{\bar A} \approx
0$ defining that gauge, there would be {\it no space-time manifold
of point-events} to be mathematically identified by one coordinate
chart over the {\it micro-space-time}: only a {\it gauge-dependent
non-commutative structure} which is likely to lack any underlying
topological space structure. However, for each Hilbert space, a
{\it coarse-grained} space-time of point-events might be
associated to each solution of the functional Schr\"odinger
equation, through the expectation values of the operators ${\hat
{\tilde F}}{}^{\bar A}$:

\beq
 {\bar \Sigma}^{\bar A} (\tau, \vec \sigma) = \langle \Psi
\Big| {\tilde F}^{\bar A}_{G}[ {\bf R}^{\bar a}(\tau ,\vec \sigma
), {\bf \Pi}_{\bar a}(\tau , \vec \sigma)] \Big| \Psi
\rangle,\quad a = 1,2,
 \label{VIII1}
  \eeq

\noindent where ${\bf R}^{\bar a}(\tau ,\vec \sigma )$ and ${\bf
\Pi}_{\bar a}(\tau , \vec \sigma)$ are {\it scalar Dirac
operators}.

Let us note that, by means of Eq.(\ref{VIII1}), the {\it
non-locality} of the {\it classical} individuation of point-events
would directly get imported at the basis of the ordinary quantum
non-locality.

Also, one could evaluate in principle the expectation values of
the operators corresponding to the lapse and shift functions of
that gauge. Since we are considering a quantization of the
3-geometry (like in loop quantum gravity), evaluating the
expectation values of the quantum 3-metric and the quantum lapse
and shift functions could permit to reconstruct a coarse-grained
foliation with coarse-grained WSW hyper-surfaces\footnote{ This
foliation is called \cite{44} the Wigner-Sen-Witten (WSW)
foliation due to its properties at spatial infinity. See also
footnotes 72 and 73 in Appendix B.}.
\medskip

\noindent 2) In order to avoid inequivalent Hilbert spaces, we
could quantize {\it before} adding any gauge-fixing (i.e.
independently of the choice of the 4-coordinates and the
individuation of point-events), using e.g., the following rule of
quantization, which respects relativistic causality: in a given
canonical basis of the conjecture, quantize the two pairs of DO=BO
observables and the matter DO, but leave the 8 gauge variables
$\zeta^{\alpha}(\tau ,\vec \sigma )$, $\alpha =1,..,8$, as {\it
c-number classical fields}. Like in Schr\"odinger's theory with
time-dependent Hamiltonian, the momenta conjugate to the gauge
variables would be represented by the functional derivatives $i
\delta / \delta \zeta^{\alpha}(\tau ,\vec \sigma )$. Assuming
that, in the chosen canonical basis of our {\it main conjecture},
7 among the eight constraints be gauge momenta, we would get 7
Schr\"odinger equations $i \delta / \delta \zeta^{\alpha}(\tau
,\vec \sigma )\, \Psi( {R}^{\bar a} | \tau; \zeta^{\alpha} ) = 0$
from them. Let ${\sl H}(new) \approx 0$ be the super-Hamitonian
constraint and $E_{ADM}(new)$ the weak ADM energy, in the new
basis. Both would become operators $\hat{\sl H}\,\, or\,\, {\hat
E}_{ADM}({\hat r}_a, {\hat \pi}_a, \zeta^{\alpha}, i \delta /
\delta \zeta^{\alpha})$. If an ordering existed such that the 8
quantum constraints ${\hat \phi}_{\alpha}$ and ${\hat E}_{ADM}$
satisfied a closed algebra $[ {\hat \phi}_{\alpha}, {\hat
\phi}_{\beta}] = {\hat C}_{\alpha\beta\gamma}\, {\hat
\phi}_{\gamma}$ and $[ {\hat E}_{ADM}, {\hat \phi}_{\alpha} ] =
{\hat B}_{\alpha\beta}\, {\hat \phi}_{\beta}$ (with the quantum
structure functions tending to the classical ones for $\hbar
\mapsto 0$), we might quantize by imposing the following 9 coupled
integrable functional Schr\"odinger equations

\bea
 &&i {{\delta} \over
{\delta \zeta_{\alpha}(\tau ,\vec \sigma )}}\, \Psi( {R}^{\bar a}
| \tau; \zeta^{\alpha} ) = 0,\quad \alpha = 1,..,7,\quad
\Rightarrow\,\, \Psi = \Psi ({R}^{\bar a}|\tau ;
\zeta^8),\nonumber \\
 &&{}\nonumber \\
 &&\hat {\sl H}(r^a, i {{\delta}\over {\delta r_a}}, \zeta^{\alpha}, i
 {{\delta}\over {\delta \zeta^{\alpha}}})\, \Psi ( {R}^{\bar a}
| \tau; \zeta^{\alpha} ) = 0,\nonumber \\
 &&{}\nonumber \\
 &&i\, {{\partial}\over {\partial \tau}}\, \Psi ( {R}^{\bar a}
| \tau; \zeta^{\alpha} ) =  {\hat E}_{ADM}(r^a, i {{\delta}\over
{\delta r^a}}, \zeta^{\alpha}, i
 {{\delta}\over {\delta \zeta^{\alpha}}})\, \Psi( {R}^{\bar a}
| \tau; \zeta^{\alpha} ),
 \label{VIII2}
 \eea

\noindent with the associated usual scalar product $ \langle \Psi
\Big| \Psi \rangle$ being independent of $\tau$ and $
\zeta^{\alpha}$'s because of Eq.(\ref{VIII2}). This is similar to
what happens in the quantization of the two-body problem in
relativistic mechanics \cite{50,51}.

If the previously described quasi-Shanmugadhasan canonical basis
existed, the wave functional would depend on 8 functional field
parameters $\zeta^{\alpha}(\tau ,\vec \sigma )$, besides the
mathematical time $\tau$ (actually only on $\zeta^8$). Each {\it
curve} in this parameter space would be associated to a
Hamiltonian gauge in the following sense: for each solution $\Psi$
of the previous equations, the classical gauge-fixings $\sigma^A -
F_G^{\bar A} \approx 0$ implying $\zeta^{\alpha} =
\zeta^{(G)\alpha}(R^a, \Pi_a)$, would correspond to expectation
values $< \Psi |\, \zeta^{(G)\alpha}(\tau ,\vec \sigma ) | \Psi >
=  {\tilde \zeta}^{(G)\alpha}(\tau ,\vec \sigma )$ defining the
{\it curve} in the parameter space. Again, we would have a {\it
mathematical micro space-time} and a {\it coarse-grained
space-time of "point-events"}. At this point, by going to {\it
coherent states}, we could try to recover classical gravitational
fields\footnote{ At the classical level, we have the ADM
Poincar\'e group at spatial infinity on the asymptotic Minkowski
hyper-planes orthogonal to the ADM 4-momentum, while the WSW
hyper-surfaces (see Appendix B) tend to such Minkowski
hyper-planes in every 4-region where the 4-curvature is
negligible, because their extrinsic curvature tends to zero in
such regions. Thus, matter and gauge fields could be approximated
there by the rest-frame relativistic fields whose quantization
leads to relativistic QFT. Since at the classical level, in each
4-coordinate system, matter and gauge field satisfy $\phi (\tau
,\vec \sigma ) = \phi (\sigma^A) \approx \tilde \phi (F^{\bar A})
= {\tilde {\tilde \phi}}(R^a, \Pi_a)$, they could be thought of as
functions of either the intrinsic pseudo-coordinates (as DeWitt
does) or the DO=BO observables of that gauge.}. The 3-geometry
(volumes, areas, lengths) would be quantized, perhaps in a way
coherent with the results of loop quantum gravity.

\bigskip

It is important to stress that, according to both of our
suggestions, {\it only the Dirac observables would be quantized}.
The upshot is that fluctuations in the gravitational field
(better, in the Dirac observables) would entail fluctuations of
the point texture that lends itself to the basic space-time scheme
of standard relativistic quantum field theory: such fluctuating
texture, however, could be recovered as a coarse-grained
structure. This would induce fluctuations in the coarse-grained
metrical relations, and thereby in the causal structure, both of
which would tend to disappear in a semi-classical approximation.
Such a situation should be conceptually tolerable, and even
philosophically appealing, as compared with the impossibility of
defining a causal structure within all of the attempts grounded
upon quantization of the full 4-geometry.

Besides, in space-times with matter, our procedure entails {\it
quantizing the tidal effects and action-at-a-distance potentials
between matter elements but not the inertial aspects of the
gravitational field}. As shown before, the latter are connected
with the gauge variables whose variations reproduce all the
possible viewpoints of local accelerated time-like observers.
Thus, quantizing the gauge variables would be tantamount to
quantizing the metric {\it and} the {\it passive observers} and
their reference frames associated to the congruences studied in
Section VI. Of course, such observers have nothing to do with the
{\it dynamical observers}, which should be realized in terms of
the DO of matter.

Finally, concerning different ways of looking at inertial forces,
consider for the sake of completeness the few known attempts of
extending non-relativistic quantum mechanics from global inertial
frames to global non-inertial ones \cite{108} by means of
time-dependent unitary transformations $U(t)$. The resulting
quantum potentials $V(t) = i\, \dot U(t)\, U^{-1}(t)$ for the
fictitious forces in the new Hamiltonian $\tilde H = U(t)\, H
U^{-1}(t) + V(t)$ for the transformed Schr\"odinger
equation\footnote{Note that as it happens with the time-dependent
Foldy-Wouthuysen transformation \cite{109}, the operator $\tilde
H$ describing the non-inertial time evolution is no more the
energy operator.}, as seen by an accelerated observer (passive
view), are often re-interpreted as action-at-a-distance Newtonian
gravitational potentials in an inertial frame (active view). This
fact, implying in general a change in the emission spectra of
atoms, is justified by invoking an extrapolation of the
non-relativistic limit of the weak equivalence principle
(universality of free fall or identity of inertial and
gravitational masses) to quantum mechanics. Our Hamiltonian
distinction among tidal, inertial and action-at-a-distance effects
supports Synge's criticism \cite{31}b of Einstein's statements
about the equivalence of {\it uniform} gravitational fields and
{\it uniform} accelerated frames. Genuine physical uniform
gravitational fields do not exist over finite regions\footnote{Nor
is their definition a unambiguous task in general \cite{110}.} and
must be replaced by tidal and action-at-a-distance effects: these,
however, are clearly not equivalent to uniform acceleration
effects. From our point of view, the latter are generated as
inertial effects whose appearance depends upon the gauge
variables. Consequently, the non-relativistic limit of our
quantization procedure should be consistent with the previous
passive view in which atom spectra are not modified by pure
inertial effects, and should match the formulation of standard
non-relativistic quantum mechanics of Ref.\cite{101}.

\vfill\eject

\appendix

\section{ADM Canonical Metric Gravity.}

Let $M^4$ be a globally hyperbolic pseudo-Riemannian 4-manifold,
asymptotically flat at spatial infinity, Let $M^4$ be foliated
(3+1 splitting or slicing) with space-like Cauchy hypersurfaces
$\Sigma_{\tau}$ through the embeddings $i_{\tau}:\Sigma
\rightarrow \Sigma_{\tau} \subset M^4$, $\vec \sigma \mapsto
x^{\mu}=z^{\mu}(\tau ,\vec \sigma )$, of a 3-manifold $\Sigma$,
assumed diffeomorphic to $R^3$, into $M^4$
 \footnote{$\tau :M^4\rightarrow R$ is a
global, time-like, future-oriented function labelling the leaves
of the foliation; $x^{\mu}$ are local coordinates in a chart of
$M^4$; $\vec \sigma =\{ \sigma^r \}$, r=1,2,3, are coordinates in
a global chart of $\Sigma$, which is diffeomorphic to $R^3$; we
shall use the notations $\sigma^A= (\sigma^{\tau}=\tau ;\vec
\sigma )$, $A=\tau ,r$,  for the coordinates of $M^4$ adapted to
the 3+1 splitting, and $z^{\mu}(\sigma )=z^{\mu}(\tau ,\vec \sigma
)$ for the embedding functions.}. The non-degenerate 4-metric
tensor ${}^4g_{\mu\nu}(x)$ has Lorentzian signature $\epsilon
(+,-,-,-)$ \footnote{Here $\epsilon =\pm 1$ according to particle
physics and general relativity conventions, respectively.}.

Let $n^{\mu}(\sigma )$ and $l^{\mu}(\sigma )= N(\sigma )
n^{\mu}(\sigma )$ be the controvariant timelike normal and unit
normal [${}^4g_{\mu\nu}(z(\sigma ))\, l^{\mu}(\sigma )
l^{\nu}(\sigma )=\epsilon$] to $\Sigma_{\tau}$ at the point
$z(\sigma )\in \Sigma_{\tau}$. The positive function $N(\sigma ) >
0$ is the lapse function: $N(\sigma )d\tau$ measures the proper
time interval at $z(\sigma )\in \Sigma_{\tau}$ between
$\Sigma_{\tau}$ and $\Sigma_{\tau +d\tau}$. The shift functions
$N^r(\sigma )$ are defined so that $N^r(\sigma )d\tau$ describes
the horizontal shift on $\Sigma_{\tau}$ such that, if
$z^{\mu}(\tau +d\tau ,\vec \sigma +d\vec \sigma )\in \Sigma_{\tau
+d\tau}$, then $z^{\mu}(\tau +d\tau ,\vec \sigma +d\vec \sigma
)\approx z^{\mu}(\tau ,\vec \sigma )+N(\tau , \vec \sigma )d\tau
l^{\mu}(\tau ,\vec \sigma )+[d\sigma^r+N^r(\tau ,\vec \sigma
)d\tau ]{{\partial z^{\mu}(\tau ,\vec \sigma )}\over {\partial
\sigma^r} }$; therefore, the so called evolution vector is
${{\partial z^{\mu}(\sigma )}\over {\partial \tau}}= N(\sigma )
l^{\mu}(\sigma )+N^r(\sigma ) {{\partial z^{\mu}(\tau ,\vec \sigma
)}\over {\partial \sigma^r}}$. The covariant unit normal to
$\Sigma_{\tau}$ is $l_{\mu}(\sigma )={}^4g _{\mu\nu}(z(\sigma ))
l^{\nu}(\sigma )=N(\sigma ) \partial_{\mu}\tau{|} _{x=z(\sigma
)}$, with $\tau =\tau (\sigma )$ a global timelike future-oriented
function.

Instead of local coordinates $x^{\mu}$ for $M^4$, we use
coordinates $\sigma^A$ on $R\times \Sigma \approx M^4$
[$x^{\mu}=z^{\mu}(\sigma )$ with inverse $\sigma^A= \sigma^A(x)$],
and the associated $\Sigma_{\tau}$-adapted holonomic coordinate
basis  $\partial_A={{\partial}\over {\partial \sigma^A}}\in
T(R\times \Sigma ) \mapsto b^{\mu}_A(\sigma ) \partial_{\mu}
={{\partial z^{\mu}(\sigma )}\over {\partial \sigma^A}}
\partial_{\mu} \in TM^4$ for vector fields, and  $dx^{\mu}\in
T^{*}M^4 \mapsto d\sigma^A=b^A _{\mu}(\sigma )dx^{\mu}={{\partial
\sigma^A(z)}\over {\partial z^{\mu}}} dx ^{\mu} \in T^{*}(R\times
\Sigma )$ for differential one-forms.

In the new basis, the induced 4-metric becomes

\beq
 {}^4g_{AB}=\left( \begin{array}{ll} {}^4g_{\tau\tau}=
\epsilon (N^2-{}^3g_{rs}N^rN^s)& {}^4g_{\tau s}=- \epsilon \,
{}^3g_{su}N^u
\\ {}^4g_{\tau r}=- \epsilon \, {}^3g_{rv}N^v&
{}^4g_{rs}=-\epsilon \, {}^3g_{rs} \end{array} \right) ,
 \label{a1}
 \eeq

where the 3-metric $\, {}^3g_{rs}=-\epsilon \, {}^4g_{rs}$ with
signature (+++),  of $\Sigma_{\tau}$ has been introduced. The line
element of $M^4$ is

\bea
 ds^2&=&{}^4g_{\mu\nu} dx^{\mu} dx^{\nu}=
\epsilon (N^2-{}^3g_{rs}N^rN^s) (d\tau )^2-2\epsilon \,
{}^3g_{rs}N^s d\tau d\sigma^r -\epsilon \, {}^3g_{rs}
d\sigma^rd\sigma^s=\nonumber \\ &=&\epsilon \Big[ N^2(d\tau )^2
-{}^3g_{rs}(d\sigma^r+N^rd\tau )(d\sigma^s+ N^sd\tau )\Big] ,
\label{a2}
 \eea

such that $\epsilon \, {}^4g_{oo} >0$, $\epsilon \, {}^4g_{ij} <
0$, $\left| \begin{array}{cc} {}^4g_{ii}& {}^4g_{ij}
\\ {}^4g_{ji}& {}^4g_{jj} \end{array} \right| > 0$, $\epsilon \,
det\, {}^4g_{ij} > 0$, hold true.

Defining $g={}^4g=|\, det\, ({}^4g_{\mu\nu})\, |$ and $\gamma
={}^3g =|\, det\, ({}^3g_{rs})\, |$, the lapse and shift functions
assume the form

\begin{eqnarray}
&&N=\sqrt{ {{}^4g\over {{}^3g}} }={1\over {
\sqrt{{}^4g^{\tau\tau}} }} = \sqrt{{g\over {\gamma}}}=
\sqrt{{}^4g_{\tau\tau}-\epsilon \, {}^3g^{rs}\, {}^4g_{\tau r}
{}^4g_{\tau s} },\nonumber \\ &&N^r=-\epsilon \, {}^3g^{rs}\,
{}^4g_{\tau s} = -{{{}^4g^{\tau r}}\over {{}^4g^{\tau\tau}}}
,\quad N_r={}^3g_{rs}N^s=-\epsilon \,\, {}^4g_{rs}N^s=-\epsilon
{}^4g_{\tau r}.
 \label{a3}
 \end{eqnarray}

Given an arbitrary 3+1 splitting of $M^4$, the ADM action
\cite{43} expressed in terms of the independent
$\Sigma_{\tau}$-adapted variables N, $N_r={}^3g_{rs}N^s$,
${}^3g_{rs}$ is

\bea
 S_{ADM}&=&\int d\tau \, L_{ADM}(\tau )= \int d\tau d^3\sigma
{\cal L}_{ADM}(\tau ,\vec \sigma )=\nonumber \\
 &=&- \epsilon k\int_{\triangle \tau}d\tau  \, \int d^3\sigma \, \lbrace
\sqrt{\gamma} N\, [{}^3R+{}^3K_{rs}\, {}^3K^{rs}-({}^3K)^2]\rbrace
(\tau ,\vec \sigma ),
 \label{a4}
  \eea

\noindent where $k={{c^3}\over {16\pi G}}$, with $G$ the Newton
constant.
\medskip \noindent

The Euler-Lagrange equations are\footnote{The symbol ${\buildrel
\circ \over =}$ means "evaluated on the extremals of the
variational principle", namely on the solutions of the equation of
motion.}

\begin{eqnarray}
L_N&=&{{\partial {\cal L}_{ADM}}\over {\partial
N}}-\partial_{\tau} {{\partial {\cal L}_{ADM}}\over {\partial
\partial_{\tau}N}}-\partial_r {{\partial {\cal L}_{ADM}}\over
{\partial \partial_rN}}=\nonumber \\
 &=&-\epsilon k \sqrt{\gamma}
[{}^3R-{}^3K_{rs}\, {}^3K^{rs}+({}^3K)^2]=-2\epsilon k\, {}^4{\bar
G}_{ll}\, {\buildrel \circ \over =}\, 0,\nonumber \\
 L^r_{\vec
N}&=&{{\partial {\cal L}_{ADM}}\over {\partial
N_r}}-\partial_{\tau} {{\partial {\cal L}_{ADM}}\over {\partial
\partial_{\tau}N_r}}-\partial_s {{\partial {\cal L}_{ADM}}\over
{\partial \partial_sN_r}}=\nonumber \\
 &=&2\epsilon k
[\sqrt{\gamma}({}^3K^{rs}-{}^3g^{rs}\, {}^3K)]_{\, |s}=2k\,
{}^4{\bar G}_l{}^r \, {\buildrel \circ \over =}\, 0,\nonumber \\
 L_g^{rs}&=& -\epsilon k \Big[ {{\partial}\over {\partial
\tau}}[\sqrt{\gamma}({}^3K^{rs}-{}^3g ^{rs}\, {}^3K)]\, -
N\sqrt{\gamma}({}^3R^{rs}- {1\over 2} {}^3g^{rs}\,
{}^3R)+\nonumber \\
 &&+2N\, \sqrt{\gamma}({}^3K^{ru}\,
{}^3K_u{}^s-{}^3K\, {}^3K^{rs})+{1\over 2}N
\sqrt{\gamma}[({}^3K)^2-{}^3K_{uv}\,
{}^3K^{uv}){}^3g^{rs}+\nonumber \\
 &&+\sqrt{\gamma} ({}^3g^{rs}
N^{|u}{}_{|u}-N^{|r |s})\Big] =-\epsilon kN \sqrt{\gamma}\,
{}^4{\bar G}^{rs}\, {\buildrel \circ \over =}\,0,
 \label{a5}
 \end{eqnarray}

\noindent and correspond to Einstein's equations in the form
${}^4{\bar G} _{ll}\, {\buildrel \circ \over =}\, 0$, ${}^4{\bar
G}_{lr}\, {\buildrel \circ \over =}\, 0$, ${}^4{\bar G}_{rs}\,
{\buildrel \circ \over =}\, 0$, respectively. The four contracted
Bianchi identities imply that only two of the six equations
$L^{rs}_g\, {\buildrel \circ \over =}\, 0$ are independent.

The canonical momenta (densities of weight -1) are

\begin{eqnarray}
&&{\tilde \pi}^N(\tau ,\vec \sigma )={{\delta S_{ADM}}\over
{\delta
\partial_{\tau}N(\tau ,\vec \sigma )}} =0,\nonumber \\
&&{\tilde \pi}^r_{\vec N}(\tau ,\vec \sigma )={{\delta
S_{ADM}}\over {\delta \partial_{\tau} N_r(\tau ,\vec \sigma )}}
=0,\nonumber \\ &&{}^3{\tilde \Pi}^{rs}(\tau ,\vec \sigma
)={{\delta S_{ADM}}\over {\delta \partial_{\tau} {}^3g_{rs}(\tau
,\vec \sigma )}}=\epsilon k\, [
\sqrt{\gamma}({}^3K^{rs}-{}^3g^{rs}\, {}^3K)](\tau ,\vec \sigma
),\nonumber \\ &&{}\nonumber \\ &&\Downarrow \nonumber \\
&&{}\nonumber \\ &&{}^3K_{rs}={{\epsilon}\over {k\sqrt{\gamma}}}
[{}^3{\tilde \Pi}_{rs}-{1\over 2}{}^3g_{rs}\, {}^3{\tilde
\Pi}],\quad\quad {}^3\tilde \Pi ={}^3g_{rs}\, {}^3{\tilde
\Pi}^{rs}=-2\epsilon k\sqrt{\gamma}\, {}^3K,
 \label{a6}
 \end{eqnarray}

\noindent and satisfy the Poisson brackets

\begin{eqnarray}
&&\lbrace N(\tau ,\vec \sigma ),{\tilde \Pi}^N(\tau ,{\vec
\sigma}^{'} ) \rbrace =\delta^3(\vec \sigma ,{\vec
\sigma}^{'}),\nonumber \\ &&\lbrace N_r(\tau ,\vec \sigma
),{\tilde \Pi}^s_{\vec N}(\tau ,{\vec \sigma} ^{'} )\rbrace
=\delta^s_r \delta^3(\vec \sigma ,{\vec \sigma}^{'}),\nonumber \\
&&\lbrace {}^3g_{rs}(\tau ,\vec \sigma ),{}^3{\tilde
\Pi}^{uv}(\tau ,{\vec \sigma}^{'}\rbrace = {1\over 2}
(\delta^u_r\delta^v_s+\delta^v_r\delta^u_s) \delta^3(\vec \sigma
,{\vec \sigma}^{'}).
 \label{a7}
 \end{eqnarray}

The Wheeler- De Witt super-metric is

\begin{equation}
{}^3G_{rstw}(\tau ,\vec \sigma )=[{}^3g_{rt}\,
{}^3g_{sw}+{}^3g_{rw}\, {}^3g _{st}-{}^3g_{rs}\, {}^3g_{tw}](\tau
,\vec \sigma ).
 \label{a8}
 \end{equation}

\noindent Its inverse is defined by the equations

\begin{eqnarray}
{}&&{1\over 2} {}^3G_{rstw}\, {1\over 2} {}^3G^{twuv} ={1\over
2}(\delta^u_r \delta^v_s+\delta^v_r\delta^u_s),\nonumber \\
{}^3G^{twuv}(\tau ,\vec \sigma )&=&[{}^3g^{tu}\,
{}^3g^{wv}+{}^3g^{tv}\, {}^3g ^{wu}-2\, {}^3g^{tw}\,
{}^3g^{uv}](\tau ,\vec \sigma ),
 \label{a9}
 \end{eqnarray}

\noindent so that

\begin{eqnarray}
{}^3{\tilde \Pi}^{rs}(\tau ,\vec \sigma )&=&{1\over 2}\epsilon k
\sqrt{\gamma}\, {}^3G^{rsuv}(\tau ,\vec \sigma )\, {}^3K_{uv}(\tau
,\vec \sigma ),\nonumber \\ {}^3K_{rs}(\tau ,\vec \sigma
)&=&{{\epsilon}\over {2k\sqrt{\gamma}}}\, {}^3G_{rsuv}(\tau ,\vec
\sigma )\, {}^3{\tilde \Pi}^{uv}(\tau ,\vec \sigma ), \nonumber \\
&&{} \nonumber \\
\partial_{\tau}\, {}^3g_{rs}(\tau ,\vec \sigma )&=&[N_{r|s}+N_{s|r}-{{\epsilon
N}\over {k\sqrt{\gamma}}}\, {}^3G_{rsuv}\, {}^3{\tilde
\Pi}^{uv}](\tau , \vec \sigma ).
 \label{a10}
 \end{eqnarray}

Since ${}^3{\tilde \Pi}^{rs}\partial_{\tau}\,
{}^3g_{rs}=$${}^3{\tilde \Pi} ^{rs} [N_{r | s}+N_{s |
r}-{{\epsilon N}\over {k\sqrt{\gamma}}} {}^3G_{rsuv} {}^3{\tilde
\Pi}^{uv}]=$$-2 N_r {}^3{\tilde \Pi}^{rs}{}_{| s}- {{\epsilon
N}\over {k\sqrt{\gamma}}}\, {}^3G_{rsuv}\, {}^3{\tilde \Pi}^{rs}
{}^3{\tilde \Pi}^{uv}+ (2N_r\, {}^3{\tilde \Pi}^{rs})_{| s}$, we
obtain the canonical Hamiltonian \footnote{Since $N_r\,
{}^3{\tilde \Pi}^{rs}$ is a vector density of weight -1, it holds
${}^3\nabla_s(N_r\, {}^3{\tilde \Pi}^{rs})=\partial_s(N_r\,
{}^3{\tilde \Pi}^{rs})$.}

\begin{eqnarray}
H_{(c)ADM}&=& \int_Sd^3\sigma \, [{\tilde
\pi}^N\partial_{\tau}N+{\tilde \pi}^r_{\vec
N}\partial_{\tau}N_r+{}^3{\tilde \Pi}^{rs}\partial_{\tau}
{}^3g_{rs}](\tau ,\vec \sigma ) -L_{ADM}=\nonumber \\
&=&\int_Sd^3\sigma \, [\epsilon N(k\sqrt{\gamma}\, {}^3R- {1\over
{2k\sqrt{\gamma}}} {}^3G_{rsuv}{}^3{\tilde \Pi}^{rs} {}^3{\tilde
\Pi}^{uv})-2N_r\, {}^3{\tilde \Pi}^{rs}{}_{| s}](\tau ,\vec \sigma
)+ \nonumber \\ &+&2\int_{\partial S}d^2\Sigma_s [N_r\,
{}^3{\tilde \Pi}^{rs}\,\, ](\tau ,\vec \sigma ).
 \label{a11}
 \end{eqnarray}

\noindent In the following discussion the surface term will be
omitted.

The Dirac Hamiltonian is \footnote{The $\lambda (\tau ,\vec \sigma
)$'s are arbitrary Dirac multipliers.}

\begin{equation}
H_{(D)ADM}=H_{(c)ADM}+\int d^3\sigma \, [\lambda_N\, {\tilde
\pi}^N + \lambda ^{\vec N}_r\, {\tilde \pi}^r_{\vec N}](\tau ,\vec
\sigma ).
 \label{a12}
 \end{equation}

The $\tau$-constancy of the primary constraints [$\partial_{\tau}
{\tilde \pi}^N(\tau ,\vec \sigma )\, {\buildrel \circ \over =}\,
\lbrace {\tilde \pi}^N(\tau ,\vec \sigma ),H_{(D) ADM}\rbrace
\approx 0$, $\partial_{\tau} {\tilde \pi}^r_{\vec N}(\tau ,\vec
\sigma )\, {\buildrel \circ \over =}\, \lbrace {\tilde
\pi}^r_{\vec N}(\tau ,\vec \sigma ),H_{(D)ADM} \rbrace \approx 0$]
generates four secondary constraints (densities of weight -1)
which correspond to the Einstein equations ${}^4{\bar G}_{ll}
(\tau ,\vec \sigma )\, {\buildrel \circ \over =}\, 0$, ${}^4{\bar
G}_{lr} (\tau ,\vec \sigma )\, {\buildrel \circ \over =}\, 0$

\begin{eqnarray}
{\tilde {\cal H}}(\tau ,\vec \sigma )&=&\epsilon [k\sqrt{\gamma}\,
{}^3R-{1\over {2k \sqrt{\gamma}}} {}^3G_{rsuv}\, {}^3{\tilde
\Pi}^{rs}\, {}^3{\tilde \Pi}^{uv}] (\tau ,\vec \sigma )=\nonumber
\\
 &=&\epsilon [\sqrt{\gamma}\, {}^3R-{1\over
{k\sqrt{\gamma}}}({}^3{\tilde \Pi} ^{rs}\, {}^3{\tilde
\Pi}_{rs}-{1\over 2}({}^3\tilde \Pi )^2)](\tau ,\vec \sigma )
\approx 0, \nonumber \\
 &&{}\nonumber \\
 {}^3{\tilde {\cal H}}^r(\tau ,\vec \sigma )&=&-2\, {}^3{\tilde
\Pi}^{rs}{}_{| s} (\tau ,\vec \sigma )=-2[\partial_s\, {}^3{\tilde
\Pi}^{rs}+{}^3\Gamma^r_{su} {}^3{\tilde \Pi}^{su}](\tau ,\vec
\sigma )=\nonumber \\ &=&-2\epsilon k \{
\partial_s[\sqrt{\gamma}({}^3K^{rs}-{}^3g^{rs}\, {}^3K)]+
{}^3\Gamma^r_{su}\sqrt{\gamma}({}^3K^{su}-{}^3g^{su}\, {}^3K) \}
(\tau ,\vec \sigma )\approx 0,
 \label{a13}
 \end{eqnarray}

\noindent so that the Hamiltonian becomes

\begin{equation}
H_{(c)ADM}= \int d^3\sigma [N\, {\tilde {\cal H}}+N_r\,
{}^3{\tilde {\cal H}}^r](\tau ,\vec \sigma )\approx 0,
 \label{a14}
 \end{equation}

\noindent with ${\tilde {\cal H}}(\tau ,\vec \sigma )\approx 0$
called the superhamiltonian constraint and ${}^3{\tilde {\cal
H}}^r(\tau ,\vec \sigma ) \approx 0$  the supermomentum
constraints. One may say that the term $-\epsilon k
\sqrt{\gamma}({}^3K_{rs}\, {}^3K^{rs}-{}^3K^2)$ in ${\tilde {\cal
H}}(\tau ,\vec \sigma )\approx 0$, is the kinetic energy and
$\epsilon k\sqrt{\gamma}\, {}^3R$ the potential energy.

All the constraints are first class, because the only
non-identically zero Poisson brackets correspond to the so called
universal Dirac algebra \cite{45}:

\begin{eqnarray}
\lbrace {}^3{\tilde {\cal H}}_r(\tau ,\vec \sigma ),{}^3{\tilde
{\cal H}}_s (\tau ,{\vec \sigma}^{'})\rbrace  &=&{}^3{\tilde {\cal
H}}_r(\tau ,{\vec \sigma}^{'} )\, {{\partial \delta^3(\vec \sigma
,{\vec \sigma}^{'})}\over {\partial \sigma^s}} + {}^3{\tilde {\cal
H}}_s(\tau ,\vec \sigma ) {{\partial \delta^3(\vec \sigma ,{\vec
\sigma}^{'})}\over {\partial \sigma^r}}, \nonumber \\ \lbrace
{\tilde {\cal H}}(\tau ,\vec \sigma ),{}^3{\tilde {\cal H}}_r(\tau
, {\vec \sigma}^{'})\rbrace &=& {\tilde {\cal H}}(\tau ,\vec
\sigma ) {{\partial \delta^3(\vec \sigma ,{\vec \sigma}^{'})}\over
{\partial \sigma^r}}, \nonumber \\ \lbrace {\tilde {\cal H}}(\tau
,\vec \sigma ),{\tilde {\cal H}}(\tau ,{\vec \sigma}^{'})\rbrace
&=&[{}^3g^{rs}(\tau ,\vec \sigma ) {}^3{\tilde {\cal H}}_s (\tau
,\vec \sigma )+\nonumber \\ &+&{}^3g^{rs}(\tau ,{\vec
\sigma}^{'}){}^3{\tilde {\cal H}}_s(\tau ,{\vec
\sigma}^{'})]{{\partial \delta^3(\vec \sigma ,{\vec
\sigma}^{'})}\over {\partial \sigma^r}},
 \label{a15}
 \end{eqnarray}

\noindent with ${}^3{\tilde {\cal H}}_r={}^3g_{rs}\, {}^3{\tilde
{\cal H}}^r$ as the combination of the supermomentum constraints
satisfying the algebra of 3-diffeomorphisms. In Ref.\cite{52} it
is shown that Eqs.(\ref{a15}) are sufficient conditions for the
embeddability of $\Sigma_{\tau}$ into $M^4$. In the second paper
in Ref.\cite{111} it is shown that the last two lines of the Dirac
algebra are the phase space equivalent of the Bianchi's identities
${}^4G^{\mu\nu}{}_{;\nu}\equiv 0$.

The Hamilton-Dirac equations are  [$\cal L$ is the notation for
the Lie derivative]

\begin{eqnarray}
\partial_{\tau}N(\tau ,\vec \sigma )\, &{\buildrel \circ \over =}\,&
\lbrace N(\tau ,\vec \sigma ),H_{(D)ADM} \rbrace =\lambda_N(\tau
,\vec \sigma ),\nonumber \\
\partial_{\tau}N_r(\tau ,\vec \sigma )\, &{\buildrel \circ \over =}\,&
\lbrace N_r(\tau ,\vec \sigma ), H_{(D)ADM}\rbrace =\lambda^{\vec
N}_r(\tau ,\vec \sigma ),\nonumber \\
\partial_{\tau}\, {}^3g_{rs}(\tau ,\vec \sigma )\,
&{\buildrel \circ \over =}\,& \lbrace {}^3g_{rs}(\tau , \vec
\sigma ),H_{(D)ADM}\rbrace =[N_{r | s}+N_{s | r}-{{2\epsilon
N}\over {k\sqrt{\gamma}}}({}^3{\tilde \Pi}_{rs}-{1\over
2}{}^3g_{rs}\, {}^3{\tilde \Pi})](\tau ,\vec \sigma )=\nonumber \\
&=&[N_{r|s}+N_{s|r}-2N\, {}^3K_{rs}](\tau ,\vec \sigma ),\nonumber
\\
\partial_{\tau}\, {}^3{\tilde \Pi}^{rs}(\tau ,\vec \sigma )
\, &{\buildrel \circ \over =}\,& \lbrace {}^3 {\tilde
\Pi}^{rs}(\tau ,\vec \sigma ),H_{(D)ADM}\rbrace =\epsilon [N\,
k\sqrt{\gamma} ({}^3R^{rs}-{1\over 2}{}^3g^{rs}\, {}^3R)](\tau
,\vec \sigma )- \nonumber \\ &-&2\epsilon [{N\over
{k\sqrt{\gamma}}}({1\over 2}{}^3{\tilde \Pi}\, {}^3{\tilde
\Pi}^{rs}-{}^3{\tilde \Pi}^r{}_u\, {}^3{\tilde \Pi}^{us})(\tau
,\vec \sigma )-\nonumber \\ &-&{{\epsilon N}\over 2}
{{{}^3g^{rs}}\over {k\sqrt{\gamma}}}({1\over 2}{}^3{\tilde
\Pi}^2-{}^3{\tilde \Pi}_{uv}\, {}^3{\tilde \Pi}^{uv})](\tau ,\vec
\sigma )+\nonumber \\ &+&{\cal L}_{\vec N}\, {}^3{\tilde
\Pi}^{rs}(\tau ,\vec \sigma )+\epsilon [k\sqrt{\gamma} (N^{| r |
s}-{}^3g^{rs}\, N^{| u}{}_{| u})](\tau ,\vec \sigma ),\nonumber \\
&&{}\nonumber \\ &&\Downarrow \nonumber \\ &&{}\nonumber \\
\partial_{\tau}\, {}^3K_{rs}(\tau ,\vec \sigma )\,
&{\buildrel \circ \over =}\,& \Big( N [{}^3R_{rs}+{}^3K\,
{}^3K_{rs}-2\, {}^3K_{ru}\, {}^3K^u{}_s]- \nonumber \\
&-&N_{|s|r}+N^u{}_{|s}\, {}^3K_{ur}+N^u{}_{|r}\, {}^3K_{us}+N^u\,
{}^3K_{rs|u} \Big) (\tau ,\vec \sigma ),\nonumber \\
 &&{}\nonumber \\
 &&with \nonumber \\
 {\cal L}_{\vec N}\, {}^3{\tilde \Pi}^{rs}&=&-\sqrt{\gamma}\,
{}^3\nabla_u({{N^u}\over {\sqrt{\gamma}}} {}^3{\tilde \Pi}^{rs})+
{}^3{\tilde \Pi}^{ur}\, {}^3\nabla_u N^s+{}^3{\tilde \Pi}^{us}\,
{}^3\nabla_u N^r.
 \label{a16}
 \end{eqnarray}

The above equation for $\partial_{\tau}\, {}^3g_{rs}(\tau ,\vec
\sigma )$ shows that the generator of space pseudo-diffeomorphisms
\footnote{The Hamiltonian transformations generated by these
constraints are the extension to the 3-metric of passive
diffeomorphisms, namely changes of coordinate charts, of
$\Sigma_{\tau}$ [$Diff\, \Sigma_{\tau}$].} $\int d^3\sigma
N_r(\tau ,\vec \sigma )$ $\, {}^3{\tilde {\cal H}}^r(\tau ,\vec
\sigma )$ produces a variation, tangent to $\Sigma_{\tau}$,
$\delta_{tangent} {}^3g_{rs}={\cal L} _{\vec N}\,
{}^3g_{rs}=N_{r|s}+N_{s|r}$ in accord with the infinitesimal
pseudo-diffeomorphisms in $Diff\, \Sigma_{\tau}$. Instead, the
gauge transformations induced by the super-hamiltonian generator
$\int d^3\sigma N(\tau ,\vec \sigma )\,$ ${\tilde {\cal H}}(\tau
,\vec \sigma )$ do not reproduce the infinitesimal diffeomorphisms
in $Diff\, M^4$ normal to $\Sigma_{\tau}$ (see Ref.\cite{111}).
For a further clarification of the connection between space-time
diffeomorphisms and Hamiltonian gauge transformations see
Ref.\cite{44}.

\vfill\eject

\section{Congruences of Time-Like Accelerated Observers.}

In this Appendix we consider the point of view of the special
(non-rotating, surface-forming) congruence of time-like
accelerated observers whose 4-velocity field is the field of unit
normals to the space-like hyper-surfaces $\Sigma_{\tau}$; as
evolution parameter of the Hamiltonian description we use the
parameter $\tau$, labeling the leaves of the
foliation\footnote{According to Ref.\cite{86}, this is the {\it
hyper-surface point of view}. The {\it threading point of view} is
instead a description involving only a rotating congruence of
observers: since the latter is rotating, it is not surface-forming
(non-zero vorticity) and in each point we can only divide the
tangent space in the direction parallel to the 4-velocity and the
orthogonal complement (the local rest frame). On the other hand,
the {\it slicing point of view}, originally adopted in ADM
canonical gravity, uses two congruences: the non-rotating one with
the normals to $\Sigma_{\tau}$ as 4-velocity fields and a second
(rotating, non-surface-forming) congruence of observers, whose
4-velocity field is the field of time-like unit vectors determined
by the $\tau$ derivative of the embeddings identifying the leaves
$\Sigma_{\tau}$ (their so-called evolution vector field).
Furthermore, as Hamiltonian evolution parameter it uses the affine
parameter describing the world-lines of this second family of
observers.}.

We want to describe this non-rotating Hamiltonian congruence, by
emphasizing its interpretation in terms of gauge variables and DO.
The field of contravariant and covariant unit normals to the
space-like hyper-surfaces $\Sigma_{\tau}$ are expressed only in
terms of the lapse and shift gauge variables (as in Sections III
and IV, we use coordinates adapted to the foliation: $l^A(\tau
,\vec \sigma ) = b^A_{\mu}(\tau ,\vec \sigma )\, l^{\mu}(\tau
,\vec \sigma )$ with the $b^A_{\mu}(\tau ,\vec \sigma ) =
{{\partial \sigma^A}\over {\partial z^{\mu}}}$ being the
transition coefficients from adapted to general coordinates )

 \bea
 &&l^A(\tau ,\vec \sigma ) =  {1\over {N(\tau ,\vec \sigma
 )}}\, \Big( 1; -N^r(\tau ,\vec \sigma )\Big),\nonumber \\
 &&l_A(\tau ,\vec \sigma ) = N(\tau ,\vec \sigma )\, \Big( 1;
 0\Big),\qquad l^A(\tau ,\vec \sigma )\, l_A(\tau ,\vec \sigma ) =
 1.
 \label{b1}
 \eea

Since this congruence is surface forming by construction, it has
zero vorticity and is non-rotating (in the sense of congruences).
According to footnote 18, in Christodoulou-Klainermann spacetimes
we have $N(\tau ,\vec \sigma ) = \epsilon + n(\tau ,\vec \sigma
)$, $N^r(\tau ,\vec \sigma ) = n^r(\tau ,\vec \sigma )$. The
specific time-like direction identified by the normal has
inertial-like nature, in the sense of being dependent on
Hamiltonian gauge variables only. Therefore the world-lines of the
observers of this foliation\footnote{ It is called the
Wigner-Sen-Witten (WSW) foliation\cite{44} due to its properties
at spatial infinity. The associated observers are called {\it
Eulerian observers} when a perfect fluid is present as dynamical
matter.} change on-shell going from a 4-coordinate system to
another. On the other hand, the embeddings $z^{\mu}_{\Sigma}(\tau
,\vec \sigma )$ of the leaves $\Sigma_{\tau}$ of the WSW foliation
in space-time\footnote{ See Ref.\cite{44}, Eqs.(12.8) and (12.9),
for the way in which the embedding of the WSW foliation
corresponding to a solution of Einstein equations has to be
determined in every 4-coordinate system corresponding on-shell to
a completely fixed Hamiltonian gauge. As said in footnote 18,
there are asymptotic inertial observers corresponding to the {\it
fixed stars} that can be endowed with a spatial triad. Then the
asymptotic tetrad formed by the ADM 4-momentum and the spatial
triad can be transported in a dynamical way (on-shell) by using
the Sen-Witten connection \cite{113} in the Frauendiener
formulation \cite{114} in every point of $\Sigma_{\tau}$. This
defines a {\it local compass of inertia} to be compared with the
local gyroscopes (whether Fermi-Walker transported or not). The
WSW local compass of inertia consists in pointing to the fixed
stars with a telescope. It is needed in a satellite like Gravity
Probe B to detect the frame dragging (or gravitomagnetic
Lense-Thirring effect) of the inertial frames by means of the
rotation of a FW transported gyroscope relative to it.} depend on
{\it both} the DO and the gauge variables.

If $x^{\mu}_{{\vec \sigma}_o}(\tau )$ is the time-like world-line
of the observer crossing the leave $\Sigma_{\tau_o}$ at ${\vec
\sigma}_o$, we have \footnote{ Note that the mathematical time
parameter $\tau$ labeling the leaves of the foliation is not in
general the proper time of any observer of the congruence.}

\bea
 x^{\mu}_{{\vec \sigma}_o}(\tau ) &=& z^{\mu}_{\Sigma}(\tau ,
 {\vec \rho}_{{\vec \sigma}_o}(\tau )),\quad with\,\, {\vec
 \rho}_{{\vec \sigma}_o}(\tau_o) = {\vec \sigma}_o,\qquad
 {\dot x}^{\mu}_{{\vec \sigma}_o}(\tau ) = {{d x^{\mu}_{{\vec
 \sigma}_o}(\tau )}\over {d \tau}},\nonumber \\
 &&{}\nonumber \\
 l^{\mu}_{{\vec \sigma}_o}(\tau ) &=& l^{\mu}(\tau , {\vec
 \rho}_{{\vec \sigma}_o}(\tau )) = {{{\dot x}^{\mu}_{{\vec
 \sigma}_o}(\tau )}\over {\sqrt{{}^4g_{\alpha\beta}(x_{{\vec \sigma}_o}(\tau ))\,
 {\dot x}^{\alpha}_{{\vec \sigma}_o}(\tau )\, {\dot x}^{\beta}_{{\vec \sigma}_o}(\tau
 )} }},\nonumber \\
 &&{}\nonumber \\
 &&a^{\mu}_{{\vec \sigma}_o}(\tau ) = {{d l^{\mu}_{{\vec
 \sigma}_o}(\tau )}\over {d \tau}},\qquad a^{\mu}_{{\vec
 \sigma}_o}(\tau )\, l_{{\vec \sigma}_o\, \mu}(\tau ) = 0.
 \label{b2}
 \eea

\noindent Here $a^{\mu}_{{\vec \sigma}_o}(\tau )$ is the
4-acceleration of the observer $x^{\mu}_{{\vec \sigma}_o}(\tau )$.

As for any congruence, we have the decomposition ($P_{\mu\nu} =
\eta_{\mu\nu} - l_{\mu}\, l_{\nu}$)

\bea
 {}^4\nabla_{\mu}\, l_{\nu} &=&  l_{\mu}\, a_{\nu} + {1\over
3}\, \Theta \, P_{\mu\nu} + \sigma_{\mu\nu} +
\omega_{\mu\nu},\nonumber \\
 &&a^{\mu} = l^{\nu}\, {}^4\nabla_{\nu}\, l^{\mu} = {\dot l}^{\mu},
 \nonumber \\
 &&\Theta = {}^4\nabla_{\mu}\, l^{\mu},\nonumber \\
 &&\sigma_{\mu\nu} = {1\over 2}\, (a_{\mu}\, l_{\nu} + a_{\nu}\,
 l_{\mu}) + {1\over 2}\, ({}^4\nabla_{\mu}\,
l_{\nu} + {}^4\nabla_{\nu}\, l_{\mu}) - {1\over 3}\, \Theta\,
P_{\mu\nu},\nonumber \\
 && with magnitude \sigma^2={1\over 2} \sigma_{\mu\nu}\sigma^{\mu\nu}),\nonumber \\
 &&\omega_{\mu\nu} = - \omega_{\nu\mu} = \epsilon_{\mu\nu\alpha\beta}\,
 \omega^{\alpha}\, l^{\beta} = {1\over 2}\, (a_{\mu}\, l_{\nu} - a_{\nu}\,
 l_{\mu}) + {1\over 2}\, ({}^4\nabla_{\mu}\, l_{\nu} - {}^4\nabla_{\nu}\,
  l_{\mu}) = 0,\nonumber \\
 && \omega^{\mu} = {1\over 2}\, \epsilon^{\mu\alpha\beta\gamma}\, \omega_{\alpha\beta}\,
l_{\gamma} = 0,
 \label{b3}
 \eea

\noindent where $a^{\mu}$ is the 4-acceleration, $\Theta$ the {\it
expansion} (it measures the average expansion of the
infinitesimally nearby world-lines surrounding a given world-line
in the congruence), $\sigma _{\mu\nu}$ the {\it shear} (it
measures how an initial sphere in the tangent space to the given
world-line, which is Lie transported along $l^{\mu}$ \footnote{It
has zero Lie derivative with respect to $l^{\mu}\,
\partial_{\mu} $.}, is distorted towards an ellipsoid with principal axes
given by the eigenvectors of $\sigma^{\mu}{}_{\nu}$, with rate
given by the eigenvalues of $\sigma^{\mu}{}_{\nu}$) and
$\omega_{\mu\nu}$ the {\it twist or vorticity} (it measures the
rotation of the nearby world-lines infinitesimally surrounding the
given one); $\sigma_{\mu\nu}$ and $\omega _{\mu\nu}$ are purely
spatial ($\sigma_{\mu\nu} l^{\nu} = \omega_{\mu\nu} l^{\nu} = 0$).
Due to the Frobenius theorem, the congruence is (locally)
hyper-surface orthogonal if and only if $\omega_{\mu\nu}=0$. The
equation ${1\over l}\, l^{\mu}\, \partial_{\mu}\, l = {1\over 3}\,
\Theta$ defines a representative length $l$ along the world-line
of $l^{\mu}$, describing the volume expansion (or contraction)
behaviour of the congruence.

While all these quantities depend on the Hamiltonian gauge
variables, the expansion and the shear depend a priori also upon
the DO, because  the covariant derivative is used in their
definition.

Yet, the ADM canonical formalism provides {\it additional
information}. Actually, on each space-like hyper-surface
$\Sigma_{\tau}$ of the foliation, there is a {\it privileged
contravariant space-like direction} identified by the lapse and
shift gauge variables \footnote{The unit vector ${\cal N}
^{\mu}(\tau ,\vec \sigma )$ contains a DO dependence in the
overall normalizing factor. The existence of this space-like gauge
direction seems to indicate that {\it synchronous or time
orthogonal} 4-coordinates with $N_r(\tau ,\vec \sigma ) = -
{}^4g_{\tau r}(\tau ,\vec \sigma ) = 0$ (absence of
gravito-magnetism) have singular nature \cite{115}. Note that the
evolution vector of the {\it slicing point of view} has $N(\tau
,\vec \sigma )\, l^{\mu}(\tau ,\vec \sigma )$ and $|\vec N(\tau
,\vec \sigma )|\, {\cal N}^{\mu}(\tau ,\vec \sigma )$ as
projections along the normal and the plane tangent to
$\Sigma_{\tau}$, respectively.}

\bea
 {\cal N}^{\mu}(\tau ,\vec \sigma ) &=& {1\over {|\vec N(\tau
,\vec \sigma )|}}\, \Big( 0; n^r(\tau ,\vec \sigma
)\Big),\nonumber \\
 {\cal N}_{\mu}(\tau ,\vec \sigma ) &=& |\vec N(\tau ,\vec \sigma
 )|\, \Big( 1; {{N_r(\tau ,\vec \sigma )}\over {|\vec N(\tau ,\vec
 \sigma )|^2}}\Big),\nonumber \\
 &&{\cal N}^{\mu}(\tau ,\vec \sigma )\, l_{\mu}(\tau ,\vec \sigma
 ) = 0,\qquad {\cal N}^{\mu}(\tau ,\vec \sigma )\, {\cal N}_{\mu}(\tau ,\vec \sigma )
 = - 1,\nonumber \\
 && |\vec N(\tau ,\vec \sigma )| = \sqrt{({}^3g_{rs}\, N^r\, N^s)(\tau ,\vec \sigma
 )}.
 \label{b4}
 \eea

If 4-coordinates, corresponding to an on-shell complete
Hamiltonian gauge fixing, exist such that the vector field defined
by ${\cal N}^{\mu}(\tau ,\vec \sigma )$ on each $\Sigma_{\tau}$ is
surface-forming (zero vorticity\footnote{ This requires that
${\cal N}_{\mu}\, dx^{\mu}$ is  a closed 1-form, namely that in
adapted coordinates we have $\partial_{\tau}\, {{N_r}\over {|\vec
N|}} =
\partial_r\, |\vec N|$ and $\partial_r\, {{N_s}\over {|\vec N|}} =
\partial_s\, {{N_r}\over {|\vec N|}} $. This requires in turn
${{N_r}\over {|\vec N|}} = \partial_r\, f$ with $\partial_{\tau}\,
f = |\vec N| + const.$}), then each $\Sigma_{\tau}$ can be
foliated with 2-surfaces, and the 3+1 splitting of space-time
becomes a (2+1)+1 splitting corresponding to the 2+2 splittings
studied by Stachel and d'Inverno \cite{116}.

We have therefore a natural candidate for {\it one} of the three
spatial vectors of each observer, namely: $E^{\mu}_{{\vec
\sigma}_o\, ({\cal N})}(\tau ) = {\cal N}^{\mu}_{{\vec
\sigma}_o}(\tau ) = {\cal N}^{\mu}(\tau ,{\vec \rho}_{{\vec
\sigma}_o}(\tau ))$. By means of $l^{\mu}_{{\vec \sigma}_o}(\tau )
= l^{\mu}(\tau ,{\vec \rho}_{{\vec \sigma}_o}(\tau ))$ and ${\cal
N}^{\mu}_{{\vec \sigma}_o}(\tau )$, we can construct two {\it null
vectors} at each space-time point

\bea
 &&{\cal K}^{\mu}_{{\vec \sigma}_o}(\tau ) = \sqrt{{{|\vec N|}\over
 2}}\, \Big( l^{\mu}_{{\vec \sigma}_o}(\tau ) + {\cal N}^{\mu}_{{\vec
\sigma}_o}(\tau ) \Big),\nonumber \\
 &&{\cal L}^{\mu}_{{\vec \sigma}_o}(\tau ) = {1\over {\sqrt{2\, |\vec
 N|}}}\, \Big( l^{\mu}_{{\vec \sigma}_o}(\tau ) - {\cal N}^{\mu}_{{\vec
\sigma}_o}(\tau ) \Big).
 \label{b5}
 \eea

\noindent and then get a {\it null tetrad} of the type used in the
Newman-Penrose formalism \cite{38}. The last two axes of the
spatial triad can be chosen as two space-like circular complex
polarization vectors $E^{\mu}_{{\vec \sigma}_o\, (\pm )}(\tau )$,
like in electromagnetism. They are built starting from the
transverse helicity polarization vectors $E^{\mu}_{{\vec
\sigma}_o\, (1,2)}(\tau )$, which are the first and second columns
of the standard Wigner helicity boost generating ${\cal
K}^{\mu}_{{\vec \sigma}_o}(\tau )$ from the reference vector
${\buildrel \circ \over {\cal K}}{}^{\mu}_{{\vec \sigma}_o}(\tau )
= |\vec N|\, \Big( 1; 001\Big)$ (see for instance the Appendices
of Ref.\cite{117}).

Let us call $E^{(ADM) \mu}_{{\vec \sigma}_o\, (\alpha )}(\tau )$
the ADM tetrad formed by $l^{\mu}_{{\vec \sigma}_o}(\tau )$,
${\cal N}^{\mu}_{{\vec \sigma}_o}(\tau )$, $E^{\mu}_{{\vec
\sigma}_o\, (1,2)}(\tau )$ \footnote{It is a tetrad in adapted
coordinates: if $E^{\mu}_{(\alpha )} = {{\partial
z^{\mu}_{\Sigma}}\over {\partial \sigma^A}}\, E^A_{(\alpha )}$,
then $E^{(ADM)\, A}_{{\vec \sigma}_o\, (\alpha )}(\tau )\,
{}^4g_{AB}(\tau , {\vec \rho}_{{\vec \sigma}_o}(\tau ))\,
E^{(ADM)\, B}_{{\vec \sigma}_o\, (\beta )}(\tau ) =
{}^4\eta_{(\alpha )(\beta )}$.}. This tetrad will not be in
general Fermi-Walker transported along the world-line
$x^{\mu}_{{\vec \sigma}_o}(\tau )$ of the observer\footnote{Given
the 4-velocity $l^{\mu}_{{\vec \sigma}_o}(\tau ) = E^{\mu}_{{\vec
\sigma}_o}(\tau )$ of the observer, the spatial triads
$E^{\mu}_{{\vec \sigma}_o\, (a)}(\tau )$, $a = 1,2,3$, have to be
chosen in a conventional way, namely by means of a conventional
assignement of an origin for the local measurements of rotations.
Usually, the choice corresponds to Fermi-Walker (FW) transported
({\it gyroscope-type transport, non-rotating observer}) tetrads
$E^{(FW)\, \mu}_{{\vec \sigma}_o\, (\alpha )}(\tau )$, such that

\begin{eqnarray*}
{D\over {D\tau}}\, E^{(FW)\, \mu}_{{\vec \sigma}_o\, (a)}(\tau )
&=& \Omega^{(FW)}_{{\vec \sigma}_o}{}^{\mu}{}_{\nu}(\tau )\,
E^{(FW)\, \nu}_{{\vec \sigma}_o\, (a)}(\tau ) = l^{\mu}_{{\vec
\sigma}_o}(\tau )\, a_{{\vec \sigma}_o\, \nu}(\tau )\, E^{(FW)\,
\nu}_{{\vec \sigma}_o\, (a)}(\tau ),\nonumber \\
 &&\Omega^{(FW)}_{{\vec \sigma}_o}{}^{\mu\nu}(\tau ) =
 a^{\mu}_{{\vec \sigma}_o}(\tau )\, l^{\nu}_{{\vec \sigma}_o}(\tau
 ) - a^{\nu}_{{\vec \sigma}_o}(\tau )\, l^{\mu}_{{\vec \sigma}_o}(\tau
 ).
 \end{eqnarray*}

\noindent The triad $E^{(FW)\, \mu}_{{\vec \sigma}_o\, (a)}(\tau
)$ is the correct relativistic generalization of {\it global
Galilean non-rotating frames} (see Ref.\cite{82}) and is defined
using only local geometrical and group-theoretical concepts. Any
other choice of the triads (Lie transport, co-rotating-FW
transport,...) is obviously also possible \cite{86}. A generic
triad $E^{\mu}_{{\vec \sigma}_o\, (a)}(\tau )$ will satisfy
${D\over {D\tau}}\, E^{\mu}_{{\vec \sigma}_o\, (a)}(\tau ) =
\Omega_{{\vec \sigma}_o}{}^{\mu}{}_{\nu}(\tau )\, E^{\nu}_{{\vec
\sigma}_o\, (a)}(\tau )$ with $\Omega^{\mu\nu}_{{\vec \sigma}_o} =
\Omega^{(FW)\, \mu\nu}_{{\vec \sigma}_o} + \Omega^{(SR)\,
\mu\nu}_{{\vec \sigma}_o}$ with the spatial rotation part
$\Omega^{(SR)\, \mu\nu}_{{\vec \sigma}_o} =
\epsilon^{\mu\nu\alpha\beta}\, l_{{\vec \sigma}_o\, \alpha}\,
J_{{\vec \sigma}_o\, \beta}$, $J^{\mu}_{{\vec \sigma}_o}\,
l_{{\vec \sigma}_o\, \mu} = 0$, producing a rotation of the
gyroscope in the local space-like 2-plane orthogonal to
$l^{\mu}_{{\vec \sigma}_o}$ and $J^{\mu}_{{\vec \sigma}_o}$.}.

Another possible (but only on-shell) choice of the spatial triad
together with the unit normal to $\Sigma_{\tau}$ is the {\it local
WSW (on-shell) compass of inertia}, namely the triads transported
with the Frauendiener-Sen-Witten transport (see footnote 73 and
Eq.(12.2) of Ref.\cite{44}) starting from an asymptotic
conventional triad (choice of the fixed stars) added to the ADM
4-momentum at spatial infinity. As shown in Eq.(12.3) of
Ref.\cite{44}, they have the expression $E^{(WSW) \mu}_{{\vec
\sigma}_o (a)}(\tau) = {{\partial z^{\mu}_{\Sigma}}\over {\partial
\sigma^s}} {|}_{x_{{\vec \sigma}_o}(\tau )}\, {}^3e^{(WSW)
s}_{{\vec \sigma}_o\, (a)}(\tau )$ where the triad
${}^3e^{(WSW)}_{{\vec \sigma}_o (a)}$ is solution of the
Frauendiener-Sen-Witten equation restricted to a solution of
Einstein equations.

Given an observer with world-line $x^{\mu}_{{\vec \sigma}_o}(\tau
)$ and tetrad $E^{\mu}_{{\vec \sigma}_o\, (\alpha )}(\tau )$, the
geometrical properties are described by the Frenet-Serret
equations \cite{118}

\bea
 &&{D\over {D\tau}}\, l^{\mu}_{{\vec \sigma}_o}(\tau ) =
 \kappa_{{\vec \sigma}_o}(\tau )\, E^{\mu}_{{\vec \sigma}_o\,
 (1)}(\tau ),\nonumber \\
 &&{D\over {D\tau}}\,  E^{\mu}_{{\vec \sigma}_o\, (1)}(\tau ) =
 a^{\mu}_{{\vec \sigma}_o}(\tau ) =
 \kappa_{{\vec \sigma}_o}(\tau )\, l^{|mu}_{{\vec \sigma}_o}(\tau
 ) + \tau_{{\vec \sigma}_o\, (1)}(\tau )\,  E^{\mu}_{{\vec \sigma}_o\,
 (2)}(\tau ),\nonumber \\
 &&{D\over {D\tau}}\,  E^{\mu}_{{\vec \sigma}_o\, (2)}(\tau ) = -
 \tau_{{\vec \sigma}_o\, (1)}(\tau )\,  E^{\mu}_{{\vec \sigma}_o\,
 (1)}(\tau ) + \tau_{{\vec \sigma}_o\, (2)}(\tau )\,  E^{\mu}_{{\vec \sigma}_o\,
 (3)}(\tau ),\nonumber \\
 &&{D\over {D\tau}}\,  E^{\mu}_{{\vec \sigma}_o\, (3)}(\tau ) = -
 \tau_{{\vec \sigma}_o\, (2)}(\tau )\,  E^{\mu}_{{\vec \sigma}_o\,
 (2)}(\tau ),
 \label{b6}
 \eea

\noindent where $\kappa_{{\vec \sigma}_o}(\tau )$, $\tau_{{\vec
\sigma}_o (a)}(\tau )$, $a = 1,2$, are the curvature and the first
and second torsion of the world-line. $ E^{\mu}_{{\vec \sigma}_o\,
(a)}(\tau )$, $a = 1,2,3$ are said the normal and the first and
second bi-normal of the world-line, respectively.
\bigskip

Let us now look at the description of a geodesics $y^{\mu}(\tau
)$, the world-line of a scalar test particle, from the point of
view of those observers $\gamma_{{\vec \sigma}_o, y(\tau )}$ of
the congruence who intersect it, namely such that at $\tau$ it
holds $x^{\mu}_{{\vec \sigma}_o, y(\tau )}(\tau ) = y^{\mu}(\tau
)$. The family of these observers is called a {\it relative
observer world 2-sheet} in Ref.\cite{86}.

Since the parameter $\tau$ labeling the leaves $\Sigma_{\tau}$ of
the foliation is not the proper time $s = s(\tau )$ of the
geodesics $y^{\mu}(\tau ) = Y^{\mu}(s(\tau ))$, the geodesics
equation ${{d^2 Y^{\mu}(s)}\over {ds^2}} +
{}^4\Gamma^{\mu}_{\alpha\beta}(Y(s))\, {{dY^{\alpha}(s)}\over
{ds}}\, {{d Y^{\beta}(s)}\over {ds}} = 0$ (or $m\, a^{\mu}(s) =
m\, {{d^2 Y^{\mu}(s)}\over {ds^2}} = F^{\mu}(s)$, where $m$ is the
mass of the test particle), becomes

\bea
 &&{{d^2 y^{\mu}(\tau )}\over {d\tau^2}} +
{}^4\Gamma^{\mu}_{\alpha\beta}(y(\tau ))\,
{{dy^{\alpha}(\tau)}\over {d\tau}}\, {{d y^{\beta}(\tau )}\over
{d\tau}} - {{dy^{\mu}(\tau )}\over {d\tau}}\, {{d^2 s(\tau
)}\over{d\tau^2}}\, \Big( {{d s(\tau )}\over {d\tau}}\Big)^{-1} =
0,\nonumber \\
 &&{}
 \label{b7}
  \eea

or

\bea
 && m\, a_y^{\mu}(\tau ) = m\, {{d^2 y^{\mu}(\tau )}\over {d\tau^2}}
 =  f^{\mu}(\tau ).
 \label{b8}
 \eea

\noindent We see that the force $f^{\mu}(\tau )$ contains an
extra-piece with respect to $F^{\mu}(s(\tau ))$, due to the change
of time parameter. \vskip 0.5truecm

Let $U^{\mu}(\tau ) = V^{\mu}(s(\tau )) = {{d Y^{\mu}(s)}\over
{ds}} {|}_{s = s(\tau )} = {{{\dot y}^{\mu}(\tau)}\over
{\sqrt{{}^4g_{\alpha\beta}(y(\tau ))\, {\dot y}^{\alpha}(\tau )\,
{\dot y}^{\beta}(\tau ) }}}$ with ${\dot y}^{\mu}(\tau ) =
{{dy^{\mu}(\tau )}\over {d\tau}}$ be the 4-velocity of the test
particle and $ds = \sqrt{{}^4g_{\alpha\beta}(y(\tau ))\, {\dot
y}^{\alpha}(\tau )\, {\dot y}^{\beta}(\tau ) }\, d\tau$ be the
relation between the two parameters. By using the {\it intrinsic
or absolute derivative} along the geodesics parametrized with the
proper time $s = s(\tau )$ , the geodesics equation becomes ${\cal
A}^{\mu}(s) = {{D V^{\mu}(s)}\over {ds}} = 0$ [or ${\tilde {\cal
A}}^{\mu}(\tau ) = {{D U^{\mu}(\tau )}\over {d\tau}} =
{{dy^{\mu}(\tau )}\over {d\tau}}\, {{d^2 s(\tau
)}\over{d\tau^2}}\, \Big( {{d s(\tau )}\over {d\tau}}\Big)^{-1} =
g^{\mu}(\tau ) $].

In non-relativistic physics {\it spatial inertial forces are
defined as minus the spatial relative accelerations}, with respect
to an accelerated {\it global Galilean frame} (see Ref.\cite{82}).
In general relativity one needs the whole relative observer world
2-sheet to define an abstract 3-path in the quotient space of
space-time by the observer-family world-lines, representing the
trajectory of the test particle in the observer {\it 3-space}.
Moreover, a well defined projected {\it time derivative} is needed
to define a relative acceleration associated to such 3-path. At
each point $P(\tau )$ of the geodesics, identified by a value of
$\tau$, we have the two vectors $U^{\mu}(\tau )$ and
$l^{\mu}_{{\vec \sigma}_o\, y(\tau )}(\tau )$. Therefore, each
vector $X^{\mu}$ in the tangent space to space-time in that point
$P(\tau )$ admits two splittings:
\bigskip

i) $X^{\mu} = X_U\, U^{\mu} + P(U)^{\mu}{}_{\nu}\, X^{\nu}$,
$P^{\mu\nu}(U) = {}^4g^{\mu\nu} - U^{\mu}\, U^{\nu}$, i.e., into a
temporal component along $U^{\mu}$ and a spatial transverse
component, living in the local rest frame $LRS_U$;

ii) $X^{\mu} = X_l\, l^{\mu}_{{\vec \sigma}_o y(\tau )} +
P(l_{{\vec \sigma}_o y(\tau )})^{\mu}{}_{\nu}\, X^{\nu}$, i.e.,
into a temporal component along $l^{\mu}_{{\vec \sigma}_o y(\tau
)}(\tau )$ and a spatial transverse component, living in the local
rest frame $LRS_l$, which is the plane tangent to the leave
$\Sigma_{\tau}$ in $P(\tau )$ for our surface-forming congruence.

The {\it measurement} of $X^{\mu}$ by the observer congruence
consists in determining the scalar $X_l$ and the spatial
transverse vector. In adapted coordinates and after a choice of
the spatial triads, the spatial transverse vector is described by
the three (coordinate independent) tetradic components $X_{(a)} =
E^{\mu}_{(a)}\, X_{\mu}$. The same holds for every tensor.
Moreover, every spatial vector like $ P(U)^{\mu}{}_{\nu}\,
X^{\nu}$ in $LRS_U$ admits a 2+1 orthogonal decomposition ({\it
relative motion orthogonal decomposition}) into a component in the
2-dimensional rest subspace $LRS_U \cap LRS_L$ transverse to the
direction of relative motion and one component in the
1-dimensional (longitudinal) orthogonal complement along the
direction of the relative motion in each such rest space.

At each point $P(\tau)$, the tangent space is split into the {\it
relative observer 2-plane} spanned by $U^{\mu}(\tau )$ and
$l^{\mu}_{{\vec \sigma}_o y(\tau )}(\tau )$ and into an orthogonal
space-like 2-plane. We have the 1+3 orthogonal decomposition

\bea
 U^{\mu}(\tau ) &=& \gamma (U,l)(\tau )\, \Big( l^{\mu}_{{\vec
 \sigma}_o\, y(\tau )}(\tau ) + \nu^{\mu}(U,l)(\tau
 )\Big),\nonumber \\
 &&\gamma (U,l) = U_{\mu}\, l^{\mu}_{{\vec \sigma}_o y(\tau
 )},\qquad \nu(U,l) = \sqrt{\nu^{\mu}(U,l)\, \nu_{\mu}(U,l)},\nonumber \\
 &&{\hat \nu}^{\mu}(U,l) = {{\nu^{\mu}(U,l)}\over {\nu
 (U,l)}},\qquad relative\,\,\, 4-velocity\,\,\, tangent\, to\, \Sigma_{\tau}.
 \label{b9}
 \eea

\bigskip

The equation of geodesics, written as $m\, {\cal A}^{\mu}(s) = 0$,
is described by the observers' family as:

i) a temporal projection along $l^{\mu}_{{\vec \sigma}_o y(\tau
)}$, leading to the evolution equation $m\, {\cal A}_{\mu}\,
l^{\mu}_{{\vec \sigma}_o y(\tau )} = 0$, for the observed energy
($E(U,l) = \gamma (U,l)$) of the test particle along its
world-line;

ii) a spatial projection orthogonal to $l^{\mu}_{{\vec \sigma}_o
y(\tau )}$ (tangent to $\Sigma_{\tau}$), leading to the evolution
equation for the observed 3-momentum of the test particle along
its world-line, with the kinematical quantities describing the
motion of the family of observers entering as {\it inertial
forces}. If, instead of writing $m\, P(l)^{\mu}{}_{\nu}\, {\cal
A}^{\mu}(s) = 0$ with $P(l)^{\mu\nu} = {}^4g^{\mu\nu} -
l^{\mu}_{{\vec \sigma}_o y(\tau )}\, l^{\nu}_{{\vec \sigma}_o
y(\tau )}$, we rescale the particle proper time $s(\tau )$ to the
sequence of observer proper times $s_{(U,l)}$ defined by ${{d
s_{(U,l)}}\over {ds}} = \gamma (U,l)$, the spatial projection of
the geodesics equation, re-scaled with the gamma factor
\footnote{Namely $m\, \gamma^{-1}(U,l)\, P(l)^{\mu}{}_{\nu}\,
{\cal A}^{\nu} = 0$.}, can be written in the form

\bea
 m\, \Big( {{D_{(FW)}(U,l)\,}\over {d s_{(U,l)}}}\Big)^{\mu}{}_{\nu}\,
 v^{\nu}(U,l) &=& m\, a^{\mu}_{(FW)}(U,l) = F_{(FW)}^{(G)\mu}(U,l),\nonumber \\
 &&{}\nonumber \\
 F^{(G)\mu}_{(FW)}(U,l) &=& - \gamma (U,l)^{-1}\,
 P^{\mu}{}_{\nu}(l)\, {{D l^{\nu}_{{\vec \sigma}_o y(\tau )}(\tau (s))}\over
 {ds}} =\nonumber \\
 &=& - \Big( {{D_{(FW)}(U,l)}\over {ds_{(U,l)}}}\Big)^{\mu}{}_{\nu} \,
 l^{\nu}_{{\vec \sigma}_o y(\tau )}(\tau (s_{(U,l)})) =\nonumber \\
 &=& \gamma (U,l)\, \Big[ - a^{\mu}(l) + \Big( -
 \omega^{\mu}{}_{\nu}(l) + \theta^{\mu}{}_{\nu}(l) \Big)
 \nu^{\nu}(U,l) \Big],
 \label{b10}
 \eea

\noindent where $v^{\mu}(U,l) = U^{\mu} - \gamma (U,l)\,
l^{\mu}_{{\vec \sigma}_o y(\tau )} = v(U,l)\, {\hat
\nu}^{\mu}(U,l)$ with $v(U,l) = \gamma (U,l)\, \nu (U,l)$, and
$P(l)^{\mu}{}_{\nu}\, {D\over {ds}} = \Big( {{D_{(FW)}(U,l)}\over
{ds}}\Big)^{\mu}{}_{\nu}$ is the spatial FW intrinsic derivative
along the test world-line and $a^{\mu}_{(FW)}(U,l)$ is the {\it FW
relative accelaration}. The term $F^{(G)\mu}_{(FW)}(U,l)$ can be
interpreted as the set of {\it inertial forces} due to the motion
of the observers themselves, as in the non-relativistic case. Such
inertial forces depend on the following congruence properties:
\bigskip

i) the acceleration vector field $a^{\mu}(l)$, leading to a {\it
gravito-electric field and a spatial gravito-electric
gravitational force};

ii) the vorticity $\omega^{\mu}{}_{\nu}(l)$ and expansion + shear
$\theta^{\mu}{}_{\nu}(l) = \sigma^{\mu}_{\nu}(l) + {1\over 3}\,
\Theta (l)\, P^{\mu}{}_{\nu}(l)$ mixed tensor fields, leading to a
{\it gravito-magnetic vector field and tensor field and a Coriolis
or gravito-magnetic force} linear in the relative velocity
$\nu^{\mu}(U,l)$. \vskip 0.5truecm

Then, by writing $v^{\mu}(U,l) = v(U,l)\, {\hat \nu}^{\mu}(U,l)$,
the FW relative acceleration can be decomposed into a longitudinal
and a transverse relative acceleration

\bea
 a^{\mu}_{(FW)}(U,l) &=& {{D_{(FW)}(U,l)\, v(U,l)}\over
 {ds_{(U,l)}}}\, {\hat \nu}^{\mu}(U,l) + \gamma (U,l)\,
 a^{(\perp)\mu}_{(FW)}(U,l),\nonumber \\
 &&{}\nonumber \\
 a^{(\perp)\mu}_{(FW)}(U,l) &=& v(U,l)\, \Big({{D_{(FW)}}\over
 {ds_{(U,l)}}}\Big)^{\mu}{}_{\nu}\, {\hat \nu}^{\nu}(U,l)
 =\nonumber \\
 &=& \nu^2(U,l)\, \Big( {{D_{(FW)}}\over
 {dr_{(U,l)}}}\Big)^{\mu}{}_{\nu}\, {\hat \nu}^{\nu}(U,l) =
 {{\nu^2(U,l)}\over {\rho_{(FW)}(U,l)}}\, {\hat
 \eta}^{\mu}_{(FW)}(U,l).
 \label{b11}
 \eea

\noindent In the second expression of the transverse FW relative
acceleration, the reparametrization ${{dr_{(U,l)}}\over
{ds_{(U,l)}}} = \nu (U,L)$ to a spatial arclength parameter has
been done. Since $\gamma (U,l)\, a^{(\perp)\mu}_{(FW)}(U,l)$ is
the transverse part of the relative acceleration, i.e. the {\it FW
relative centripetal acceleration}, $- m\, \gamma (U,l)\,
a^{(\perp)\mu}_{(FW)}(U,l)$ may be interpreted as a {\it
centrifugal force}, so that the geodesics equation is rewritten as
$m\, {{D_{(FW)}(U,l)\, v(U,l)}\over {ds_{(U,l)}}}\, {\hat
\nu}^{\mu}(U,l) = F^{(G)\mu}_{(FW)}(U,l) - m\, \gamma (U,l)\,
a^{(\perp)\mu}_{(FW)}(U,l)$, with the first member called
sometimes {\it Euler force}.

The 3-path in the abstract quotient space can be treated as an
ordinary 3-curve in a 3-dimensional Riemann space. Its tangent is
${\hat \nu}^{\mu}(U,l)$, while its normal and bi-normal are
denoted ${\hat \eta}^{\mu}_{(FW)}(U,l)$ and ${\hat
\xi}^{\mu}_{(FW)}(U,l)$ respectively. The 3-dimensional
Frenet-Serret equations are then

\bea
 \Big( {{D_{(FW)}(U,l)}\over {dr_{(U,l)}}}\Big)^{\mu}{}_{\nu}\,
 {\hat \nu}^{\nu}(U,l) &=& \kappa_{(FW)}(U,l)\, {\hat
 \eta}^{\mu}_{(FW)}(U,l),\nonumber \\
 \Big( {{D_{(FW)}(U,l)}\over {dr_{(U,l)}}}\Big)^{\mu}{}_{\nu}\,
 {\hat \eta}^{\nu}_{(FW)}(U,l) &=& - \kappa_{(FW)}(U,l)\, {\hat
 \nu}^{\mu}(U,l) + \tau_{(FW)}(U,l)\, {\hat
 \xi}^{\mu}_{(FW)}(U,l),\nonumber \\
  \Big( {{D_{(FW)}(U,l)}\over {dr_{(U,l)}}}\Big)^{\mu}{}_{\nu}\,
 {\hat \xi}^{\nu}_{(FW)}(U,l) &=& - \tau_{(FW)}(U,l)\, {\hat
 \eta}_{(FW)}^{\mu}(U,l),
 \label{b12}
 \eea

\noindent where $\kappa_{(FW)}(U,l) = 1/ \rho_{(FW)}(U,l)$ and
$\tau_{(FW)}(U,l)$ are the curvature and torsion of the 3-curve,
respectively.

The main drawback of the 1+3 ({\it threading}) description,
notwithstanding its naturality from a locally operational point of
view, is the use of a rotating congruence of time-like observers:
this introduces an element of non-integrability and, as yet, no
formulation of the Cauchy problem for the 1+3 reformulation of
Einstein's equations has been worked out.

\vfill\eject

\section{Axiomatic Foundations and Theory of Measurement in General Relativity.}

In this Appendix we review the axiomatic approach to the theory of
measurement in general relativity by means of idealized test
matter.

After a critique of the Synge's {\it chronometric} axiomatic
approach\cite{31} \footnote{Synge accepts as basic primitive
concepts {\it particles} and {\it (standard) clocks}. Then he
introduces the 4-metric as the fundamental structure, postulating
that whenever $x$, $x + dx$ are two nearby events contained in the
world line or history of a clock, then the separation associated
with $(x, x+dx)$ equals the time interval as measured by that (and
by other suitably scaled) clock. These axioms are good for the
{\it deduction} of the subsequent theory, but are not a good {\it
constructive} set of axioms for relativistic space-times
geometries. The Riemannian line element cannot be derived by
clocks alone without the use of light signals. The chronometric
determination of the 4-metric components does not compellingly
determine the behaviour of freely falling particles and light rays
and Synge has to add a further axiom (the geodesic hypothesis). On
the basis of this axiom it is then possible (Marzke \cite{99},
Kundt-Hoffmann \cite{119}) to construct clocks by means of freely
falling particles and light rays (i.e. to give a physical
interpretation of the 4-metric in terms of time). Therefore the
chronometric axioms appear either as redundant or, if the term
{\it clock} is interpreted as {\it atomic clock}, as a link
between macroscopic gravitation theory and atomic physics: these
authors claim for the equality of gravitational and atomic time.
It should be better to test this equality experimentally (in radar
tracking of planetary orbits atomic time has been used {\it only}
as an {\it ordering parameter}, whose relation to gravitational
time was to be determined from the observations) or to derive it
eventually from a theory that embraces both gravitational and
atomic phenomena, rather than to postulate it as an
axiom.},Ehlers, Pirani and Schild (see Ref.\cite{40}), reject {\it
clocks} as basic tools for setting up the space-time geometry and
propose to use {\it light rays} and {\it freely falling
particles}. The full space-time geometry can then be synthesized
from a few local assumptions about light propagation and free
fall.
\bigskip

a) The propagation of light determines at each point of space-time
the infinitesimal null cone and thus establishes its {\it
conformal structure} ${\cal C}$. In this way one introduces the
notions of being space-like, time-like and null and one can single
out as {\it null geodesics} the null curves contained in a null
hyper-surface (the light rays).

b) The motions of freely falling particles determine a family of
preferred ${\cal C}$-time-like curves. By assuming that this
family satisfies a generalized law of inertia (existence of local
inertial frames in free fall, equality of inertial and passive
gravitational mass), it follows that free fall defines a {\it
projective structure} ${\cal P}$ in space-time such that the world
lines of freely falling particles are the ${\cal C}$-time-like
geodesics of ${\cal P}$.

c) Since, experimentally, an ordinary particle (positive rest
mass), though slower than light, can be made to chase a photon
arbitrarily close, the conformal and projective structures of
space-time are {\it compatible}, in the sense that every ${\cal
C}$-null geodesic is also a ${\cal P}$-geodesic. This makes $M^4$
a {\it Weyl space} $(M^4, {\cal C}, {\cal P})$. A Weyl space
possesses a unique {\it affine structure} ${\cal A}$ such that
${\cal A}$-geodesics coincide with ${\cal P}$-geodesics and ${\cal
C}$-nullity of vectors is preserved under ${\cal A}$-parallel
displacement. In conclusion, light propagation and free fall
define a Weyl structure $(M^4, {\cal C}, {\cal A})$ on space-time
(this is equivalent to an {\it affine connection} due to the
presence of both the projective and the conformal structure).

d) In a Weyl space-time, one can define an {\it arc length}
(unique up to linear transformations) along any non-null curve.
Applying such definition to the time-like world line of a particle
$P$ (not necessarily freely falling), we obtain a {\it proper
time} (= arc length) $t$ on $P$, provided two events on $P$ have
been selected as {\it zero point} and {\it unit point of time}.
The (idealized) Kundt-Hoffmann experiment \cite{119} designed to
measure proper time along a time-like world line in Riemannian
space-time by means of light signals and freely falling particles
can be used without modifications to measure the proper time $t$
in a Weyl space-time.

e) In absence of a {\it second clock effect} \footnote{The first
clock effect is essentially the twin paradox effect. On the other
hand, if the time unit cannot be fixed for all standard clocks
simultaneously in a consistent way, Perlick \cite{94} speaks of a
{\it second clock effect}} a Weyl space $(M^4, {\cal C}, {\cal
A})$ becomes a Riemannian space, in the sense that there exists a
Riemannian 4-metric ${\cal M}$ compatible with ${\cal C}$ (i.e.
having the same null-cones) and having ${\cal A}$ as its metric
connection. The Riemannian metric is necessarily unique up to a
constant positive factor. Since ${\cal A}$ determines a {\it
curvature tensor} $R$, the use of the equation of geodesic
deviation shows that $(M^4, {\cal C}, {\cal A})$ is Riemannian if
and only if the proper times $t$, $t^{'}$ of two arbitrary,
infinitesimally close, freely falling particles $P$, $P^{'}$ are
{\it linearly} related (to first order in the distance) by {\it
Einstein simultaneity} (see Ref.\cite{40}). In Newtonian
space-time the role of ${\cal C}$ is played by the {\it absolute
time}. It is also easy to add a physically meaningful axiom that
singles out the space-time of special relativity, either by
requiring homogeneity and isotropy of $M^4$ with respect to
$({\cal C}, {\cal A})$, or by postulating vanishing relative
accelerations between arbitrary, neighboring, freely falling
particles.

Now, Perlick \cite{94} states that experimental data on standard
atomic clocks confirm the absence of the {\it second clock
effect}, so that our actual space-time is not Weyl but
pseudo-Riemannian and it is possible to introduce a notion of {\it
rigid rod}.

Let us note that the previous axiomatic approach should be
enlarged to cover tetrad gravity, because of the need of test
gyroscopes to define the triads of the tetrads of time-like
observers. Then the axiomatics would include the possibility of
measuring gravito-magnetism and would have to face the question of
whether or not the free fall of macroscopic test gyroscopes is
geodesic.

An associated theory of the measurement of time-like and
space-like intervals has been developed by Martzke-Wheeler
\cite{88,99}, using Schild geodesic clock (if it is a standard
clock, Perlick's definition of rigid rod can be used): {\it the
axiomatics is replaced by the empirical notion of a fiducial
interval as standard}. Pauri and Vallisneri \cite{120} have
further developed the Martzke-Wheeler approach, showing that,
given the {\it whole} world-line of an accelerated time-like
observer, it is possible to build an associated space-time
foliation with simultaneity space-like non-overlapping 3-surfaces.
This is to be contrasted with the local construction of Fermi
coordinate systems: it requires only a local knowledge of the
observer world-line but its validity is limited to a neighborhood
of the observer, determined by the acceleration radii,

As already said, material (test) reference fluids were introduced
by various authors \cite{90, 92,93} for simulating the axioms.

\vfill\eject

\end{document}